\documentclass[a4paper,12pt]{article}
\usepackage{amsmath,amsfonts,amssymb}

\usepackage[pdftex]{graphicx}
\usepackage{cancel}
\usepackage{bm}
\usepackage{colortbl}
\usepackage{color}
\usepackage{caption}
\numberwithin{equation}{section}

\def\beq#1\eeq{\begin{equation}#1\end{equation}}
\def\bes#1\ees{\begin{equation}\begin{split}#1
               \end{split}\end{equation}}
\def\bea#1\eea{\begin{align}#1\end{align}}
            
\newcommand{\m}{\hspace{-0.05cm}}
\newcommand{\timesm}{\m \times \m}

\newcommand{\ps}{\m + \m}
\newcommand{\ms}{\mbox{\scriptsize s}}

\font\mybb=msbm10 at 12pt

\font\mybbsmall=msbm10 at 10pt

\def\bb#1{\hbox{\mybb#1}}

\def\bbsmall#1{\hbox{\mybbsmall#1}}

\def\ZZ {\bb{Z}}

\def\PP {\bb{P}}
\def\PPsmall {\bbsmall{P}}

\newcommand\beqa{\begin{eqnarray}}
\newcommand\eeqa{\end{eqnarray}}
\newcommand\n{\nonumber\\}

\newcommand\fres{f_{\mbox{\scriptsize res}}}
\newcommand\gres{g_{\mbox{\scriptsize res}}}
\newcommand\Dres{\Delta_{\mbox{\scriptsize res}}}

\allowdisplaybreaks[0]

\begin{document} 
\baselineskip=5.3mm
\begin{titlepage}
\renewcommand{\thefootnote}{\fnsymbol{footnote}}
\nopagebreak
\vskip 5mm
\begin{flushright}
KEK-TH 2068
\end{flushright}

\vskip 10mm
\begin{center}
\baselineskip=10mm
{\Large
\textbf{%
Non-Cartan Mordell-Weil lattices of rational
elliptic surfaces and heterotic/F-theory
compactifications}}\\
\end{center}
\begin{center}
\vskip 10mm
Shun'ya \textsc{Mizoguchi}$^1$
\footnote{mizoguchi@post.kek.jp}
and Taro \textsc{Tani}$^2$
\footnote{tani@kurume-nct.ac.jp}
\vskip 5mm
\textsl{%
$^1$Theory Center, Institute of Particle and Nuclear Studies, KEK \\
Tsukuba, Ibaraki 305-0801, Japan \\\vspace{10pt}
$^2$National Institute of Technology, Kurume College \\
Kurume, Fukuoka 830-8555, Japan\\
}
\vspace{10pt}
\texttt{%
\footnotesize
$^1$mizoguchi@post.kek.jp \\
$^2$tani@kurume-nct.ac.jp
}
\end{center}
\vskip 7mm 

\begin{abstract}

The Mordell-Weil lattices (MW lattices) associated to 
rational elliptic surfaces are classified into 74 types.
Among them, there are cases in which the MW lattice is none
of the weight lattices of simple Lie algebras or direct sums thereof. 
We study how such ``non-Cartan MW lattices"
are realized in the six-dimensional heterotic/F-theory
compactifications.
In this paper, we focus on non-Cartan MW lattices that are
torsion free and whose associated singularity lattices are sublattices of $A_7$.
For the heterotic string compactification, a non-Cartan MW lattice yields
an instanton gauge group $H$ with one or more $U(1)$ group(s).
We give a method for computing massless spectra
via the index theorem and show that the $U(1)$ instanton number is limited
to be a multiple of some particular non-one integer.
On the F-theory side, we examine whether we can construct the corresponding threefold geometries, {\it i.e.},
rational elliptic surface fibrations over \scalebox{0.917}{$\displaystyle \PP^1$}.
Except for some cases, we obtain such geometries for specific distributions of instantons.
All the spectrum derived from those geometries completely match with the heterotic results.

\end{abstract}

\vfill
\end{titlepage}


\renewcommand{\thefootnote}{\arabic{footnote}}
\setcounter{footnote}{0}

\section{Introduction}
F-theory \cite{Vafa,MV1,MV2} has a unique feature 
in modern particle physics model building based on string theory.  
The $SU(5)$ GUT, which  
naturally explains the 
hypercharges of the observed quarks and leptons, 
and matter in the spinor representation of $SO(10)$, into which all the 
quarks and leptons of a single generation are successfully 
incorporated---both are readily achieved in F-theory.  
F-theory models have an advantage 
over the $E_8\times E_8$ heterotic models
as they may evade the issue 
of the GUT vs. Planck scales first addressed in \cite{Witten96}. 
F-theory can also generate 
Yukawa couplings that are perturbatively forbidden in D-brane models \cite{Dbranemodels}.

Almost a decade after the first formulation of F-theory, there was considerable 
development in understanding the local models \cite{DonagiWijnholt,BHV,BHV2,DonagiWijnholt2} 
in terms of twisted super Yang-Mills theories or the Higgs bundles \cite{HKTW,DWHiggsBundles}.
One of the notable findings in the development was 
the mechanism of the GUT gauge symmetry breaking 
by the fluxes turned on the brane. 
A large number of studies have been carried out in local models. 
An incomplete list includes \cite{localmodel1,localmodel2,localmodel3,localmodel4}.

Rather soon after this development, 
the Higgs particle was found 
at LHC in 2012, and the subsequent experiments showed that there 
was no low-energy supersymmetry. Later, the PLANCK data  
became also available to reveal that the energy of the inflation can be 
very high, even close to the GUT scale. These two new sources of 
knowledge have turned the focus of F-theory model building 
to global models. 
%
%
It is also known \cite{MV2} that $U(1)$ 
gauge symmetries in F-theory arise when the Mordell-Weil rank 
is nonzero, that is, when there are nontrivial global sections. 
This is in sharp contrast to nonabelian 
gauge symmetries, which can be solely determined by the singularity 
in the local model. Recent works on 
global models include \cite{globalmodel1,
AndreasCurio,
globalmodel2,
Collinucci,
BlumenhagenGrimmJurkeWeigand,
MarsanoSaulinaSchafer-NamekiThree-Generation,
BlumenhagenGrimmJurkeWeigand2,
MarsanoSaulinaSchaferNamekiU(1)PQ,
GrimmKrauseWeigand,
CveticGarcia-EtxebarriaHalverson,
ChenKnappKreuzerMayrhofer,
ChenChungE8point,
GrimmWeigand,
KnappKreuzerMayrhoferWalliser,
DolanMarsanoSaulinaSchaferNameki,
MarsanoSchaferNameki,
GrimmKerstanPaltiWeigand,
globalmodel3,globalmodel4,globalmodel5,globalmodel6,globalmodel7,
globalmodel8,globalmodel9,
CveticKleversPiragua,
BorchmannMayrhoferPaltiWeigand,
globalmodel10,globalmodel11,Oehlmann1,globalmodel12,
globalmodel13,globalmodel14,globalmodel15,Kimura:2018qsp,Oehlmann2}.

For K3 surfaces, the Mordell-Weil rank varies depending on its Picard 
number. In contrast, the Mordell-Weil rank of a rational 
elliptic surface is always 10. 
Its Mordell-Weil group is known to be endowed with 
a lattice structure, and the possible pairs of the singularity and 
the corresponding Mordell-Weil lattice 
have been classified into 74 types \cite{OguisoShioda}. 
Roughly speaking, the singularity lattice and 
the Mordell-Weil lattice are the orthogonal compliment of each other 
in the $E_8$ root lattice. 
%
In a typical case, the Mordell-Weil lattice coincides 
with a weight lattice of some semi-simple gauge group of 
the instantons in the dual heterotic string theory, and the singularity 
lattice is that of the unbroken gauge group. 
However, it is interesting to note that 
in the other cases the inner product matrix of the Mordell-Weil 
lattice is none of the (inverse of the) Cartan matrices of simple Lie algebras, 
nor is it their direct sum. 
It is these non-Cartan type Mordell-Weil lattices that 
we focus on in this paper. In fact,  these are the cases where 
the gauge instanton includes some $U(1)$ factor(s).\footnote{As there are 
no such things as  ``$U(1)$  instantons" in the ordinary 
four-dimensional noncompact Euclidean space, that might sound bizarre, but 
in a complex compact space they are nothing but line bundles with 
a nonzero first Chern class. } 

We are particularly interested in the explicit forms of the Weierstrass equations  
of this special class of rational elliptic surfaces with a section\,\footnote{
Incidentally, the very same objects were studied in 
late 90's as Seiberg-Witten curves for 
E-strings \cite{Lerche:1996ni,Ganor:1996pc,Minahan:1997ch,Minahan:1997ct,Yamada:1999xr,Eguchi:2002fc,Eguchi:2002nx},
though these rational elliptic surfaces were not supposed to be further 
fibered over anything then. A recent work on non-Cartan 
MW lattices for rational elliptic surfaces (and not their fibrations) is \cite{Yamadastudent}.}      
{\em which are fibered} over $\PP^1$ to form a complex threefold.
This means that the parameters of the Weierstrass equations are {\em sections}
of some line bundles over $\PP^1$ \cite{MV2}.\footnote{
Although we consider in this paper 
rational elliptic surfaces fibered over $\PPsmall^1$, the Weierstrass equations we obtain 
can be readily converted to those for a fibration over a complex two-fold $B_2$ 
by simply replacing a degree $an+b$ polynomial in the affine coordinate $z'$ of 
the $\PPsmall^1$, with a section  of  ${\cal L}^{-d}\otimes {\cal N}^s$ with 
$d =6a-\frac b2$ and $s = a$, where ${\cal L}$ is the anti-canonical bundle 
of $B_2$ and ${\cal N}$ is the twisting line bundle determining the 
normal bundle of the {\em fiber} $\PPsmall^1$, which is the base of the rational elliptic surface. 
For further explanation of this correspondence, see \cite{MTLooijenga}.
}
Each of these geometries is regarded as a part of an elliptic K3 fibration 
in the stable degeneration limit~\cite{Aspinwall:1997ye,Aspinwall:1997eh,Aspinwall:1998bw}\,,
and hence as a ``1/2 CY threefold'' since two such 
complex manifolds can be glued together into a K3-fibered Calabi-Yau threefold (CY${}_3$).

In this paper, we will specifically consider a class of rational elliptic surfaces, 
and their fibrations over $\PP^1$, which satisfy the following criteria:  
\begin{itemize}
\item[1.] The Mordel-Weil lattice $E(K)$ is neither a weight lattice of some semi-simple 
Lie algebra, nor is it a direct sum of such a weight lattice and a torsion.  
\item[2.] The singularity lattice $T$ is a sublattice of $A_7$. 
\end{itemize}
%
There are ten types of such rational elliptic surfaces 
in the Oguiso-Shioda classification, which are summarized in Table \ref{tab:MWLsu7su8} (See section \ref{sec:nonCartan}).

To find an equation representing a ``rational-elliptic-surface\,(RES)-fibered" 
complex threefold over $\PP^1$ whose fiber rational elliptic surface belongs to this class, 
our strategy is to realize in the geometry the singularity $G$ of lattice $T$ associated to a given $E(K)$.
%
%
First, we start from No.7 with singularity $G=SU(2)\times SU(2)\times SU(2)$, which is not a 
non-Cartan type, and construct the corresponding CY${}_3$ with K3 fibration.
We then successively tune the complex structures (``unHiggsing") 
to achieve the necessary singularities $G$ of the respective types given in Table\,\ref{tab:MWLsu7su8}.
Next, we map the obtained K3-fibered CY${}_3$'s to RES-fibered geometries.
We carefully construct this map so that it does not change the structure of the singularity.
Through this strategy, we can successfully obtain 
the equations for all the $SU(7)$ series (upper rows in Table \ref{tab:MWLsu7su8}) and 
some (No.22 and No.29) of the $SU(8)$ series (lower rows in Table \ref{tab:MWLsu7su8}). We will also discuss why it is 
hard to find the equations for the remaining cases. 
We note that the equations for $SU(8)$ series cannot be 
obtained by using Tate's form and we need to work in the Weierstrass form.
This is because, if the coefficients $a_i$ $(i=1,2,3,4,6)$ of Tate's form 
are assumed to have the necessary factors required from each singularity of the
$SU(8)$ series, then the total degrees of some $a_i$ (including $a_3$ and $a_6$) would exceed $i$,
leading additional unwanted singularities.

We also compute the massless spectra of 
the dual heterotic string compactifications whose vector bundles 
are supposed to be determined by the RES-fibered spaces 
above \cite{MV1,MV2,FMW}. 
For all cases except No.45 in Table\,\ref{tab:MWLsu7su8}, 
the instanton gauge group $H$ 
in the dual heterotic string theory 
has two or more irreducible group 
factors. In particular, we will see that $H$ specified by a non-Cartan 
Mordel-Weil lattice is typically a product of a semi-simple group and one or more $U(1)$ group(s). 
Thus one can distribute the total $12+n$ instanton numbers to each group factor.
Although in principle there is no problem in applying the index theorem to 
these cases, the subtlety is that the multiplicities of some massless 
hypermultiplet then become fractional {\em unless} the $U(1)$ 
instanton number is a multiple of some particular non-one integer.  
We will show why the $U(1)$ instanton number cannot take an 
arbitrary integral value but must be a multiple by examining the 
orthogonal decomposition of the $E_8$ root lattice.  This gives us a 
consistent integral number of hypermultiplets. 
For the cases of $\mbox{rank}(E(K)) > 2$, more than one choices 
of the number of $U(1)$ direction(s) are possible.
In such cases, we obtain more than one spectra for a given $E(K)$ (and hence for a given $G$).
We find that the spectrum is more general when the number of $U(1)$ direction(s) is larger.

While it is possible to compute heterotic indices for the cases of 
distributed instantons, it is a nontrivial problem to 
obtain the equations for the RES-fibered spaces corresponding 
to such particularly distributed instantons.
We will explain for the No.7 case, which is a Cartan type, how we can obtain the equations 
for an arbitrary distribution of instantons, and show the complete 
match of the six-dimensional massless spectra  
read off from the Weierstrass 
equations on the F-theory side and those obtained by the index 
computations on the heterotic side.
On the other hand, for every case of the non-Cartan type, where we have succeeded 
to find an equation for the RES-fibered space, we also show that 
the spectrum read off from the equation agrees with that of the dual 
heterotic theory for a special choice of instanton distribution.

The outline of this paper is as follows: In section 2, we begin with an  
introduction to the non-Cartan type Mordell-Weil lattices. In section 3, 
we first review a general method to compute heterotic indices 
in six dimensions and demonstrate how it works in particular 
examples. We also discuss there why the instanton numbers must be 
a multiple of some particular non-one integer in general. 
In section 4 we review the basic facts on the six-dimensional F-theory/
heterotic duality, and explain our strategy to obtain the equation
for a RES-fibered threefold having a non-Cartan Mordell-Weil lattice. 
In section 5, we start the construction by the No.7 case of the 
Oguiso-Shioda classification to obtain the equation for the case of a 
particular instanton distribution.  We then deform this equation 
in an appropriate way to find the equations for threefolds with      
arbitrarily distributed instantons. The match of the spectra is 
also verified there. Sections 6 and 7 are devoted to the considerations 
of the RES-fibered threefolds for the the cases of the
$SU(7)$ series and the $SU(8)$ series, respectively.
Finally we summarize our conclusions in section 8.
In appendix A and B, we present the detail of the heterotic 
index computations for the $SU(7)$ and the $SU(8)$ series, 
respectively. Appendix C shows the explicit forms of the 
functions $f$, $g$ of the Weierstrass equations and the 
discriminant $\Delta$ for various cases considered in the text. 
%

\section{Models with non-Cartan type Mordell-Weil lattices}

\label{sec:nonCartan}

It is known that a rational elliptic surface $S$ possesses a lattice structure, 
called the Mordell-Weil lattice $E(K)$. ($K$ is the field over which $S$ is defined.) 
When $S$ has no singularity, $E(K)$ is the self dual $E_8$ lattice.
When $S$ has a singularity of an $ADE$ type with root lattice $T$, $E(K)$ is reduced,
roughly speaking, to the orthogonal complement $T^{\perp}$ of $T$ in $E_8$.
More precisely, $E(K)$ is the dual lattice of $T^{\perp}$ accompanied with a 
torsion part:
\beq
   E(K) \cong (T^{\perp})^* \oplus \mbox{torsion},
\label{eq:EK}
\eeq 
where $^*$ denotes the dual lattice. $T^{\perp}$ satisfies
\beq
    T\oplus T^{\perp} \subset {\mathfrak e}_8.
\label{eq:TTperp}
\eeq
In the context of the duality between F-theory and heterotic string, 
the root lattice $T$ of the $ADE$ singularity in $S$ corresponds to the gauge symmetry $G$, 
while the orthogonal complement $T^{\perp}$ corresponds to the gauge bundle $H$ of the heterotic string.
The decomposition \eqref{eq:TTperp} is then interpreted as  
\beq
  G\times H \subset E_8.
\label{eq:GH}
\eeq
In other words (apart from the torsion part), 
\bes
 T & \leftrightarrow \mbox{gauge symmetry~}G,   \\
 E(K) & \leftrightarrow \mbox{dual gauge bundle~} H^*. 
\ees

The Mordell-Weil lattices $E(K)$ are classified into 74 patterns~\cite{OguisoShioda}. 
In many cases, $(T^{\perp})^*$ is a weight lattice of some $ADE$ type Lie algebra 
or a direct sum thereof, but there are some special cases where $(T^{\perp})^*$ is not a weight lattice 
of any $ADE$ type Lie algebra.
In these cases $T^{\perp}$ is also not a root lattice of any $ADE$ type, and neither is 
the gauge bundle $H$ of the heterotic string.
Let us call these Mordell-Weil lattices $E(K)$ and 
the corresponding gauge bundles $H$ ``non-Cartan type".
In this paper, among them, we will study the cases listed in Table \ref{tab:MWLsu7su8}, 
where $T$ is a subalgebra of $A_7$ and
$E(K)$ is torsion free.

\begin{table}[htb]
\captionsetup{format=hang,margin=30pt}
\caption{The non-Cartan type Mordell-Weil lattices $E(K)$ that are
torsion free and associated $T$ are sublattices of $A_7$~\cite{OguisoShioda}.}
\begin{center}
\begin{tabular}{|c|c|c|}
\hline 
$\mbox{No.}$  & $T \leftrightarrow G$                                       & $E(K)\leftrightarrow H^*$   \\ \hline \hline
$25$           & $A_6$                                    & $\Lambda_{(25)}$ \\ 
$17$           & $A_4\oplus A_1$                          & $\Lambda_{(17)}$ \\ 
$19$           & $A_3\oplus A_2$                          & $\Lambda_{(19)}$ \\ 
$12$           & $A_2\oplus A_1\oplus A_1$                & $\Lambda_{(12)}$ \\  \hline
$45$           & $A_7$                                    & $\left< 1/8 \right>$ \\ 
$29$           & $A_5\oplus A_1$                          & $A^*_1\oplus \left< 1/6 \right>$ \\ 
$31$           & $A_4\oplus A_2$                          & $\Lambda_{(31)}$ \\ 
$36$           & $A_3\oplus A_3$                          & $\left< 1/4 \right> \oplus \left< 1/4 \right>$ \\ 
$20$           & $A_2\oplus A_2\oplus A_1$                & $A^*_2\oplus \left< 1/6 \right>$ \\ 
$22$           & $A_3\oplus A_1\oplus A_1$                & $A^*_1\oplus A^*_1 \oplus \left< 1/4 \right>$ \\  \hline
\end{tabular}   
\label{tab:MWLsu7su8}
\end{center}
\end{table}
In the table, $\Lambda_{(N)}$ is a matrix representing the lattice $E(K)\cong (T^{\perp})^*$.
The inverse $\Lambda_{(N)}^{-1}$ representing the dual lattice $T^{\perp}$, or equivalently, the gauge bundle $H$, takes one of the following non-Cartan forms: 
\bes
 \Lambda_{(25)}^{-1} & = \left(
                     \begin{array}{cc}
                        4 & -1 \\
                       -1 & 2       
                     \end{array} 
                     \right),~
 \Lambda_{(17)}^{-1} = \left(
                     \begin{array}{ccc}
                        4 & -1 & 1 \\
                       -1 & 2  & -1 \\ 
                        1 & -1 & 2
                     \end{array} 
                     \right),~ 
 \Lambda_{(19)}^{-1}  = \left(
                     \begin{array}{ccc}
                        2 & 0 & -1 \\
                        0 & 2  & -1 \\ 
                       -1 & -1 & 4
                     \end{array} 
                     \right),~  \\
 \Lambda_{(12)}^{-1} &= \left(
                     \begin{array}{cccc}
                        4 & -1 & 0 & 1 \\
                       -1 & 2  & -1& 0 \\ 
                        0 & -1 & 2 & -1 \\
                        1 & 0 & -1 & 2
                     \end{array} 
                     \right),~
 \Lambda_{(31)}^{-1}  = \left(
                     \begin{array}{cc}
                        8 & -1 \\
                       -1 & 2       
                     \end{array} 
                     \right).
\ees
$A_n^*,~D_n^*$ are the weight lattices of $A_n,~D_n$.  $\left< 1/k \right>$ denotes the one-dimensional lattice with lattice spacing $\sqrt{\frac{1}{k}}$. Its dual lattice has a 
non-Cartan form unless $k=2$, in which the lattice becomes the $A_1$ weight lattice. 

In terms of heterotic string, the two series in Table \ref{tab:MWLsu7su8} are obtained by Higgsing the gauge group $G = SU(7)$ (No.25)
and $G=SU(8)$ (No.45), respectively.
Their Higgsing chains are summarized as follows:\footnote{$G = SU(2)\times SU(2)\times SU(2) \times SU(2)$ (No.14 of \cite{OguisoShioda}) 
belongs to the Higgsing chain of $G = SU(8)$,
but we excluded it from Table \ref{tab:MWLsu7su8} because it has $E(K) = A_1\oplus A_1 \oplus A_1 \oplus A_1$, 
which is not the non-Cartan type.} (the subscripts are the numbers in Table \ref{tab:MWLsu7su8})
\beq
\begin{array}{llllcll}
   &              
     &  \hspace{-0.3cm} \scalebox{1.0}{$\displaystyle SU(5) \timesm SU(2)$}_{(17)}   
       &              
         &   
           &                         \\  
   & \hspace{-0.35cm}  \nearrow     
     &                                
       & \hspace{-0.35cm}  \searrow     
         &  
           &                          \\
\scalebox{1.0}{$\displaystyle SU(7)$}_{(25)}     
   & \hspace{-0.35cm} \rightarrow  
     &  \hspace{-0.3cm} \scalebox{1.0}{$\displaystyle SU(3) \timesm SU(4)$}_{(19)} 
       & \hspace{-0.35cm}\rightarrow   
         & \hspace{-0.3cm} \scalebox{1.0}{$\displaystyle SU(3) \timesm SU(2) \timesm SU(2)$}_{(12)} 
           &     \\   
   &           &                             
     &           
       &  \hspace{-0.8cm} \uparrow     
         &                            \\ 
   &           
     &  \hspace{-0.3cm} \scalebox{1.0}{$\displaystyle SU(4) \timesm SU(4)$}_{(36)} 
       & \hspace{-0.35cm} \rightarrow 
         &  \hspace{-0.3cm} \scalebox{1.0}{$\displaystyle SU(4) \timesm SU(2) \timesm SU(2)$}_{(22)}
           & \hspace{-0.3cm}     \\
   & \hspace{-0.35cm} \nearrow  
     &                             
       &  \hspace{-0.35cm} \nearrow       
         &         
           &     \\
\scalebox{1.0}{$\displaystyle SU(8)$}_{(45)} 
   & \hspace{-0.35cm}\rightarrow 
     & \hspace{-0.3cm}  \scalebox{1.0}{$\displaystyle SU(6) \timesm SU(2)$}_{(29)} 
       &  
         &   
           &    \\
   & \hspace{-0.35cm}\searrow  
     &                             
       & \hspace{-0.35cm} \searrow  
         &         
           &    \\
   &         
     &  \hspace{-0.3cm}  \scalebox{1.0}{$\displaystyle SU(3) \timesm SU(5)$}_{(31)}  
       & \hspace{-0.35cm} \rightarrow 
         &  \hspace{-0.3cm} \scalebox{1.0}{$\displaystyle SU(3) \timesm SU(3) \timesm SU(2)$}_{(20)}    
           &    
\label{eq:su7su8Higgs}
\end{array}   
\eeq

\section{Heterotic index computations}
\label{sec:heteroindex}

\subsection{General method}
We will first quickly review the general method to compute 
the heterotic spectrum in six dimensions.

The Dirac index for a six-dimensional compactification of heterotic 
string theory on a complex 2-fold is given by 
\beqa
\mbox{index}&=&\left.\int \hat A(TZ)ch(V)\right|_4,
\eeqa
where $\hat A(TZ)$ is the A-roof genus of the tangent bundle of the 
complex 2-fold $Z$ on which the heterotic string is compactified, 
and $ch(V)$ is the Chern character of the vector bundle $V$ over $Z$.  
The number of hypermultiplets is given by $-1/2$ of the index, where 
the overall minus sign is a convention and the factor of 1/2 is due 
to the fact that the heterotic gaugino is a Majorana-Weyl spinor in 
ten dimensions.

We consider the cases where the instanton 
takes values in a subgroup $H$ of, say, the first factor of $E_8$ to 
leave the centralizer subgroup $G$ of $H$ in $E_8$ unbroken. 
We are particularly interested 
in the cases where $H$ contains some $U(1)$ factors. As in \cite{GSW}, 
let the $\bf 248$ 
of $E_8$ be decomposed into the representations of $G\times H$ as
\beqa
{\bf 248}&=&
{\oplus}_i ~L_i\otimes C_i,
\eeqa
where $L_i$ and $C_i$ are irreducible representations of $G$ and $H$,
respectively. Using the fact that $\int_{\mbox{\scriptsize K3}}p_1=-48$, the number of 
hypermultiplet $n_{L_i,C_i}$ in a representation $L_i$ of $G$ (and 
$C_i$ of $H$) is given by \cite{GSW}
\beqa
n_{L_i,C_i}&\equiv&-\frac12 \mbox{index}\n
&=&
-\mbox{dim}C_i 
+\frac12\cdot\frac1{8\pi^2} \int_{\mbox{\scriptsize K3}}\mbox{Tr}_{C_i}F \wedge F,
\label{n_{L_i,C_i}}
\eeqa
where $F$ is the instanton gauge field 2-form taking values in 
(the Lie algebra of) $H$. The trace is taken in the representation $C_i$.

The equation (\ref{n_{L_i,C_i}}) is still a correct formula 
even for the cases where $H$ contains some $U(1)$ factors. 
Since 
\beqa
\sum_{a=1,2}
\sum_i \mbox{dim}L_i^{(a)}\cdot \frac1{8\pi^2} \int\mbox{Tr}_{C_i^{(a)}} F^{(a)}\wedge F^{(a)}
&=&\frac1{8\pi^2}\int_{\mbox{\scriptsize K3}}\mbox{Tr}_{{\bf 248}\oplus {\bf 248}}F\wedge F\n
&=&-30\int_{\mbox{\scriptsize K3}} p_1
\n
&=&30\cdot 48
\eeqa
due to the anomaly cancellation condition, where  $a=1,2$ are the 
labels to distinguish the $E_8$ factors, we see that the second term of 
(\ref{n_{L_i,C_i}}) is nothing but $-30$ times the contribution of each representation 
$C_i$ to the total instanton number (=24, taking the factor of 1/2 into account). 
The normalizations of the traces and $U(1)$ charges are determined 
once the instanton number of each irreducible group factor of $H$ 
is specified.

\subsection{Example 1~: $G=SU(7)$ $(T = A_6)$ (No.25)}
\label{sec:No25hetero}
In this case one can take  $SU(2)\times U(1)$ as $H$. Let the instanton numbers 
of $SU(2)$ and $U(1)$ be $12+n-r$ and $r$, respectively, so that the total instanton 
number of $H$ is $12+n$. Then we have\,\footnote{The subscript ``K3" for the integral 
is omitted below.}
\beqa
\frac1{8\pi^2}\int F^A\wedge F^B~\mbox{Tr}_{\bf 248}\;{\rm ad}T^A\;{\rm ad}T^B
&=&60(12+n).
\label{60(12+n)}
\eeqa
The adjoint representation of $E_8$ is 
decomposed as
\beqa
E_8&\supset&SU(7)\times SU(2) \times U(1),\n
{\bf 248}&=&
({\bf 48},{\bf 1})_0~\oplus~
({\bf 1},{\bf 1})_0~\oplus~
({\bf 7},{\bf 1})_{-4}~\oplus~
({\bf \overline 7},{\bf 1})_4\n
&&
\oplus~({\bf 35},{\bf 1})_{2}~\oplus~
({\bf \overline{35}},{\bf 1})_{-2}
~\oplus~({\bf 1},{\bf 3})_0     \n
&&
\oplus~({\bf 7},{\bf 2})_{3}~\oplus~
({\bf \overline 7},{\bf 2})_{-3}~\oplus~
({\bf 21},{\bf 2})_{-1}~\oplus~
({\bf \overline{21}},{\bf 2})_{1},
\eeqa
where the subscripts denote the $U(1)$ charges.
This yields 
\beqa
\frac1{8\pi^2}\int F^A\wedge F^B~\mbox{Tr}_{\bf 248}~{\rm ad}T^A~{\rm ad}T^B
&=&
\frac1{8\pi^2}\int F^a\wedge F^b \cdot 60~\mbox{Tr}\tau^a\tau^b\n
&&+\frac1{8\pi^2}\int F^{U(1)}\wedge F^{U(1)}\cdot 14\cdot 60,
\eeqa
where $\tau^a$ are the ${\bf 2}$ representation matrices of $SU(2)$, 
and $14\cdot 60$ is the sum of the $U(1)$ charge squares.
As we assumed above, the first term is equal to $60(12+n-r)$, whereas  
the second term is $60r$. Therefore we have 
\beqa
\frac1{8\pi^2}\int F^a\wedge F^b~\mbox{Tr}\tau^a\tau^b&=&12+n-r,
\n
\frac1{8\pi^2}\int F^{U(1)}\wedge F^{U(1)}&=&\frac r{14},
\eeqa 
which means that (Note the factor of $-1/2$ in eq.(\ref{n_{L_i,C_i}})) 
a ${\bf 2}$ representation of $SU(2)$ contributes $\frac{12+n-r}2$, 
whereas each $U(1)$-charge $Q$ 
component contributes $\frac{Q^2}2\cdot\frac r{14}$,
to the multiplicities of hypermultiplets transforming 
in the corresponding $G$ representation.

As an illustration let us compute $n_{{\bf 21},{\bf 2}_{-1}}$. 
%
This is computed as 
\beqa
n_{{\bf 21},{\bf 2}_{-1}}&=&-\mbox{dim}\;{\bf 2}
+\frac12\cdot\frac1{8\pi^2} \int F^a\wedge F^b~\mbox{Tr}_{{\bf 2}}\tau^a\tau^b\n
&&+\frac12\cdot\frac1{8\pi^2} \int F^{U(1)}\wedge F^{U(1)}(-1)^2\cdot 2\n
&=&-2+\frac{12+n-r}2+\frac{(-1)^2}2\cdot\frac r{14}\cdot 2\n
&=&\frac12\left(
n-\frac67 r+8
\right).
\label{n_21_2_-1}
\eeqa
Note that the 3rd ($U(1)$ instanton) term is multiplied by 2 because 
each component of ${\bf 2}_{-1}$ contributes to the index.

Since $n_{{\bf \overline{21}},{\bf 2}_{1}}$ is equal to $n_{{\bf 21},{\bf 2}_{-1}}$
(\ref{n_21_2_-1}), and there is no distinction between ${\bf 21}$ and ${\bf \overline{21}}$ 
in six dimensions, the total multiplicity of  ${\bf 21}$ is in all
\beqa
n_{\bf 21}&=&n-\frac67 r+8.
\eeqa
This becomes an integer if and only if the instanton number $r$ is a multiple of 7.
So writing $r=7r'$, we have
\beqa
n_{\bf 21}&=&n-6 r'+8.
\eeqa
The computations of the multiplicities for other representations 
can be worked out similarly. The result is summarized in Table \ref{spectrumNo.25}.

\begin{table}[htp]
\caption{The spectrum for the configuration No.25.}
\begin{center}
\begin{tabular}{|c|c|}
\hline
Representation&Multiplicity\\
\hline
\hline
${\bf 48}$&$-1$($SU(7)$ vector)\\
${\bf 35}$&$2 r' -2$\\
${\bf 21}$&$n-6 r' +8$\\
${\bf 7}$&$n+10 r' +6$\\
${\bf 1}$&$2n-14 r' +20$\\
\hline
\end{tabular}
\end{center}
\label{spectrumNo.25}
\end{table}%

Finally, we would like to comment on the computation in \cite{AGRT} 
of the heterotic string matter spectrum with an $SU(7)$ gauge group, 
which is different from ours but led them to the same result as that derived here. 
Since the multiplicities are not integers when $r$ is a multiple of 7, they consider 
$U(2)$ instead of $SU(2)$ with an extra $U(1)$ and assumed a contribution 
of this additional $U(1)$ to the multiplicity of the singlet (See the top row of 
Table 8 of \cite{AGRT}, where the multiplicity of {\bf 1} contains 
the term 
$4c_2({\cal V}_2)\underline{-c_1({\cal L})^2}-6$
despite that the $SU(7)$ singlets (the first two terms of (4.8) in \cite{AGRT}) are 
not charged under the $U(1) \subset SU(7)\times SU(2) \times U(1)$).  
This $U(1)$ must commute with both the $SU(7)$ and the $SU(2)$ 
and must be different from the original $U(1)$ factor in $E_8$, but obviously, 
there is no room for such an {\em extra} $U(1)$ in $E_8$ as the rank is 
already exhausted.

\subsection{Orthogonal decompositions of the $E_8$ root lattice: Why a multiple of seven?}

The fact that the U(1) instanton number must be a multiple of seven 
can be understood by an orthogonal decomposition of the $E_8$ root lattice.
Let ${\bf e}_i$ (i=1,\ldots,9) be a set of orthonormal vectors of nine-dimensional 
flat Euclidean space with inner product
\beqa
{\bf e}_i\cdot {\bf e}_j&=& \delta_{ij}.
\eeqa
Then the following set of $72+84+84=240$ vectors on 
an eight-dimensional hyperplane normal to $\sum_{j=1}^9 {\bf e}_j$ 
form the set of root vectors of $E_8$:
\beqa
&&~~~~{\bf e}_i-{\bf e}_j ~~~(1\leq i\neq j \leq 9),
\label{SU(9)roots}\\
&&\pm\left({\bf e}_i+{\bf e}_j+{\bf e}_k-\frac13\sum_{l=1}^9 {\bf e}_l\right)
~~~(1\leq i< j<k\leq 9).
\label{3formroots}
\eeqa
This fact can be most easily verified by considering Freudenthal's realization 
of the $E_8$ algebra \cite{MizoguchiE10,MizoguchiSchroeder}.
The first line (\ref{SU(9)roots}) is the set of root vectors of $SU(9)$, while 
the second line (\ref{3formroots}) is the root vectors corresponding to 
the rank-3 tensors of $SU(9)$.
Using this presentation of $E_8$ roots, one can easily see where the 
roots of $SU(7)\times SU(2)$ are embedded and which root vector is 
the one corresponding to the $U(1)$ generator.  As the root vectors of 
$SU(7)$ one can take 
\beqa
&&{\bf e}_i-{\bf e}_j ~~~(1\leq i\neq j \leq 7),
\label{SU(7)roots}
\eeqa 
and  the roots of $SU(2)\subset H$ is then 
\beqa
&&\pm({\bf e}_8-{\bf e}_9).
\label{SU(2)roots}
\eeqa 
In the $E_8$ root lattice generated by the vectors (\ref{SU(9)roots})
and (\ref{3formroots}),
the orthogonal lattice normal to the $SU(7)$ and $SU(2)$ lattices spanned by 
 (\ref{SU(7)roots}) and (\ref{SU(2)roots}) is one-dimensional, 
 generated by 
 \beqa
 \frac13 \left(
 2\sum_{l=1}^7 {\bf e}_l 
 -7({\bf e}_8 +{\bf e}_9)
 \right).
 \label{U(1)rootvector}
 \eeqa
 The length square of (\ref{U(1)rootvector}) is 14, which is 7 times 
 as large as the simple-root length square. This explains why the instanton number 
 of $U(1)$ is a multiple of 7. 
 
 Note that 
 the junction lattice \cite{FYY}, 
 which is the inverse of the matrix of section parings, is  
 given in the present case as 
\beqa
\Lambda_{(25)}^{-1}&=&
\left(
\begin{array}{cr}
  4& -1      \\
  -1& 2 
\end{array}
\right).
\eeqa
This can be made diagonal by the change of basis
\beqa
\left(
\begin{array}{cc}
  2& ~~1      \\
  0& ~~1 
\end{array}
\right)
\left(
\begin{array}{cr}
  4& -1      \\
  -1& 2 
\end{array}
\right)
\left(
\begin{array}{cc}
  2& ~~0      \\
  1& ~~1 
\end{array}
\right)
&=&
\left(
\begin{array}{cr}
  14& ~0     \\
  0& ~2 
\end{array}
\right),
\eeqa
which agrees with the above consideration.

\subsection{Example 2~: $G=SU(5)\times SU(2)$ $(T=A_4\oplus A_1)$ (No.17)}
\label{sec:No17hetero}

In this case, $\mbox{rank}(E(K))=3$ and we can take one or two $U(1)$ direction(s) in $H$.
Namely, $H = SU(3)\times U(1)$ or $H = SU(2)\times U(1)\times U(1)$.
The gauge symmetry is the same ($G = SU(5)\times SU(2)$) for these two gauge bundles,
but the resulting spectra are different. 
As we will explain below, the spectrum for $H = SU(2)\times U(1)\times U(1)$ 
is more general than the one for $H = SU(3)\times U(1)$.

\subsubsection{The case when $H=SU(2)\times U(1)\times U(1)$}
The junction lattice is 
\beqa
\Lambda_{(17)}^{-1}&=&
\left(
\begin{array}{crr}
  4& -1  &1    \\
  -1& 2 &-1 \\
  1 & -1 & 2
\end{array}
\right).
\eeqa
As $H$, we allow the instantons to live also in a different $U(1)(\equiv \tilde U(1))$  
as well as the 
$SU(2)\times U(1)$ for No.25 in the previous section, 
so that the $SU(7)$ 
is further broken to $SU(5)\times SU(2)$. 
The instanton numbers are assumed to be 
$12+n-r-\tilde r$, $r$ and $\tilde r$ for 
$SU(2)$, $U(1)$  and  $\tilde U(1)$, respectively.
The decomposition of ${\bf 248}$ in representations of 
these subgroups is shown in Table \ref{No17decomposition}.

\begin{table}[htp]
\caption{Decomposition of ${\bf 248}$.}
\begin{center}
\begin{tabular}{|c|ccc|}
\hline
Rep. of $SU(5)\times SU(2)$ &Rep. of $SU(2)$ (in $H$)&$U(1)$charge&$\tilde U(1)$charge\\
\hline
\hline
$({\bf 24},{\bf 1})$&{\bf 1}&0&0\\
$({\bf 1},{\bf 3})$&{\bf 1}&0&0\\
$({\bf 5},{\bf 2})$&{\bf 1}&0&7\\
$({\overline{\bf 5}},{\bf 2})$&{\bf 1}&0&$-7$\\
$({\bf 1},{\bf 1})$ ($\tilde U(1)$)&{\bf 1}&0&0\\
$({\bf 5},{\bf 1})$&{\bf 2}&$3$&$2$\\
$({\bf 5},{\bf 1})$&{\bf 1}&$-4$&$2$\\
$({\bf 1},{\bf 2})$&{\bf 2}&$3$&$-5$\\
$({\bf 1},{\bf 2})$&{\bf 1}&$-4$&$-5$\\
$(\overline{\bf 5},{\bf 1})$&{\bf 2}&$-3$&$-2$\\
$(\overline{\bf 5},{\bf 1})$&{\bf 1}&$4$&$-2$\\
$({\bf 1},{\bf 2})$&{\bf 2}&$-3$&$5$\\
$({\bf 1},{\bf 2})$&{\bf 1}&$4$&$5$\\
$(\overline{\bf 10},{\bf 1})$&{\bf 1}&$2$&$6$\\
$({\bf 10},{\bf 2})$&{\bf 1}&$2$&$-1$\\
$({\bf 5},{\bf 1})$&{\bf 1}&$2$&$-8$\\
$({\bf 10},{\bf 1})$&{\bf 1}&$-2$&$-6$\\
$(\overline{\bf 10},{\bf 2})$&{\bf 1}&$-2$&$1$\\
$(\overline{\bf 5},{\bf 1})$&{\bf 1}&$-2$&$8$\\
$({\bf 10},{\bf 1})$&{\bf 2}&$-1$&$4$\\
$({\bf 5},{\bf 2})$&{\bf 2}&$-1$&$-3$\\
$({\bf 1},{\bf 1})$&{\bf 2}&$-1$&$-10$\\
$(\overline{\bf 10},{\bf 1})$&{\bf 2}&$1$&$-4$\\
$(\overline{\bf 5},{\bf 2})$&{\bf 2}&$1$&$3$\\
$(\overline{\bf 1},{\bf 1})$&{\bf 2}&$1$&$10$\\
$({\bf 1},{\bf 1})$ ($U(1)$)&{\bf 1}&0&0\\
$({\bf 1},{\bf 1})$&{\bf 3}&0&0\\

\hline
\end{tabular}
\end{center}
\label{No17decomposition}
\end{table}%

In the present case 
\beqa
\sum_{E_8}(U(1)
\mbox{~charge}
)^2&=&14\cdot 60
\eeqa
and
\beqa
\sum_{E_8}(\tilde U(1)
\mbox{~charge})^2&=&70\cdot 60,
\eeqa
so that 
\beqa
\frac1{8\pi^2}\int F^a\wedge F^b~\mbox{Tr}\tau^a\tau^b&=&12+n-r-\tilde r,
\n
\frac1{8\pi^2}\int F^{U(1)}\wedge F^{U(1)}&=&\frac r{14},
\n
\frac1{8\pi^2}\int F^{\tilde U(1)}\wedge F^{\tilde U(1)}&=&\frac{\tilde r}{70}.
\eeqa 

The spectrum for No.17 is similarly obtained 
as shown in Table \ref{spectrumNo.17}.
\begin{table}[htp]
\caption{The spectrum for the configuration No.17.}
\begin{center}
\begin{tabular}{|c|c|}
\hline
Representation&Multiplicity\\
\hline
\hline
$({\bf 24},{\bf 1})$&$-1$($SU(5)$ vector)\\
$({\bf 1},{\bf 3})$&$-1$($SU(2)$ vector)\\
$({\bf 1},{\bf 1})$&$3n+27-10\left(\frac27 r + \frac1{70}{\tilde r}\right)$\\
$({\bf 5},{\bf 2})$&$n+6-3\left(\frac27 r + \frac1{70}{\tilde r}\right)$\\
$({\bf 5},{\bf 1})$&$n+4+6\left(\frac27 r + \frac1{70}{\tilde r}\right)$\\
$({\bf 1},{\bf 2})$&$n+6+5\left(\frac27 r + \frac1{70}{\tilde r}\right)$\\
$({\bf 10},{\bf 2})$&$-2+\left(\frac27 r + \frac1{70}{\tilde r}\right)$\\
$({\bf 10},{\bf 1})$&$n+6-2\left(\frac27 r + \frac1{70}{\tilde r}\right)$\\
\hline
\end{tabular}
\end{center}
\label{spectrumNo.17}
\end{table}%

For $\frac27 r + \frac1{70}{\tilde r}$ to be integer, 
the general solution is 
\beqa
(r,\tilde r)&=&(3,10)k + (-1,20)l~~~(k,l\in{\ZZ}),
\label{eq:rrtNo17}
\eeqa
so that
\beqa
\frac27 r + \frac1{70}{\tilde r}&=&k~\in{\ZZ},
\eeqa
therefore $r$ or $\tilde r$ need not necessarily be 
a multiple of $7$ or $70$ in general.
The orthogonal decomposition of the junction lattice is
\beqa
\left(
\begin{array}{ccc}
  2& ~~1& ~~0      \\
  0& ~~1& ~~0     \\
1& -3& -7  
\end{array}
\right)
\left(
\begin{array}{crr}
  4& -1  &1    \\
  -1& 2 &-1 \\
  1 & -1 & 2
\end{array}
\right)
\left(
\begin{array}{ccc}
  2& ~~0& ~~1      \\
  1& ~~1& -3     \\
0& ~~0& -7  
\end{array}
\right)
&=&
\left(
\begin{array}{ccc}
  14& ~0  & ~0   \\
  0& ~2 & ~0\\
0& ~0 & ~70
\end{array}
\right),
\n
\eeqa
which implies that the minimal $\tilde U(1)$ charge square 
(measured by the $E_8$ root space inner product) is 
35 times (and not 70 times) as large as that of the $SU(2)$ 
factor. This is almost equivalent to the condition for 
the integrality of the multiplicities, but the latter is ``twice as" severe.

\subsubsection{The case when $H=SU(3)\times  U(1)$ }
We can see the $SU(3)$ Cartan matrix in the right-down block of 
\beqa
\Lambda_{(17)}^{-1}&=&
\left(
\begin{array}{crr}
  4& -1  &1    \\
  -1& 2 &-1 \\
  1 & -1 & 2
\end{array}
\right), \nonumber
\eeqa
and we can also take $SU(3)\times  U(1)(\equiv U(1)_{(17)})$ as $H$. 
Orthogonal decomposition of the junction lattice is
\beqa
\left(
\begin{array}{ccc}
  3& ~~1& -1      \\
  0& ~~1& ~0     \\
 0& ~~0& ~1  
\end{array}
\right)
\left(
\begin{array}{crr}
  4& -1  &1    \\
  -1& 2 &-1 \\
  1 & -1 & 2
\end{array}
\right)
\left(
\begin{array}{ccc}
  3& ~0& ~~0      \\
  1& ~1& ~~0     \\
 -1& ~0& ~~1  
\end{array}\right)
&=&
\left(
\begin{array}{ccc}
  30& ~0  & ~0   \\
  0& ~2 & -1\\
0& -1 & ~2
\end{array}
\right).\n
\eeqa

In representations of this $G\times H$,  the $E_8$ adjoint is decomposed as shown in
Table \ref{No17decomposition2}.

\begin{table}[htp]
\caption{Decomposition of ${\bf 248}$.}
\begin{center}
\begin{tabular}{|c|cc|}
\hline
Rep. of $SU(5)\times SU(2)$ &Rep. of $SU(3)$ &$U(1)_{(17)}$charge\\
\hline
\hline
$({\bf 24},{\bf 1})$&{\bf 1}&0\\
$({\bf 1},{\bf 3})$&{\bf 1}&0\\
$({\bf 1},{\bf 2})$&{\bf 3}&$5$\\
$({\bf 1},{\bf 2})$&$\overline{\bf 3}$&$-5$\\
$({\bf 1},{\bf 1})$ &{\bf 8}&0\\
$({\bf 1},{\bf 1})$ &{\bf 1}&0\\
&&\\
$({\bf 5},{\bf 2})$&$\overline{\bf 3}$&$1$\\
$({\bf 5},{\bf 1})$&{\bf 3}&$-4$\\
$({\bf 5},{\bf 1})$&{\bf 1}&$6$\\
$(\overline{\bf 5},{\bf 2})$&{\bf 3}&$-1$\\
$(\overline{\bf 5},{\bf 1})$&$\overline{\bf 3}$&$4$\\
$(\overline{\bf 5},{\bf 1})$&{\bf 1}&$-6$\\
&&\\
$({\bf 10},{\bf 2})$&{\bf 1}&$3$\\
$({\bf 10},{\bf 1})$&$\overline{\bf 3}$&$-2$\\

$(\overline{\bf 10},{\bf 2})$&{\bf 1}&$-3$\\
$(\overline{\bf 10},{\bf 1})$&{\bf 3}&$2$\\

\hline
\end{tabular}
\end{center}
\label{No17decomposition2}
\end{table}%

We assume $12+n-r_{(17)}$ instantons in $SU(3)$ and $r_{(17)}$ 
in $U(1)_{(17)}$.  Then since
\beqa
\sum_{E_8}( U(1)_{(17)}\mbox{~charge})^2&=&30\cdot 60,
\eeqa
we obtain 
\beqa
\frac1{8\pi^2}\int F^a\wedge F^b~\mbox{Tr}\lambda^a\lambda^b&=&12+n-r_{(17)},
\n
\frac1{8\pi^2}\int F^{U(1)_{(17)}}\wedge F^{U(1)_{(17)}}&=&\frac{ r_{(17)}}{30},
\eeqa 
where $\lambda^a$ ($a=1,\ldots,8$) are the Gell-Mann matrices.
Incidentally, 
$U(1)_{(17)}$ is  the ``hypercharge" $U(1)$ that breaks one of $SU(5)$ of 
$SU(5)\times SU(5)\subset E_8$ into $SU(3)\times SU(2)$.
The spectrum is as shown in Table \ref{spectrumNo.17(2)}.

\begin{table}[htbp]
\caption{The spectrum for the configuration No.17. Another derivation.}
\begin{center}
\begin{tabular}{|c|c|}
\hline
Representation&Multiplicity\\
\hline
\hline
$({\bf 24},{\bf 1})$&$-1$($SU(5)$ vector)\\
$({\bf 1},{\bf 3})$&$-1$($SU(2)$ vector)\\
$({\bf 1},{\bf 1})$&$3n+27-30 r'_{(17)}$\\
$({\bf 5},{\bf 2})$&$n+6-9r'_{(17)}$\\
$({\bf 5},{\bf 1})$&$n+4+18r'_{(17)}$\\
$({\bf 1},{\bf 2})$&$n+6+15r'_{(17)}$\\
$({\bf 10},{\bf 2})$&$-2+3r'_{(17)}$\\
$({\bf 10},{\bf 1})$&$n+6-6r'_{(17)}$\\
\hline
\end{tabular}
\end{center}
\label{spectrumNo.17(2)}
\end{table}%

Here we have defined $r_{(17)}=10r'_{(17)}$ so that the multiplicities become 
all integers when the $U(1)_{(17)}$ instanton number 
$r_{(17)}$ is a multiple of $10$ (which is when $r'_{(17)}$ is an integer). 

Note that this spectrum coincides with the spectrum shown in 
Table \ref{spectrumNo.17} provided that the replacement
\beqa
\frac27 r + \frac1{70}{\tilde r}=k=3r'_{(17)}
\eeqa
is made. Thus the spectrum for $H=SU(3)\times U(1)$  is ``three times" 
as restrictive as that for $H=SU(2)\times U(1)^2$.

\medskip
The index computations for other cases can be done similarly. We summarize the relevant 
results (No.19 and No.12 for $SU(7)$ series and No.29 and No.22 for $SU(8)$ series)
in Appendices A and B.

\section{Geometries for non-Cartan Mordell-Weil lattices}

\label{sec:CY3}

Suppose that a non-Cartan Mordell-Weil lattice $E(K)$ (non-Cartan gauge bundle $H$) is given.
To construct the corresponding geometry, which is a rational elliptic surface fibered over $\PP^1$,
we first construct a K3-fibered CY${}_3$ with the singularity $T$ (gauge symmetry $G$) paired with the given 
$E(K)$.
Then, we map the CY${}_3$ to a RES-fibered geometry.

\subsection{CY${}_3$ in six-dimensional F-theory/heterotic duality}
\label{sec:WandT}

We first review the construction of a CY${}_3$ for six-dimensional F-theory compactification \cite{MV1,MV2,BIKMSV}.
$E_8 \times E_8$ heterotic string/K3 contains 24 instantons.
Suppose $12+n$ of them take values in a gauge bundle $H$ 
in the first $E_8$ and the gauge symmetry is broken to $G$,  
while the other $12-n$ instantons take values in another gauge 
bundle $H'$ in the second $E_8$
and the gauge symmetry is broken to $G'$.
It has been known \cite{MV1} that the corresponding dual F-theory geometry is a K3-fibered CY${}_3$ over $\PP^1$ and at the same time
is an elliptic fibered CY${}_3$ over the Hirzebruch surface $F_n$
with singularities $G$ and $G'$.
The Weierstrass form is given by
\beq
  0  = y^2+x^3+f(z,z')x+g(z,z')
\label{eq:Weierstrassform}
\eeq
with
\bes
 f(z,z') & = \sum_{i=0}^{8} f_{8+(4-i)n}z^i, \\
 g(z,z') & = \sum_{j=0}^{12} g_{12+(6-j)n}z^j.
\label{eq:Weierstrass}
\ees
$F_n$ is a $\PP^1$ fibration over the base $\PP^1$, whose coordinates are $z$ and $z'$, respectively.
$f_k$ and $g_k$ are polynomials of $z'$ with homogeneous order $k$. 
Let us write the fiber coordinate $z$ in a homogeneous form $(u:v)$.
Under the $C^*$ action $(u:v)\rightarrow (\mu u:\mu v)$, the other coordinates transform as
\beq
 (x,y,z') \rightarrow (\mu^4 x, \mu^6 y,z').
\label{eq:xyzzp0}
\eeq
Also, for the base coordinate $z'\equiv (u':v')$, the $C^*$ action $(u':v')\rightarrow (\lambda u':\lambda v')$
gives 
\beq
    (x,\, y,\, z) \rightarrow (\lambda^{4+2n}x,\, \lambda^{6+3n}y,\, \lambda^n z).
\label{eq:xyzzp}
\eeq
The Weierstrass form \eqref{eq:Weierstrassform} is homogeneous under these $C^*$ actions with degrees $12$ and $12+6n$.

Singularities $G$ and $G'$ are realized by demanding that $f$, $g$ and the discriminant $\Delta$, 
\beq
 \Delta(z,z') = 4f(z,z')^3+27 g(z,z')^2,
\label{eq:WeierDelta}
\eeq
have suitable vanishing orders near $z=0$ and $1/z=0$ (see Table \ref{tab:Kodaira}).
\begin{table}[htb]
\caption{The Kodaira classification of singularities}
\begin{center}
\begin{tabular}{r@{\hspace{2.0cm}}r@{\hspace{2.0cm}}c@{\hspace{1.5cm}}cc}
\hline 
$\mbox{ord}(f)$     &  ord$(g)$      &  ord$(\Delta)$    &   Fiber type  & \quad  Singularity type   \\  \hline
  $\geq 0$  &  $\geq 0$    &      $0$        &   smooth     &     none              \\ 
  $0$      &    $0$       &      $n$        &   $I_n$      &    $A_{n-1}$          \\
 $\geq 1$  &    $1$       &      $2$        &   $II$       &     none              \\
  $1$      &    $\geq 2$  &      $3$        &   $III$      &    $A_1$              \\
 $\geq 2$  &    $2$       &      $4$        &   $IV$       &    $A_2$              \\  
  $2$      &    $\geq 3$  &      $n+6$      &   $I_n^{*}$  &    $D_{n+4}$          \\
  $\geq 2$ &    $3$       &      $n+6$      &   $I_n^{*}$  &    $D_{n+4}$          \\
  $\geq 3$ &    $4$       &      $8$        &   $IV^{*}$   &    $E_6$              \\
  $3$      &    $\geq 5$  &      $9$        &   $III^{*}$  &    $E_7$              \\
 $\geq 4$  &    $5$       &      $10$       &   $II^{*}$   &    $E_8$              \\ 
 $\geq 4$  &    $\geq 6$  &   \hspace{-0.6cm}  $\geq 12$  &   non-min    &       --- \\ \hline
\end{tabular}   
\label{tab:Kodaira}
\end{center}
\end{table}

For realizing $ADE$ gauge symmetry in six dimensions, these Kodaira fibers should satisfy the so-called split conditions \cite{BIKMSV},
which we will require in our construction since all the gauge symmetries that we would like to achieve (Table~\ref{tab:MWLsu7su8})
are of the $A$ type.
(The explicit form of the split condition can be seen in Table \ref{tab:Tate} below.) 

The last row of Table \ref{tab:Kodaira} expresses the singularity that cannot be resolved by any blow up of the fiber.
To resolve it, blowing up the base is required 
and an additional tensor multiplet appears in the spectrum. 
Such a theory does not have a heterotic dual in the perturbative regime, 
and hence is not the subject of our study.

The elliptic fibration over $F_n$ can also be written in Tate's form \cite{BIKMSV,KatzMorrisonNamekiSully}
\beq
  0 = y^2+x^3+a_1 x y +a_2 x^2 +a_3 y + a_4 x +a_6,
\label{eq:Tate0}
\eeq
where $a_i$ are polynomials of $z$ and $z'$.
Because of the homogeneity under \eqref{eq:xyzzp0}, $a_i$ are degree $2i$ polynomials in $z$:
\beq
  a_i(z,z') = \sum_{j=0}^{2i}a_{ij} z^j.
\label{eq:Tateform}
\eeq
Here $a_{ij}$ are polynomials of $z'$. Their degrees are determined by the homogeneity under \eqref{eq:xyzzp} as
\beq
 \mbox{deg}\,(a_{ij}) =  2i+(i-j)n.
\label{eq:polydegree}
\eeq
Discriminant $\Delta$ is given by
\beq
  \Delta = -b_2^2\, b_8 + 27\, b_6^2 + 4\, b_4^3 - 18\, b_2 b_4 b_6,
\label{eq:Deltab}
\eeq
where $b_n$ ($n=2,4,6,8$) are defined in terms of $a_i$ as follows:
\bes
 b_2 & = -\frac{a_1^2}{4}+a_2, \\
 b_4 & = -\frac{a_1a_3}{2}+a_4,  \\
 b_6 & = a_6-\frac{a_3^2}{4}, \\
 b_8 & = b_4^2 - 4 b_2 b_6.
\label{eq:b2b4b6b8}
\ees

Singularities $G$ and $G'$ are realized by requiring $a_i$ ($i=1,2,3,4,6$) to vanish at suitable orders of $z$,
known as Tate's algorithm (see Table \ref{tab:Tate}).
\begin{table}[htb]
\caption{Tate's algorithm}
\begin{center}
\begin{tabular}{l@{\hspace{0.2cm}}l@{\hspace{0.2cm}}l@{\hspace{0.2cm}}l@{\hspace{0.2cm}}l|l@{\hspace{0.7cm}}l|l@{\hspace{0.1cm}}l}
\hline 
{\scriptsize ord$(a_1)$}&{\scriptsize ord$(a_2)$}
                 &{\scriptsize ord$(a_3)$}
                          & {\scriptsize ord$(a_4)$}
                                   &{\scriptsize ord$(a_6)$}
                                            &{\scriptsize ord$(\Delta)$}
                                                     &  Coefficient of $z^{\mbox{\scriptsize ord}(\Delta)}$      & Fiber       &  Group          \\ \hline
 $0$    &   $0$  &  $1$   &  $1$   &  $1$   &  $1$      &  \,$4 \, b_{20}^3 \, a_{61}$                         &  $I_1$ & \quad --- \\ 
 $0$    &   $0$  &  $1$   &  $1$   &  $2$   &  $2$      &  \hspace{-0.1cm}\,$-\,b_{20}^2 \, b_{82}$               &  $I_2$      &  $SU(2)$               \\ 
 $0$    &   $1$  &  $1$   &  $2$   &  $3$   &  $3$      &  \hspace{-0.1cm}$\frac{1}{16}\, a_{10}^3 (a_{31}^3 - a_{10} b_{83})$
                                                                                                                  &  $I_3^{\ms}$ &  $SU(3)$     \\ 
 $0$    &   $1$  &  $k$   &  $k$   &  $2k$  &  $2k$     &  \hspace{-0.4cm}$-\frac{1}{16}\, a_{10}^4 \, b_{8,2k}$  &  $I_{2k}^{\ms}$ &  $SU(2k)$     \\  
 $0$    &   $1$  &  $k$   &$k\ps1$&$2k\ps1$&$ 2k\ps1$   &  \hspace{-0.4cm}$-\frac{1}{16}\, a_{10}^4 \, b_{8,2k+1}$&  $I_{2k+1}^{\ms}$ & $SU(2k\ps1)$ \\ \hline
 $1$    &   $1$  &  $1$   &  $1$   &  $1$   &  $2$      &  \hspace{-0.3cm}\, $27\,a_{61}^2$                      &  $II$  & \quad --- \\
 $1$    &   $1$  &  $1$   &  $1$   &  $2$   &  $3$      &  \hspace{0.0cm}\,$4 \, a_{41}^3$                       &  $III$       &  $SU(2)$     \\ 
 $1$    &   $1$  &  $1$   &  $2$   &  $3$   &  $4$      &  \hspace{-0.1cm}$\frac{27}{16}\, a_{31}^4$             &  $IV^{\ms}$  &  $SU(3)$     \\
 $1$    &   $1$  &  $2$   &  $2$   &  $4$   &  $6$      &  \hspace{-0.4cm}$-\frac{1}{16}\,p^2(p+a_{21})^2(p-a_{21})^2$   
                                                                                                                 &  $I_0^{*\,\ms}$&  $SO(8)^{\dagger}$ \\ 
 $1$    &   $1$  &  $k$   & $k\ps1$ &$2k\ps1$&  $2k\ps3$ &  \hspace{-0.1cm}\,$-\,a_{21}^3 \, a_{3k}^2$            &  $I_{2k-3}^{*\,\ms}$&  $SO(4k\ps2)$    \\ 
 $1$    &   $1$  & $k\ps1$&  $k\ps1$&$2k\ps1$&  $2k\ps4$ &  \hspace{-0.1cm}\,$-\,q^2\,a_{21}^2$            &  $I_{2k-2}^{*\,\ms}$&  $SO(4k\ps4)^{\dagger}$ \\ 
 $1$    &   $2$  &  $2$   &  $3$   &  $5$   &  $8$      &  \hspace{-0.1cm}$\frac{27}{16}\, a_{32}^4$             &  $IV^{*\,\ms}$ &  $E_6$       \\ 
 $1$    &   $2$  &  $3$   &  $3$   &  $5$   &  $9$      &  \hspace{0.0cm}\,$4 \, a_{43}^3$                       &  $III^{*\,\ms}$&  $E_7$       \\ 
 $1$    &   $2$  &  $3$   &  $4$   &  $5$   &  $10$     &  \hspace{-0.3cm}\, $27\,a_{65}^2$                      &  $II^{*\,\ms}$ &  $E_8$       \\ 
 $1$    &   $2$  &  $3$   &  $4$   &  $6$   &  $12$     &  [$\Delta/z^{12}]_{z=0}$                   &  non-min  & \quad  --- \\ \hline
\end{tabular}   
\label{tab:Tate}
\end{center}
\end{table} 
The vanishing orders of $a_i$ not only determine the type of the local singularity but also control its global structure.
We listed in Table \ref{tab:Tate} the fibers
for which the split condition is satisfied (the subscript ``s" is 
attatched\,\footnote{The fibers without subscript ``s" in the Table
have at most only one exceptional divisor, and hence
are necessarily of the split type.}).
The first six columns of the table represent the lowest orders of $a_i$ and $\Delta$ in $z$.
The next column is the coefficient of $z^{\mbox{\scriptsize ord}(\Delta)}$,
where $b_{nj}$ (or $b_{n,j}$) represents the coefficient of $z^j$ in $b_n$.
The ``split conditions" for Kodaira fibers are the conditions that the 
discriminants \eqref{eq:WeierDelta} should be the same forms as in this column.
Last two columns are the corresponding fiber degeneracy and the singularity type.\footnote{
To realize the groups with dagger, one more condition should be fulfilled. 
There are some polynomials $p$ and $q$ such that (here $a_{n,j}$ denotes $a_{nj}$)
\bes
   SO(8)~&:~ a_{21}^2-4a_{42}=p^2, \\
   SO(4k+4)~&:~ a_{4, k+1}^2 - 4 a_{21} a_{6, 2k+1} = q^2. \nonumber
\ees
}
The last row of the table expresses the singularity that never be resolved by blowing up the fiber.

The relation between the Weierstrass form \eqref{eq:Weierstrass} and Tate's form \eqref{eq:Tateform} is given by 
\bes
 f & = -\frac{b_2^2}{3}+b_4,  \\
 g & = \frac{2}{27}b_2^3-\frac{b_4 b_2}{3} +b_6. \\
\label{eq:TateWeierstrass}
\ees

\subsection{Mapping CY${}_3$ to RES-fibered geometry}
\label{sec:conditions}

The Weierstrass form of a rational elliptic surface is given by 
\beq
   0  = y^2+x^3+f(z)x+g(z),
\eeq
where $f$ and $g$ are sections of $\mathcal{O}(4)$ and $\mathcal{O}(6)$ on the base $\PP^1$
and have the form 
\beq
  f(z) = \sum_{i=0}^{4} f_i z^i,~~ g(z) = \sum_{j=0}^{6} g_j z^j.
\eeq

Suppose that we are given a CY${}_3$ with the Weierstrass form \eqref{eq:Weierstrassform}, whose 
$f$ and $g$ are sections of $\mathcal{O}(8)$ and $\mathcal{O}(12)$ on $\PP^1$ as in \eqref{eq:Weierstrass}, 
yielding a K3 fibration.
By using their coefficients $f_{8+(4-i)n}$ for $i\leq 4$ and $g_{12+(6-j)n}$ for $j\leq 6$, one can construct a geometry 
with a Weierstrass form 
\beq
 0 = y^2 + x^3 + f'(z,z')x+g'(z,z'),
\label{eq:RES0}
\eeq
where
\bes
 f'(z,z') & = \sum_{i=0}^{4} f_{8+(4-i)n}z^i, \\
 g'(z,z') & = \sum_{j=0}^{6} g_{12+(6-j)n}z^j
\label{eq:fdgd}
\ees
with discriminant
\beq
 \Delta' = 4f'^3+27g'^2.
\eeq
Here we regard $f'$ and $g'$ as sections of $\mathcal{O}(4)$ and $\mathcal{O}(6)$ on $\PP^1$ (with coordinate $z$).
Then the resulting geometry is a rational elliptic surface fibered over $\PP^1$ (with coordinate $z'$).
This gives the map from the K3-fibered CY${}_3$ to a RES-fibered geometry:
\beq
 \mbox{CY}_3 \rightarrow \mbox{RES-fibered geometry}.
\label{eq:CYtoRES}
\eeq 

When the rank of $G$ is large, we have to do a slight modification to $f'$ and $g'$
in order that the map does not change the singularity $G$.
For example, for $G=SU(n)$ with $n\leq 5$, the above procedure maps $G$ of a CY${}_3$ to the same $G$ in a RES-fibered geometry,
but for $G=SU(6)$, the na\"{i}ve mapping changes the singularity.
The explicit form of a CY${}_3$ with $G=SU(6)$ is given by~\cite{AGRT}
\bes
  f& =-\frac{\alpha^4 \beta^4}{48} - \frac{\alpha^2 \beta^3}{6}  \nu z 
      -\frac{\beta}{6}  (\alpha^2 \phi_2 + 2 \beta \nu^2) z^2 - (3 \beta \lambda + \frac{1}{3}\phi_2 \nu) z^3 
      + f_8 z^4 \\
   & \,\,+ f_{8-n} z^5 +\cdots f_{8-4n}z^8,  \\
  g& = \frac{\alpha^6 \beta^6}{864} + \frac{\alpha^4 \beta^5}{72} \nu z 
      +\frac{\alpha^2 \beta^3}{72} (\alpha^2 \phi_2 + 4 \beta \nu^2) z^2 
      +\frac{\beta^2}{108}\big(8 \beta \nu^3 + 9 \alpha^2 (3 \beta \lambda + \phi_2 \nu)\big) z^3 \\
   &\,\, + \frac{1}{36} \big( \alpha^2 (\phi_2^2 - 3 \beta^2 f_8) 
                             +4 \beta \nu (9 \beta \lambda + \phi_2 \nu)\big) z^4 
         + \frac{1}{12}(12 \phi_2 \lambda -4 \beta \nu f_8 - \alpha^2 \beta^2 f_{8-n}) z^5 
         + g_{12} z^6 \\
   &\,\,+g_{12-n}z^7+\cdots +g_{12-6n}z^{12}.
\ees
In this CY${}_3$, $\mbox{ord}(\Delta)=6$ and the gauge symmetry is $SU(6)$.
After mapping to $f'$ and $g'$ and calculating $\Delta'$, one can see that $\mbox{ord}(\Delta')=5$,
{\it i.e.}, the singularity is reduced by the map.
The source of this reduction is $f_{8-n}$.
Although $f_{8-n}$ is the coefficient of the term higher than $o(z^4)$ in $f$, it also appears in the $o(z^5)$ term in $g$.
Thus, after the map $g\rightarrow g'$, it is still contained in $g'$.
By setting this polynomial to zero in $g'$, one recovers $\mbox{ord}(\Delta')=6$ and 
the singularity remains to be $SU(6)$.

In summary, the map \eqref{eq:CYtoRES} is obtained by first replacing $f$ and $g$ to $f'$ and $g'$,
and then regarding $f'$ and $g'$ as sections of $\mathcal{O}(4)$ and $\mathcal{O}(6)$ of $\PP^1$,
and finally setting to zero the polynomials which constitute the coefficients of the terms higher 
than $o(z^4)$ in $f$ and $o(z^6)$ in $g$
and are still contained in  $f'$ and $g'$ even after the map $f,g\rightarrow f',g'$. 
Hereafter, the resulting $f'$, $g'$ and $\Delta'$ will be denoted by 
$\fres$, $\gres$ and $\Dres$.

This last step, however, does not work for some CY${}_3$ with sufficiently large rank $G$.
In such cases, one cannot recover the original singularity $G$ of CY${}_3$ in the RES-fibered geometry
and the map \eqref{eq:CYtoRES} does change the singularity. 

There are two such cases.
In the first case, additional singularities other than $G$ are produced by the map.
As an example, let us consider a CY${}_3$ with $G=SU(8)$ constructed by using Tate's algorithm.
The orders of $a_i$ in \eqref{eq:Tateform} are set to be $\mbox{ord}(a_i)=(0,1,4,4,8)$ (see Table \ref{tab:Tate}).
Its Weierstrass form is obtained by using \eqref{eq:b2b4b6b8} and \eqref{eq:TateWeierstrass}.
As a result, $\mbox{ord}(\Delta)= 8$ is realized.
Mapping to $f'$ and $g'$, one finds that $\mbox{ord}(\Delta') = 5$.
One can see that $f'$ and $g'$ contain the following  polynomials which constitute the coefficients of the terms 
higher than $o(z^4)$ in $f$ and $o(z^6)$ in $g$: 
\beq
 a_{12},a_{23},a_{24},a_{34},a_{35},a_{36},a_{45},a_{46}.
\label{eq:su8Tatezero}
\eeq
Setting these polynomials to zero, one recovers $\mbox{ord}(\Dres) = 8$,
but this $SU(8)$ singularity is accompanied by additional singularities. Explicitly, $\Dres$
has a factorized form
\beq
 \Dres = a_{44}^2 z^8 D_{8+4n} 
\label{eq:su8factorization0}
\eeq
with 
\bes
 D_{8+4n} &= -\frac{1}{16} \Big[ a_{10}^4 + 4 a_{10}^2(a_{10} a_{11} - 2 a_{21}) z 
                               + \big\{ a_{10}^2 (6a_{11}^2-8a_{22}) - 16 a_{21}(a_{10} a_{11} - a_{21}) \big\}  z^2  \\
                       & \quad \,\,    
                               + 4 (a_{11}^2-4a_{22})(a_{10} a_{11} - 2a_{21}) z^3 
                               + \big\{ a_{11}^2(a_{11}^2 - 8 a_{22}) + 16 (a_{22}^2 - 4 a_{44}) \big\}z^4 \Big].
\ees
Since the degree of $a_{44}$ is 8 , we can write $a_{44}(z') = \prod_{i=1}^8 (z'-z'_i)$.
Then we obtain
\beq
   \Dres =  z^8\, \prod_{i=1}^8(z'-z'_i)^2 \, D_{8+4n}.
\label{eq:su8factorization}
\eeq
At $z'=z'_i$, $\mbox{ord}(\Dres)$ in $z'$ is enhanced to 2.
Also, one can see that $\mbox{ord}(\fres)=\mbox{ord}(\gres)=0$ at these loci,
yielding $I_2$ fibers.
Thus an additional gauge symmetry $SU(2)$ appears along each line $z'=z'_i$
perpendicular to $z=0$ where the $SU(8)$ singularity exists.
To explain the origin of such additional singularities, let us go back to Tate's algorithm and focus on the orders of $a_3$ and $a_6$, 
which are 4 and 8 for $G=SU(8)$.
It means that $a_3(z)$ contains $a_{34},a_{35},a_{36}$ and $a_6(z)$ contains $a_{68},a_{69},\ldots,a_{6,12}$.
In $f'$ and $g'$, $a_{6j}$ are not contained since they only appear in $f$ and $g$ in the terms higher than $o(z^4)$ and $o(z^6)$.
Also, $\fres$ and $\gres$ do not contain $a_{3j}$,
since they are set to zero \eqref{eq:su8Tatezero}.
It means $a_3(z)=a_6(z)=0$ in $\fres$ and $\gres$. 
Then, from \eqref{eq:Deltab} and \eqref{eq:b2b4b6b8}, discriminant has a factorized form 
$\Dres = a_4^2(4a_4-b_2^2)$,
leading to unwanted additional singularities \eqref{eq:su8factorization0}.
In general, when the orders of $a_3$ and $a_6$ in Tate's algorithm exceed $3$ and $6$ simultaneously, 
additional singularities appear in the resulting RES-fibered geometry.

In the second case, the map \eqref{eq:CYtoRES} changes the singularity type of the fiber at $z=0$. 
This occurs when the singularity $G$ in a CY${}_3$ is not contained within one $E_8$ ($z=0$) 
but spreads into the other $E_8$ ($z=\infty$).
For example, let us consider $G=SU(8)$ again, 
but take another CY${}_3$ which is different from the one obtained from Tate's algorithm presented above 
and is constructed by working in the Weierstrass form directly.
As seen from \eqref{eq:TateWeierstrass}, a CY${}_3$ in Tate's form is always rewritten in the 
Weierstrass form, but the inverse is not the case in general.
This means that there may exist models that are described only in the Weierstrass form and have 
no Tate's counterpart (see {\it e.g.}, \cite{KleversTaylor}).
Hence the appearance of additional singularities \eqref{eq:su8factorization} in RES-fibered geometry
may be an artifact in Tate's form.
There may exist another $SU(8)$ model that cannot be reached by Tate's algorithm 
and for that CY${}_3$ additional singularity may not arise after the map \eqref{eq:CYtoRES}.
A candidate of such a CY${}_3$ is the one constructed in \cite{AGRT}.
The explicit form is given by\,\footnote{This is the $r=0$ case given by setting $\delta=1$ and $\zeta_4=0$ in \cite{AGRT}.}
\bes
\scalebox{0.7}{$\displaystyle f $} 
   & \scalebox{0.7}{$\displaystyle = -\frac{1}{48}\tau^4  -  \frac{1}{6}\tau^2 \zeta_3 z  -  \frac{1}{6}(2\zeta_3^2  +  \tau^2 \omega_1)z^2
    -  \frac{1}{3}(9\tau^2\lambda_2  +  2\zeta_3\omega_1)z^3
    +  \Big( \phi_4  -  \frac{1}{3}(18\zeta_3\lambda_2  +  \omega_1^2)\Big)z^4  +  \psi_5 z^5  + f_6 z^6+f_7z^7+f_8z^8, $}\\
\scalebox{0.7}{$\displaystyle g $} 
   & \scalebox{0.7}{$\displaystyle =\frac{1}{864}\tau^6+\frac{1}{72}\tau^4 \zeta_3 z+\frac{1}{72}\tau^2(4\zeta_3^2+\tau^2 \omega_1)z^2
   +\frac{1}{108}(8\zeta_3^3+12\tau^2 \zeta_3 \omega_1+27\tau^4\lambda_2)z^3  
   +\frac{1}{36}\Big( -3\tau^2\phi_4+ 2(4\zeta_3^2\omega_1 + 27\tau^2\zeta_3\lambda_2 + \tau^2\omega_1^2)\Big)z^4 $} \\
   & \,\, \scalebox{0.7}{$\displaystyle +\frac{1}{36}\Big( -3(4\zeta_3\phi_4 + \tau^2\psi_5) + 4(18\zeta_3^2\lambda_2
                + 2\zeta_3\omega_1^2 + 9\tau^2\lambda_2\omega_1)\Big) z^5 
                +\Big(-\frac{1}{12}\tau^2f_6-\frac{1}{3}(\omega_1\phi_4+\zeta_3\psi_5)
               +\frac{2}{27}\omega_1^3+2\zeta_3\lambda_2\omega_1+9\tau^2\lambda_2^2 \Big)z^6  $}\\
   & \,\, \scalebox{0.7}{$\displaystyle +\Big( -\frac{1}{3}(18\lambda_2 \phi_4+\omega_1 \psi_5)-\frac{1}{3}\zeta_3f_6
                                        -\frac{1}{12}\tau^2 f_7 \Big)z^7+ g_8 z^8 +g_9 z^9+g_{10}z^{10} +g_{11} z^{11}+g_{12}z^{12},   $} \\
\scalebox{0.7}{$\displaystyle \Delta $} & = \scalebox{0.7}{$\displaystyle z^8 \, E_{24+4n} $}
\label{eq:su8fg0}
\ees
with an irreducible polynomial
\beq
\scalebox{0.7}{$\displaystyle
 E_{24+4n} = -\frac{1}{192} \tau^4(12 \phi_4^2 + 144 \phi_4 \zeta_3 \lambda_2 + 432 \zeta_3^2 \lambda_2^2 - 12 g_8 \tau^2 - 4 f_7 \zeta_3 \tau^2 - 
 72 \lambda_2 \psi_5 \tau^2 - f_8 \tau^4 - 4 f_6 \tau^2 \omega_1 - 432 \lambda_2^2 \tau^2 \omega_1)+o(z) $}.
\eeq
The important difference from the CY${}_3$ constructed by Tate's algorithm is that the terms higher than 
$o(z^4)$ in $f$ or $o(z^6)$ in $g$ contain a term written by only the polynomials that are needed to 
express $\fres$ and $\gres$.
It is $-6\lambda_2 \phi_4 z^7$ contained in $o(z^7)$ of $g$. 
It is the sign that this singularity $G=SU(8)$ can not be realized in a rational elliptic surface fibration 
but we need a full-fledged K3 fibration.
As a result, the map \eqref{eq:CYtoRES} changes the singularity as follows.
Mapping $f$ and $g$ to $f'$ and $g'$, one obtains $\mbox{ord}(\Delta') =5$.
The $o(z^5)$ and $o(z^6)$ terms of $g'$ contain the polynomials
\beq
 \psi_5, f_6,
\label{eq:su8zero}
\eeq
which are the coefficients of the terms higher than $o(z^4)$ in $f$ and $o(z^6)$ in $g$.
Setting them to zero, we have $\mbox{ord}(\Dres)=7$.
Thus, the singularity is reduced by the map \eqref{eq:CYtoRES}.\footnote{If we set the term $-6\lambda_2 \phi_4 z^7$ of $g$ 
to zero in advance, {\it i.e.}, if we set $\lambda_2=0$ or $\phi_4=0$ in \eqref{eq:su8fg0}, the $SU(8)$ singularity is contained within one $E_8$. 
In these cases, mapping to $f'$ and $g'$ gives $\mbox{ord}(\Delta')=5$
and setting $\psi_5$ and $f_6$ \eqref{eq:su8zero} to zero gives $\mbox{ord}(\Dres)=8$.
However, the resulting $\Dres$ has a factorized form and additional singularities arise.
That is, these cases result in Tate's case described above.
Explicitly, $\Dres = z^8 \phi_4^2 \tilde{E}_{8+4n}$
for $\lambda_2=0$ and $\Dres = z^8 \lambda_2^2 \tilde{E}_{16+6n}$
for $\phi_4=0$. 
}

\section{Geometry for $G = SU(2)\times SU(2) \times SU(2)$ (No.7)}
\label{sec:su2su2su2}

The starting point of our construction is the CY${}_3$ for the $G=SU(2)\times SU(2)\times SU(2)$ model.
It is No.7 of \cite{OguisoShioda} and its Mordell-Weil lattice is $E(K) = D_4^*\oplus A_1^*$.
As we claimed in the previous section, there are models that can be written in the Weierstrass form but cannot be written in Tate's form.
Therefore, we use the Weierstrass form throughout this paper, but for the $G=SU(2)\times SU(2)\times SU(2)$ model,
we start from Tate's form and convert it to the Weierstrass form.\footnote{We can alternatively start from the Weierstrass form for $G=SU(6)$
and Higgsing it to derive the $SU(2)\times SU(2)\times SU(2)$ model, but the construction by using Tate's algorithm is more straightforward.}
This is because Tate's algorithm is more convenient for realizing product gauge groups. 

\subsection{The heterotic spectrum}

In this model, since $H= SO(8) \times SU(2)$ ($E(K) = D_4^*\oplus A_1^*$) is the Cartan type, the heterotic spectrum is 
determined in a standard way. 
Let us divide the $12+n$ instantons taking values in $H$ into each factor $SO(8)$ and $SU(2)$ such that 
\bes
  SO(8)~ & :~ n+8-r,  \\
  SU(2)~ & :~ 4+r.
\label{eq:D4A1instdist}
\ees
By using the index theorem, we obtain the spectrum as in Table~\ref{tab:su2su2su2spec}.
\begin{table}[htb]
\caption{The spectrum for $G=SU(2)\times SU(2)\times SU(2)$}
\begin{center}
\begin{tabular}{|c|c|}
\hline 
Representation  & Multiplicity  \\ \hline \hline
$(\bf{3},\bf{1},\bf{1}),~(\bf{1},\bf{3},\bf{1}),~(\bf{1},\bf{1},\bf{3})$   &  $-1$ ($SU(2)\times SU(2)\times SU(2)$ vector)  \\
$(\bf{1},\bf{1},\bf{1})$           &  $6n+25-4r$  \\
$(\bf{2},\bf{1},\bf{1}),(\bf{1},\bf{2},\bf{1}),(\bf{1},\bf{1},\bf{2})$  & $2n+16+2r$   \\
$(\bf{2},\bf{2},\bf{1}),(\bf{2},\bf{1},\bf{2}),(\bf{1},\bf{2},\bf{2})$  & $n-r$ \\
$\frac{1}{2}(\bf{2},\bf{2},\bf{2})$               &   $r$ \\ \hline                                                          
\end{tabular}   
\label{tab:su2su2su2spec}
\end{center}
\end{table}
The number $n(H)$ of the hypermultiplets is given by
\beq
 n(H) = (6n+25-4r)+(2n+16+2r)\cdot 3\cdot 2+(n-r)\cdot 3\cdot 2^2+r\cdot \frac{1}{2}2^3=30\,n+121,
\eeq
while the number of vector multiplets reads $n(V) = 3\times 3 = 9$. 
Therefore, anomaly cancellation condition is satisfied as
\beq
 n(H)-n(V) = 30\,n+112.
\eeq

\subsection{$r=0$ case}

It is necessary to put the $A_1$ singularities on three different lines.
We first put the $A_1$ singularity at $z=0$. 
From Tate's algorithm (Table \ref{tab:Tate}), the orders $\mbox{ord}(a_i)$ are given by $(0,0,1,1,2)$, and hence
$a_{ij}$ should satisfy
\beq
a_{30}=a_{40}=a_{60}=a_{61}=0.
\label{eq:A1cond}
\eeq
Next we put the second $A_1$ on another line.
This line is taken to be
\beq
   \tilde{z} \equiv z+h_{n}=0.
\eeq
$z$ is shifted by an order $n$ polynomial $h_n(z')$, so that $\tilde{z}$ is homogeneous under the transformation \eqref{eq:xyzzp}.
To realize the $A_1$ singularity at $\tilde{z}=0$, we re-expand $a_i$ in terms of $\tilde{z}$ as
\beq
  a_i(z)= a_i(\tilde{z}-h_n) = \sum_{j=0}^{2i} \tilde{a}_{ij}\tilde{z}^j
\eeq
and impose on $\tilde{a}_{ij}$ the same conditions as in \eqref{eq:A1cond},
\beq
\tilde{a}_{30}=\tilde{a}_{40}=\tilde{a}_{60}=\tilde{a}_{61}=0.
\eeq
Finally, we place the third $A_1$ at
\beq
 \hat{z} \equiv z+h'_n = 0
\eeq 
and re-expand $a_i$ as
\beq
  a_i(z)= a_i(\hat{z}-h'_n) = \sum_{j=0}^{2i} \hat{a}_{ij}\hat{z}^j
\eeq
and impose 
\beq
\hat{a}_{30}=\hat{a}_{40}=\hat{a}_{60}=\hat{a}_{61}=0.
\eeq
The resulting form of $a_i$'s are given as follows:
\bes
 a_1(z) & =a_{10} + a_{11} z + a_{12}z^2, \\
 a_2(z) & =a_{20} + a_{21} z + a_{22} z^2 +a_{23}z^3+a_{24}z^4, \\
 a_3(z) & = z (z + h_n) (z + h'_n)\{(a_{33}+\cdots)+(a_{34}+\cdots)z+\cdots +(a_{36}+\cdots)z^3\}, \\
 a_4(z) & = z (z + h_n) (z + h'_n) \{(a_{43}+\cdots)+(a_{44}+\cdots)z+\cdots +(a_{48}+\cdots)z^5\}, \\
 a_6(z) & = z^2 (z + h_n)^2 (z + h'_n)^2 \{(a_{66}+\cdots)+(a_{67}+\cdots)z+\cdots +(a_{6,12}+\cdots)z^{6}\},
\label{eq:A1A1A1curveTate}
\ees
where $(\cdots)$ are written by $a_{ij}$, $h_n$ and $h'_n$.
To translate this Tate's form into the Weierstrass form, we calculate $b_i$ \eqref{eq:b2b4b6b8}.
Although $b_i$ have complicated dependence of $a_{ij}$, $h_n$ and $h'_n$, one can redefine $a_{ij}$ 
so that they are arranged to have simple forms:
\bes
 b_2 & = A_{20}+A_{21}z+\cdots+A_{24}z^4 , \\
 b_4 & = z (z+h_n) (z+h'_n) \{A_{43}+A_{44}z+\cdots +A_{48}z^5\}, \\
 b_6 & = z^2 (z+h_n)^2 (z+h'_n)^2 \{A_{66}+A_{67}z+\cdots +A_{6,12}z^6 \}.
\label{eq:su2su2su2b2b4b6}
\ees
Substituting them into \eqref{eq:TateWeierstrass}, we obtain $f$ and $g$ in terms of $A_{ij}$.
To arrange so that the middle polynomials $f_8$ and $g_{12}$ have simple form, we further redefine as
\bes
  \tilde{A}_{44} & \equiv  A_{44}-\frac{1}{3} A_{22}^2,  \\
  \tilde{A}_{66} & \equiv  A_{66}+\frac{1}{27}(2A_{22}^3-9A_{22}A_{44}). 
\label{eq:A44A66redef}
\ees
Similarly, we redefine $A_{6j}$ with $j=7,8,9,10$ by subtracting $A_{22}$ dependent terms: 
\bes
 \tilde{A}_{6j}  & \equiv A_{6j}-\frac{1}{3} A_{22} A_{4\,j-2} \quad \quad (j=7,8,9,10).
\label{eq:A6jredef}
\ees
Moreover, we found that $A_{23}$ and $A_{24}$ can be discarded from $f$ and $g$ without any effect on the singularity structure.
As a result, we obtain the following form:
\bes
 \scalebox{0.7}{$\displaystyle f $} 
   & = \scalebox{0.7}{$\displaystyle -\frac{1}{3}A_{20}^2+\Big(-\frac{2}{3}A_{20}A_{21}+A_{43}\sigma_2 \Big)z
                                     +\frac{1}{3}\Big(-A_{21}^2-2A_{20}A_{22}+3 A_{43}\sigma_1+(A_{22}^2+3\tilde{A}_{44})\sigma_2 \Big)z^2  $} \\
   & \,\, \scalebox{0.7}{$\displaystyle  +\frac{1}{3}\Big(-2A_{21}A_{22}+3 A_{43}+(A_{22}^2+3\tilde{A}_{44})\sigma_1+3A_{45}\sigma_2 \Big)z^3 
                                       +\Big(\tilde{A}_{44}+A_{45}\sigma_1+A_{46}\sigma_2  \Big)z^4
                                       +(A_{45}+A_{46}\sigma_1+A_{47}\sigma_2)z^5  $} \\
   & \,\, \scalebox{0.7}{$\displaystyle  +(A_{46}+A_{47}\sigma_1+A_{48}\sigma_2)z^6+(A_{47}+A_{48}\sigma_1)z^7 +A_{48}z^8  $}  \\
 \scalebox{0.7}{$\displaystyle g $} 
   & = \scalebox{0.7}{$\displaystyle \frac{2}{27}A_{20}^3+\frac{1}{9}\Big( 2A_{20}^2 A_{21} - 3A_{20}A_{43}\sigma_2 \Big)z  $} \\
   & \,\, \scalebox{0.7}{$\displaystyle   + \frac{1}{27}\Big( 6(A_{20}A_{21}^2+A_{20}^2A_{22})-9A_{20}A_{43}\sigma_1
                                          -3(3A_{21}A_{43}+A_{20}A_{22}^2+3A_{20}\tilde{A}_{44})\sigma_2 
                                          +(A_{22}^3+9A_{22} \tilde{A}_{44}+27\tilde{A}_{66}) \sigma_2^2 \Big)z^2 $} \\
   & \,\, \scalebox{0.7}{$\displaystyle + \frac{1}{27}\Big( 2A_{21}^3+12 A_{20}A_{21}A_{22}-9A_{20}A_{43}
                                                     -3(3A_{21}A_{43}+A_{20}A_{22}^2+3 A_{20}\tilde{A}_{44})\sigma_1
                                                     -3(3A_{22}A_{43}+A_{21}A_{22}^2+3A_{21}\tilde{A}_{44}+3A_{20}A_{45})\sigma_2 $} \\
   & \quad \quad \scalebox{0.7}{$\displaystyle  +2(A_{22}^3+9A_{22}\tilde{A}_{44}+27 \tilde{A}_{66})\sigma_1\sigma_2 
                                                +9 (A_{22} A_{45}+3 \tilde{A}_{67})\sigma_2^2 \Big)z^3  $}  \\
   & \,\, \scalebox{0.7}{$\displaystyle + \frac{1}{27}\Big( 3(2 A_{21}^2 A_{22} +  A_{20} A_{22}^2 - 3 A_{21} A_{43} -3 A_{20} \tilde{A}_{44})
                                                       -3 (3 A_{22} A_{43} + A_{21}A_{22}^2+3A_{21} \tilde{A}_{44} + 3A_{20} A_{45})\sigma_1 
                                                       +(A_{22}^3+9A_{22}\tilde{A}_{44}+27 \tilde{A}_{66})\sigma_1^2  $}\\
   & \quad \quad \scalebox{0.7}{$\displaystyle -(A_{22}^3-9A_{22} \tilde{A}_{44} +9 A_{21} A_{45} + 9A_{20} A_{46} - 54 \tilde{A}_{66})\sigma_2
                                      +18(A_{22}A_{45}+3 \tilde{A}_{67}) \sigma_1\sigma_2 + 9(A_{22} A_{46}+3 \tilde{A}_{68})\sigma_2^2 \Big)z^4  $}  \\      
   & \,\, \scalebox{0.7}{$\displaystyle +\frac{1}{27} \Big( 3( A_{21} A_{22}^2 - 3 A_{22} A_{43} - 3 A_{21} \tilde{A}_{44} - 3 A_{20} A_{45}) 
                                                         -(A_{22}^3-9A_{22} \tilde{A}_{44} + 9A_{21} A_{45} +9 A_{20} A_{46} - 54 \tilde{A}_{66})\sigma_1 
                                                         + 9 ( A_{22} A_{45}+ 3 \tilde{A}_{67}) \sigma_1^2 $} \\
   & \quad \quad \scalebox{0.7}{$\displaystyle  +9 ( 6 \tilde{A}_{67}+A_{22} A_{45} - A_{21} A_{46} - A_{20} A_{47})\sigma_2 
                                     +18 (A_{22} A_{46} + 3 \tilde{A}_{68})\sigma_1\sigma_2 +9(A_{22} A_{47}+3 \tilde{A}_{69})\sigma_2^2  \Big)z^5    $}  \\
   & \,\, \scalebox{0.7}{$\displaystyle +\frac{1}{3}\Big(3 \tilde{A}_{66}-A_{21}A_{45}-A_{20}A_{46} 
                                                          +(6 \tilde{A}_{67}+A_{22} A_{45} - A_{21} A_{46} - A_{20} A_{47})\sigma_1 
                                                          +(A_{22} A_{46}+3 \tilde{A}_{68})\sigma_1^2   $}\\
   & \quad \quad \scalebox{0.7}{$\displaystyle  +( 6 \tilde{A}_{68}+A_{22} A_{46} - A_{21} A_{47} - A_{20} A_{48})\sigma_2 
                                                +2(A_{22} A_{47}+3 \tilde{A}_{69})\sigma_1\sigma_2 
                                                +(A_{22} A_{48}+3 \tilde{A}_{6,10})\sigma_2^2       \Big)z^6    $}   \\
   & \,\, \scalebox{0.7}{$\displaystyle +\frac{1}{3} \Big( 3\tilde{A}_{67} - A_{21} A_{46} - A_{20} A_{47}
                                                         + (6\tilde{A}_{68}+A_{22} A_{46} - A_{21} A_{47} - A_{20} A_{48})\sigma_1 
                                                         +(A_{22} A_{47}+ 3 \tilde{A}_{69})\sigma_1^2  $} \\
   & \quad \quad \scalebox{0.7}{$\displaystyle +(6\tilde{A}_{69}+A_{22} A_{47} - A_{21} A_{48})\sigma_2 
                                               +2(A_{22} A_{48} +3 \tilde{A}_{6,10})\sigma_1\sigma_2 +3A_{6,11}\sigma_2^2      \Big)z^7 $}  \\
   & \,\, \scalebox{0.7}{$\displaystyle +\frac{1}{3} \Big( 3\tilde{A}_{68}- A_{21} A_{47} - A_{20} A_{48}
                                                         +(6\tilde{A}_{69}+A_{22} A_{47} - A_{21} A_{48})\sigma_1 
                                                         +(A_{22} A_{48} + 3\tilde{A}_{6,10})\sigma_1^2   $}  \\
   & \quad \quad \scalebox{0.7}{$\displaystyle +(6\tilde{A}_{6,10}+A_{22} A_{48})\sigma_2 + 6A_{6,11}\sigma_1\sigma_2 +3A_{6,12}\sigma_2^2 \Big) z^8  $}   \\
   & \,\, \scalebox{0.7}{$\displaystyle +\frac{1}{3} \Big(3 \tilde{A}_{69} - A_{21} A_{48} + (6 \tilde{A}_{6,10}+A_{22} A_{48})\sigma_1
                                                         +3A_{6,11}(\sigma_1^2+2\sigma_2) +6A_{6,12}\sigma_1\sigma_2 \Big)z^9  $}  \\
   & \,\, \scalebox{0.7}{$\displaystyle +\Big(\tilde{A}_{6,10}+2A_{6,11}\sigma_1+A_{6,12}(\sigma_1^2+2\sigma_2) \Big)z^{10} 
                                        +\Big( A_{6,11}+2A_{6,12}\sigma_1 \Big) z^{11} + A_{6,12}z^{12}$}. 
\label{eq:su2su2su2fg}
\ees
Here the dependence on $h_n$ and $h'_n$ is included only through the symmetric polynomials  $\sigma_1, \sigma_2$ as it should be because 
the lines of $A_1$ singularity can be replaced with each other: 
\bes
 \sigma_1 & = h_n+h'_n, \\
 \sigma_2 & = h_n h'_n.
\label{eq:sigma12}
\ees
The discriminant has the form
\beq
 \Delta = z^2(z+h_n)^2(z+h'_n)^2\,D_{6n+24},
\label{eq:su2su2su2Delta}
\eeq
where $D_{6n+24}$ is an irreducible polynomial written by $A_{ij}$, $h_n$ and $h'_n$.
The leading expansion is given by
\bes
 \scalebox{0.7}{$\displaystyle D_{6n+24} $}
&  \scalebox{0.7}{$\displaystyle \, =\frac{1}{27} \, A_{20}^2 \, (4 A_{20} A_{22}^3 - 27 A_{43}^2 
                                                               + 36 A_{20} A_{22} \tilde{A}_{44} + 108 A_{20} \tilde{A}_{66}) $}  \\
&  \,\, \scalebox{0.7}{$\displaystyle+\frac{2}{9} \Big( A_{20}\, \big\{- 3 A_{43} (A_{20} A_{22}^2 +3 A_{21} A_{43} +3 A_{20}\tilde{A}_{44})
    +2 A_{20} A_{21} (A_{22}^3+ 9 A_{22} \tilde{A}_{44}+27 \tilde{A}_{66})  
    +6 A_{20}^2 A_{22} A_{45} + 18 A_{20}^2 \tilde{A}_{67} \big\} $} \\
& \quad \quad \scalebox{0.7}{$\displaystyle  -3 A_{43} \big\{ - 6 A_{43}^2 
                                             +A_{20} (A_{22}^3 + 9 A_{22} \tilde{A}_{44}+ 27\tilde{A}_{66}) \big\} \, \sigma_2   \Big)\, z
    +o(z^2) $} .
\label{eq:su2su2su2Dfactor}
\ees

We map these $f$ and $g$ to $f'$ and $g'$.
Then $f'$ and $g'$ contain the polynomials
\bes
 & A_{45},A_{46},A_{47},A_{48},  \\
 & \tilde{A}_{67},\tilde{A}_{68},\tilde{A}_{69},\tilde{A}_{6,10},
\label{eq:truncatedsu2su2su2}
\ees
which constitute the coefficients of the terms higher than $o(z^4)$ in $f$ and $o(z^6)$ in $g$.
Setting them to zero, one obtains a RES-fibered geometry.
One can see from the explicit forms of $\fres$, $\gres$ and $\Dres$ that the singularity remains to be $G=SU(2)\times SU(2)\times SU(2)$
(see \eqref{eq:fsu2su2su2exp}, \eqref{eq:gsu2su2su2exp}
and \eqref{eq:Dsu2su2su2exp} below).

From this geometry, let us extract the matter spectrum.
Singlets correspond to the moduli space $\mathcal{M}(H)$ of the gauge bundle $H$, which 
is identified with the complex moduli of the geometry, except for the middle polynomials $f_8$ and $g_{12}$ belonging to the
geometric moduli of the heterotic K3.
The number of singlets, therefore, is the number of degrees of independent polynomials contained in 
$\fres$ and $\gres$ except $f_8$ and $g_{12}$.
Charged matters are localized at codimension two loci of singularities on the base space. 
Suppose the generic codimension one singularity is $G$ and it enhances to $G'$ at a
codimension two locus.
Then there is a matter in a representation corresponding to the off diagonal part of $G\subset G'$.\footnote{It is justified in the 
type IIB picture by using the notion of string junctions \cite{TT}.}

In the present case, $\fres$ and $\gres$
contain the following six independent polynomials (apart from $\tilde{A}_{44}$ and $\tilde{A}_{66}$, which are not counted because 
they correspond to the middle polynomials $f_8$ and $g_{12}$):
\beq
\begin{array}{|c|c|}
\hline 
\mbox{Polynomial}  & \mbox{Degree}  \\ \hline \hline
h_n            &   n      \\
h'_n           &   n      \\
A_{20}         &   4+2n   \\
A_{21}         &   4+n    \\
A_{22}         &   4      \\
A_{43}         &   8+n    \\ \hline                                                    
\end{array}   
\label{eq:su2su2su2poly}
\eeq
The degree of each polynomial is given by \eqref{eq:polydegree}. The number $n(H_0)$ of the singlet (neutral hypermultiplet) is thus calculated as
\beq
 n(H_0) = n+n+(4+2n)+(4+n)+4+(8+n)+6-1 = 6n+25,
\label{eq:su2su2su2singlet}
\eeq
where $(-1)$ is the overall rescaling.

To see where and how the singularity enhances, 
we expand $\fres,\gres$ and $\Dres$ near each of the three lines 
$z=0,\tilde{z}=0$ and $\hat{z}=0$. The results are
\bes
\fres&=-\frac{1}{3}A_{20}^2+o(z) \\
 &= -\frac{1}{3}{\tilde{A}_{20}}^2+o(\tilde{z}) \\
 &= -\frac{1}{3}{\hat{A}_{20}}^2+o(\hat{z}), 
\label{eq:fsu2su2su2exp}
\ees
\bes
\gres&=\frac{2}{27}A_{20}^3+(\cdots)A_{20}\,z+o(z^2) \\
 &=\frac{2}{27}{\tilde{A}_{20}}^3+(\cdots)\tilde{A}_{20}\, \tilde{z}+o(\tilde{z}^2) \\
 & =\frac{2}{27}{\hat{A}_{20}}^3+(\cdots)\hat{A}_{20}\, \hat{z}+o(\hat{z}^2), 
\label{eq:gsu2su2su2exp}
\ees
\bes
\Dres
       & = A_{20}^2\, h_n^2 \, {h'_n}^2 P_{16+2n}\, z^2+(\cdots)h_n\, {h'_n}\,z^3+o(z^4) \\
       & = {\tilde{A}_{20}}^2 \, h_n^2 \, (h_n-h'_n)^2 \tilde{P}_{16+2n}\, \tilde{z}^2+(\cdots)h_n\, (h_n-h'_n)\, \tilde{z}^3+o(\tilde{z}^4) \\
       & = {\hat{A}_{20}}^2 \,{h'_n}^2\, (h_n-h'_n)^2 \hat{P}_{16+2n}\,  \hat{z}^2+(\cdots)h'_n\, (h_n-h'_n)\, \hat{z}^3+o(\hat{z}^4).
\label{eq:Dsu2su2su2exp}
\ees
Here
\bes
 \tilde{A}_{20} &= A_{20}-A_{21}h_n +A_{22}h_n^2,  \\
 \hat{A}_{20} &= A_{20}-A_{21}h'_n +A_{22}{h'_n}^2 
\ees
and $P_{16+2n},\tilde{P}_{16+2n}$ and $\hat{P}_{16+2n}$ are degree $16+2n$ irreducible polynomials.
We represented as $(\cdots)$ the factors irrelevant to the symmetry enhancement for simplicity.
From these expansions, one can read the loci and types of enhancements, and then one finds what kind of representations of matters appear at those points.
The result is given by
\beq
\begin{array}{l|c|c@{\hspace{0.2cm}}c@{\hspace{0.2cm}}c|c||c} \hline 
\mbox{\quad \, Zero}  & \mbox{Degree} & {\small \mbox{ord$(\fres)$}}
                                             & {\small \mbox{ord$(\gres)$}}
                                                   & {\small \mbox{ord$(\Dres)$}} & \mbox{Enhancement}  & \mbox{Matter} \\  \hline
A_{20},\tilde{A}_{20},\hat{A}_{20} & 4+2n    & 1   &  2   &  3    & \hspace{0.0cm} A_1 \rightarrow A_1 (I_2\rightarrow III)    &  \mbox{none} \\ \hline
P_{16+2n}         & 16+2n   & 0   &  0   &  3   & \hspace{0.0cm} A_1 \rightarrow A_2 &  (\bf{2},\bf{1},\bf{1}) \\
\tilde{P}_{16+2n} & 16+2n   & 0   &  0   &  3   & \hspace{0.0cm} A_1 \rightarrow A_2 &  (\bf{1},\bf{2},\bf{1}) \\
\hat{P}_{16+2n}   & 16+2n   & 0   &  0   &  3   & \hspace{0.0cm} A_1 \rightarrow A_2 &  (\bf{1},\bf{1},\bf{2}) \\ \hline
h_n               &   n     & 0   &  0   &  4   & \hspace{0.0cm} A_1\oplus A_1 \rightarrow A_3 &  (\bf{2},\bf{2},\bf{1}) \\  
h'_n              &   n     & 0   &  0   &  4   & \hspace{0.0cm} A_1\oplus A_1 \rightarrow A_3 &  (\bf{2},\bf{1},\bf{2}) \\   
h_n-h'_n          &   n     & 0   &  0   &  4   & \hspace{0.0cm} A_1\oplus A_1 \rightarrow A_3 &  (\bf{1},\bf{2},\bf{2}) \\ \hline
\end{array}   
\label{eq:su2su2su2charge}
\eeq
This table is obtained as follows. We first pay attention on the leading term of $\Dres$ in \eqref{eq:Dsu2su2su2exp}.
The coefficient is a product of several factors.
If one of them vanishes, the order of $\Dres$ enhances. 
The list of such factors are shown in the first column and its degree is shown in the second column.
When a factor vanishes, not only the order of $\Dres$ but also the orders of $\fres$ and $\gres$ enhance in general.
To what extent they will enhance can be read from \eqref{eq:fsu2su2su2exp}, \eqref{eq:gsu2su2su2exp}
and \eqref{eq:Dsu2su2su2exp} and given in the next three columns.
The resulting enhancement can be read from Table \ref{tab:Kodaira} and 
is shown in the next column.
The associated matter representation is given in the last column,
where the three entries correspond to the three $A_1$ at $z=0$, $\tilde{z}=0$ and $\hat{z}=0$ in this order.
As we can see, unresolvable singularity does not appear anywhere.

Let us give some comments.
Consider $A_{20}=0$ in the first row.
In this case, the fiber degeneracy is enhanced from $I_2$ to $III$,
but the singularity does not change and remains to be $A_1$. 
Therefore, there exists no matter at these points.
The same is true for $\tilde{A}_{20}=0$ and $\hat{A}_{20}=0$.
When $h_n=0$ (the fifth row), the orders of expansions around $z=0$ and $\tilde{z}=0$ simultaneously 
enhance to $A_3$, while the orders of $\hat{z}$ do not change. 
This reflects the fact that the two lines  $z=0$ and $\tilde{z}=0$ intersect at $h_n=0$,
giving rise to matter in the bi-fundamental representation $(\bf{2},\bf{2},\bf{1})$.
A similar thing is true for $h'_n=0$ and $h_n-h'_n=0$. 

Together with the singlets~\eqref{eq:su2su2su2singlet}, the total spectrum is given by 
\bes
 6n+25~ &:~ (\bf{1},\bf{1},\bf{1})  \\
 2n+16~ &:~ (\bf{2},\bf{1},\bf{1}), (\bf{1},\bf{2},\bf{1}), (\bf{1},\bf{1},\bf{2}) \\
     n~ &:~ (\bf{2},\bf{2},\bf{1}), (\bf{2},\bf{1},\bf{2}), (\bf{1},\bf{2},\bf{2}) 
\ees
This F-theoretic spectrum coincides with the heterotic spectrum (Table~\ref{tab:su2su2su2spec}) with $r=0$.

\subsection{$r\neq0$ case}
\label{sec:No7instCY}

We next construct a CY${}_3$ for general distributions of instantons and then map it to a RES-fibered geometry.
As seen in the previous section, Tate's algorithm gives the $r=0$ case only. 
This means that the $r\neq 0$ case is not obtained by Tate's algorithm and 
we have to use the Weierstrass form. 
So we start from the Weierstrass form \eqref{eq:su2su2su2fg} for the $r=0$ case 
and deform the equation appropriately. 
In order to know how to deform it, we use one particular information 
from the heterotic spectrum in Table~\ref{tab:su2su2su2spec} as an input.
The only input data is the appearance of $r$ half-hypermultiplets in the tri-fundamental representation $\frac{1}{2}(\bf{2},\bf{2},\bf{2})$.
We determine the CY${}_3$ so that they are contained.
As we will show below, it turns out that this requirement uniquely determines the CY${}_3$.
After mapping to a RES-fibered geometry, we derive the full spectrum.
The resulting F-theoretic full spectrum perfectly matches with the heterotic full spectrum
(not only the part we used as an input).

It is expected that tri-fundamental representation is localized at a triple intersection point of three lines of
$A_1$ singularity. For a triple intersection point to exist, 
$z=0,\tilde{z}=0$ and $\hat{z}=0$ need to share 
a common solution, that is, 
\beq
 z=0,\quad h_n = h'_n =0.
\eeq
Therefore, $h_n$ and $h'_n$ have a common factor. Writing this factor as $t_r$, we have
\bes
h_n & =h_{n-r} \, t_r, \\
h'_n & = h'_{n-r} \, t_r.
\label{eq:insthnhdn}
\ees
Three lines intersect at $r$ points satisfying $z=0$ and $t_r(z')=0$.

The next question is how much the gauge symmetry  is enhanced 
where a tri-fundamental appears.
As we argue below, it should be
\beq
 SU(2)\times SU(2)\times SU(2) \rightarrow SO(8).
\label{eq:su2su2su2so8enhance}
\eeq
Let us first look at the branching
of $SO(8)\supset SU(2)\times SU(2)\times SU(2)$:
\beq
\bf{28}=(\bf{3},\bf{1},\bf{1})\oplus(\bf{1},\bf{3},\bf{1})\oplus(\bf{1},\bf{1},\bf{3})
        \oplus 3(\bf{1},\bf{1},\bf{1}) \oplus 2(\bf{2},\bf{2},\bf{2}).
\eeq
The tri-fundamental is contained in the off-diagonal elements.
Moreover, there is a maximal embedding $SO(8)\supset SU(2)\times SU(2)\times SU(2)\times SU(2)$, whose branching is given by
\beq
\bf{28}=(\bf{3},\bf{1},\bf{1},\bf{1})\oplus(\bf{1},\bf{3},\bf{1},\bf{1})\oplus(\bf{1},\bf{1},\bf{3},\bf{1})
        \oplus(\bf{1},\bf{1},\bf{1},\bf{3})\oplus(\bf{2},\bf{2},\bf{2},\bf{2}).
\label{eq:so8branching}
\eeq
In the branching of the maximal embedding $G\supset G' \times H'$ with $H'=SU(2)$, the representation of $G'$ which is combined with $\bf{2}$ of $H'$ 
is a pseudo-real representation.
In the present case \eqref{eq:so8branching}, the tri-fundamental representation of $G'=SU(2)\times SU(2)\times SU(2)$ is pseudo-real,
forming a half-hypermultiplet $\frac{1}{2}(\bf{2},\bf{2},\bf{2})$.\footnote{We list below the other examples of arising half-hypermultiplets 
given in \cite{BIKMSV}. Why half-hypermultiplets appear in these cases has been explained from string junctions' point of view \cite{TT}.  
\vspace{0.3cm}
\\ \hspace{1.0cm}
$
\begin{array}{l@{\hspace{0.8cm}}l@{\hspace{0.8cm}}l@{\hspace{0.8cm}}l}  \hline
\mbox{Enhancement}  & \mbox{Maximal embedding} & \mbox{Branching} & \mbox{Matter}  \\ \hline
E_7 \rightarrow E_8   & E_8 \supset E_7\times SU(2) & \bf{248}=(\bf{133},\bf{1})\oplus(\bf{1},\bf{3})
                                                                                \oplus(\bf{56},\bf{2}) & \frac{1}{2}\bf{56} \\
SO(12) \rightarrow E_7 & E_7 \supset SO(12)\times SU(2) & \bf{133}=(\bf{66},\bf{1})\oplus(\bf{1},\bf{3})
                                                                    \oplus(\bf{32},\bf{2}) & \frac{1}{2}\bf{32} \\ \hline \nonumber 
\end{array}   
$
}

In order that the singularity is enhanced to $SO(8)$ at $t_r=0$, we have to tune the geometry so that
$(\mbox{ord}(f),\mbox{ord}(g),\mbox{ord}(\Delta))=(2,3,6)$ at $t_r=0$.
Let us first notice that $\sigma_1=\sigma_2=0$ at $t_r=0$.
One can then easily see from \eqref{eq:su2su2su2fg} that $(\mbox{ord}(f),\mbox{ord}(g))=(2,3)$ at $t_r=0$ if and only if $A_{20}$ is factorized as
(recall that the degree of $A_{20}$ is given by \eqref{eq:polydegree})
\beq
 A_{20} = p_{4+2n-r}\, t_{r}.
\label{eq:instA20}
\eeq
Calculating $\Delta$ explicitly, however, we find that it is over-enhanced to $\mbox{ord}(\Delta)=8$ at $t_r=0$.
We have to suppress it to $\mbox{ord}(\Delta)=6$ while keeping $(\mbox{ord}(f),\mbox{ord}(g))=(2,3)$. 
To absorb the excessive factors of $t_r$ in $\Delta$, we have to factorize $t_r^{-1}$ for at least one polynomial as
\beq
 A = \frac{A^{\mbox{\scriptsize new}}}{t_r}.
\label{eq:inverse}
\eeq
But as a side effect, this replacement would cause poles $t_r^{-n}$ in several terms of $f$ and $g$.
To keep the orders of $f$ and $g$ fixed, it should be accompanied with the factorization of other polynomial
\beq
 B = B^{\mbox{\scriptsize new}}\, t_r.
\eeq
Take a term in $f$ or $g$. If $A$ is contained in the coefficient of that term as a product $AB$,
we can ``\,interchange the factor $t_r$" to absorb the pole as \cite{AGRT}
\beq
 AB = \frac{A^{\mbox{\scriptsize new}}}{t_r} B = A^{\mbox{\scriptsize new}}B^{\mbox{\scriptsize new}}.
\label{eq:trinterchange}
\eeq
On the other hand, if $A$ is contained as $A^n$ for some $n$, the pole cannot be cancelled.

Among the polynomials contained in the CY${}_3$ for $r=0$ case \eqref{eq:su2su2su2fg}, only those listed in \eqref{eq:su2su2su2poly} can 
change the singularity type near $z=0$.
$h_n,h'_n$ and $A_{20}$ are already factorized.
The remainings are $A_{21}$, $A_{22}$ and $A_{43}$.
Among them, $A_{22}$ is the one that should be divided by $t_r$. The reason is as follows.
Substituting \eqref{eq:insthnhdn} and \eqref{eq:instA20} into \eqref{eq:su2su2su2fg},
we obtain 
\bes
  \scalebox{0.7}{$\displaystyle f$} 
     &   \scalebox{0.7}{$\displaystyle = -\frac{1}{3}p_{4+2n-r}^2 t_r^2+\Big( -\frac{2}{3}\,p_{4+2n-r}\,A_{21}\, t_r+A_{43}\,\hat{\sigma}_2\,t_r^2\Big)z
         + \frac{1}{3}\Big(-A_{21}^2 -2A_{22}\,p_{4+2n-r}\,t_r+3A_{43}\,t_r \, \hat{\sigma}_1+(A_{22}^2+3\tilde{A}_{44})\hat{\sigma}_2\,t_r^2 \Big) z^2 $} \\
     &\,\,  \scalebox{0.7}{$\displaystyle +\frac{1}{3}\Big(-2A_{21}A_{22}+3A_{43}+(A_{22}^2+3\tilde{A}_{44})\, \hat{\sigma}_1 \, t_r
                                                           +3A_{45}\,\hat{\sigma}_2\, t_r^2 \Big) z^3
                                                           +(\tilde{A}_{44}+A_{45}\hat{\sigma}_1t_r+A_{46}\hat{\sigma}_2 t_r^2)z^4
                                          + o(z^5) $},
\label{eq:ftr}
\ees
where we defined the reduced symmetric polynomials as
\bes
 \hat{\sigma}_1 & = h_{n-r}+h'_{n-r}, \\
 \hat{\sigma}_2 & = h_{n-r}\, h'_{n-r}.
\label{eq:hatsigma12}
\ees
Let us look at the $o(z^2)$ term. It contains $A_{21}$ as $-\frac{1}{3}A_{21}^2$.
This term cannot be eliminated by any redefinitions of other polynomials in the $o(z^2)$ term.
Since it has the form $A^n$, the interchange of $t_r$ \eqref{eq:trinterchange} does not work.
We therefore cannot remove the negative power $t_r^{-2}$ that would arise if we imposed $A_{21}={A}^{\mbox{\scriptsize new}}_{21}/t_r$.
The same argument holds for the polynomial $A_{43}$ contained in the $o(z^3)$ term.
As a result, we are forced to set $A_{22}={A}^{\mbox{\scriptsize new}}_{22}/t_r$.
In this case, the procedure of interchanging $t_r$ does work.
To apply it, we have to do in advance some redefinitions of polynomials so that $A_{22}$ appears in $f$ and $g$ as the form $A_{22}\times B$.   

One can check from \eqref{eq:su2su2su2fg} the following fact:
if $A_{22}={A}^{\mbox{\scriptsize new}}_{22}/t_r$ was 
imposed, negative power terms would arise only at $o(z^3)$ in $f$ and $o(z^4)$, $o(z^5)$ in $g$. 
Consider the $o(z^5)$ terms in $g$ first.
The negative powers of $t_r$ arise from 
\bes
   g=\cdots+\frac{1}{3}\Big[-A_{22}\Big(A_{43}-\frac{1}{3}A_{22}A_{21}+\frac{1}{9}A_{22}^2\,\hat{\sigma}_1\, t_r\Big)+\cdots \Big]z^5+\cdots.
\ees
In order to have the form $A_{22}\times B$, we should redefine
\beq
 A'_{43} \equiv A_{43}-\frac{1}{3}A_{22}A_{21}+\frac{1}{9}A_{22}^2\,\hat{\sigma}_1 \, t_r.
\label{eq:A43d}
\eeq
Next, we rewrite the $o(z^3)$ terms in $f$ by using $A'_{43}$ instead of $A_{43}$. 
The negative powers of $t_r$ come from 
\bes
   f=\cdots+\frac{1}{3}\Big[-A_{22}\Big(A_{21}-\frac{2}{3}A_{22}\,\hat{\sigma}_1\, t_r\Big)+\cdots\Big]z^3+\cdots.
\ees
It requires the redefinition
\beq
 A'_{21} \equiv A_{21}-\frac{2}{3}A_{22}\,\hat{\sigma}_1\, t_r.
\label{eq:A21d}
\eeq
Now we are ready to perform the interchange of $t_r$. It is given by
\bes
 A_{22} & = \frac{q_{4+r}}{t_r} \, ,  \\ 
 A'_{43} & = j_{8+n-r}\, t_r \, , \\
 A'_{21} & = k_{4+n-r}\, t_r \, . 
\label{eq:instA22}
\ees
The remaining sources of negative powers of $t_r$ are contained in the $o(z^4)$ terms in $g$.
Substituting all the factorizations above into \eqref{eq:su2su2su2fg}, one finds that the worst terms are proportional to $t_r^{-1}$ and given by
\beq
 g = \cdots +\frac{1}{27}\Big[\, \frac{1}{t_r}\,3q_{4+r}^2 \Big\{p_{4+2n-r}-\frac{1}{3}q_{4+r}\, \hat{\sigma}_2\Big\}+\mbox{(regular terms)} \Big]z^4+\cdots.
\eeq
The negative power is cancelled if and only if $\{\cdots\}$ is proportional to $t_r$, {\it i.e.},
\beq
 p_{4+2n-r}=s_{4+2n-2r}\, t_r + \frac{1}{3}q_{4+r}\,\hat{\sigma}_2.
\label{eq:instp}
\eeq

One can show that all these factorizations and redefinitions not merely cancel the $t_r^{-n}$ terms,\footnote{Via the replacement 
$A_{22}= \frac{q_{4+r}}{t_r}$ \eqref{eq:instA22}, no pole arises in higher order terms.
It is because we subtracted the $A_{22}$ dependent terms in advance in \eqref{eq:A6jredef}.} 
but indeed realize just the desired order $(\mbox{ord}(f),\mbox{ord}(g),\mbox{ord}(\Delta))=(2,3,6)$ at $t_r=0$.
Also, $\Delta$ has the form
\beq
 \Delta = z^2(z+h_{n-r}t_r)^2(z+h'_{n-r}t_r)^2\, {D}_{6n+24},
\eeq
where $D_{6n+24}$ is the same polynomial as given in \eqref{eq:su2su2su2Delta} and is still irreducible after 
the above factorizations and redefinitions.
To summarize, we obtained a CY${}_3$ with enhancement $SU(2)\times SU(2)\times SU(2)\rightarrow SO(8)$ at $r$ points by imposing on \eqref{eq:su2su2su2fg} 
the factorizations \eqref{eq:insthnhdn}, \eqref{eq:instA20}, \eqref{eq:instA22}, \eqref{eq:instp}
with the redefinitions \eqref{eq:A43d}, \eqref{eq:A21d}.

After mapping $f$ and $g$ to $f'$ and $g'$ and then setting the polynomials \eqref{eq:truncatedsu2su2su2}
to zero, one obtains a RES-fibered geometry with $G=SU(2)\times SU(2)\times SU(2)$.
The explicit forms of $\fres$, $\gres$ and $\Dres$ are given in Appendix~C.1.

We have derived a RES-fibered geometry whose spectrum contains $r$ half-hypermultiplet $\frac{1}{2}(\bf{2},\bf{2},\bf{2})$.
However, it is not obvious whether the geometry reproduces the other part of the spectrum in Table~\ref{tab:su2su2su2spec}.
We will show this is the case.
The number of singlets is determined by the independent polynomials. 
They are the following 7 polynomials defined above: 
\beq
\begin{array}{|c|c|}
\hline 
\mbox{Polynomial}  & \mbox{Degree}  \\ \hline \hline
t_r            &   r        \\
h_{n-r}        &   n-r      \\
h'_{n-r}       &   n-r      \\
s_{4+2n-2r}    &   4+2n-2r  \\
k_{4+n-r}      &   4+n-r    \\
q_{4+r}        &   4+r      \\
j_{8+n-r}      &   8+n-r    \\ \hline                                                    
\end{array}   
\label{eq:su2su2su2instpoly}
\eeq
The number of the singlets are given by
\bes
 n(H_0) & = r+(n \m-\m r)+(n \m-\m r)+(4 \m+\m 2n \m-\m 2r)+(4 \m+\m n \m-\m r)+(4 \m+\m r)+(8 \m+\m n \m-\m r)+7-1-1  \\
     & =6n+25-4r.
\label{eq:su2su2su2instsinglet}
\ees
Apart from the $-1$ corresponding to the overall rescaling, one more $-1$ is performed,
because we can choose the leading coefficient of $t_r$ as 1 when factorizing $h_n=h_{n-r}t_r$ and $h'_n=h'_{n-r}t_r$.
If we counted the degrees of freedom of $h_{n-r}$ as $n-r+1$ and those of $t_r$ as $r+1$, it would be an overcounting.

The charged matter spectrum is derived in Appendix~C.1 by using the series expansions of $\fres,\gres$ and $\Dres$ near each of the three lines 
$z=0,\tilde{z}=0$ and $\hat{z}=0$.
The resulting charged matter multiplets \eqref{eq:su2su2su2instcharge} together with the singlets~\eqref{eq:su2su2su2instsinglet} 
give the total spectrum as
\bes
 6n+25-4r~ &:~ (\bf{1},\bf{1},\bf{1})  \\
 2n+16+2r~ &:~ (\bf{2},\bf{1},\bf{1}), (\bf{1},\bf{2},\bf{1}), (\bf{1},\bf{1},\bf{2}) \\
     n-r ~ &:~ (\bf{2},\bf{2},\bf{1}), (\bf{2},\bf{1},\bf{2}), (\bf{2},\bf{2},\bf{1}) \\
       r ~ &:~ \frac{1}{2}(\bf{2},\bf{2},\bf{2})
\ees
which perfectly reproduces the heterotic result (Table~\ref{tab:su2su2su2spec}).\footnote{The procedure of constructing geometry 
for general instanton distribution described in this subsection is applicable to other cases where 
the gauge bundle is divided into components $H=\otimes_i H_i$.
We have studied models whose $H_i$'s are all Cartan types. There are 7 such cases out of 74 Oguiso-Shioda classification:
Nos.7,10,11,14,15,18 and 26. Among them, CY${}_3$ of No.15 and No.26 have been constructed in \cite{MTLooijenga}. 
We have examined the other cases and have constructed the CY${}_3$ of Nos.7,10,11 and 18 with general instanton distributions. 
For each case, the F-theoretic spectrum 
precisely coincides with the heterotic spectrum.
However, we have not succeeded to construct CY${}_3$ for No.14 yet.}

\begin{figure}[htbp]
  \begin{center}
   \captionsetup{format=hang,margin=50pt}
   \caption{Discriminant locus for $G=SU(2)\times SU(2) \times SU(2)$ (No.7) : $r=0$ (left) and $r\neq 0$ (right)}
    \begin{tabular}{c}
      \begin{minipage}{0.5\hsize}
        \begin{center}
           \vspace{-0.4cm}
           \hspace{-0.6cm}
           \includegraphics[clip,width=8.7cm]{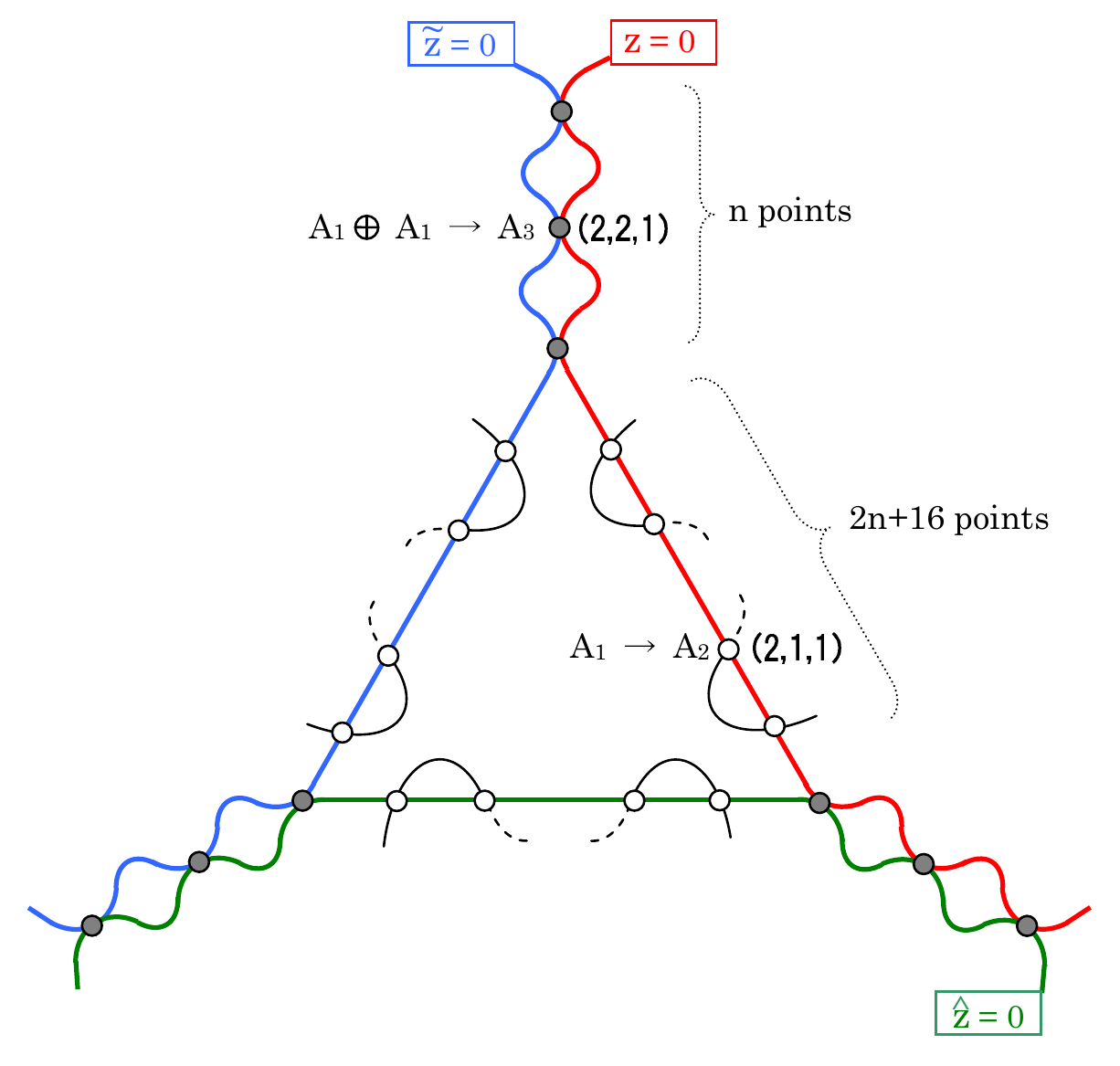}
        \end{center}
      \end{minipage}
      \begin{minipage}{0.5\hsize}
        \begin{center}
          \vspace{1.0cm}
          \includegraphics[clip,width=7.8cm]{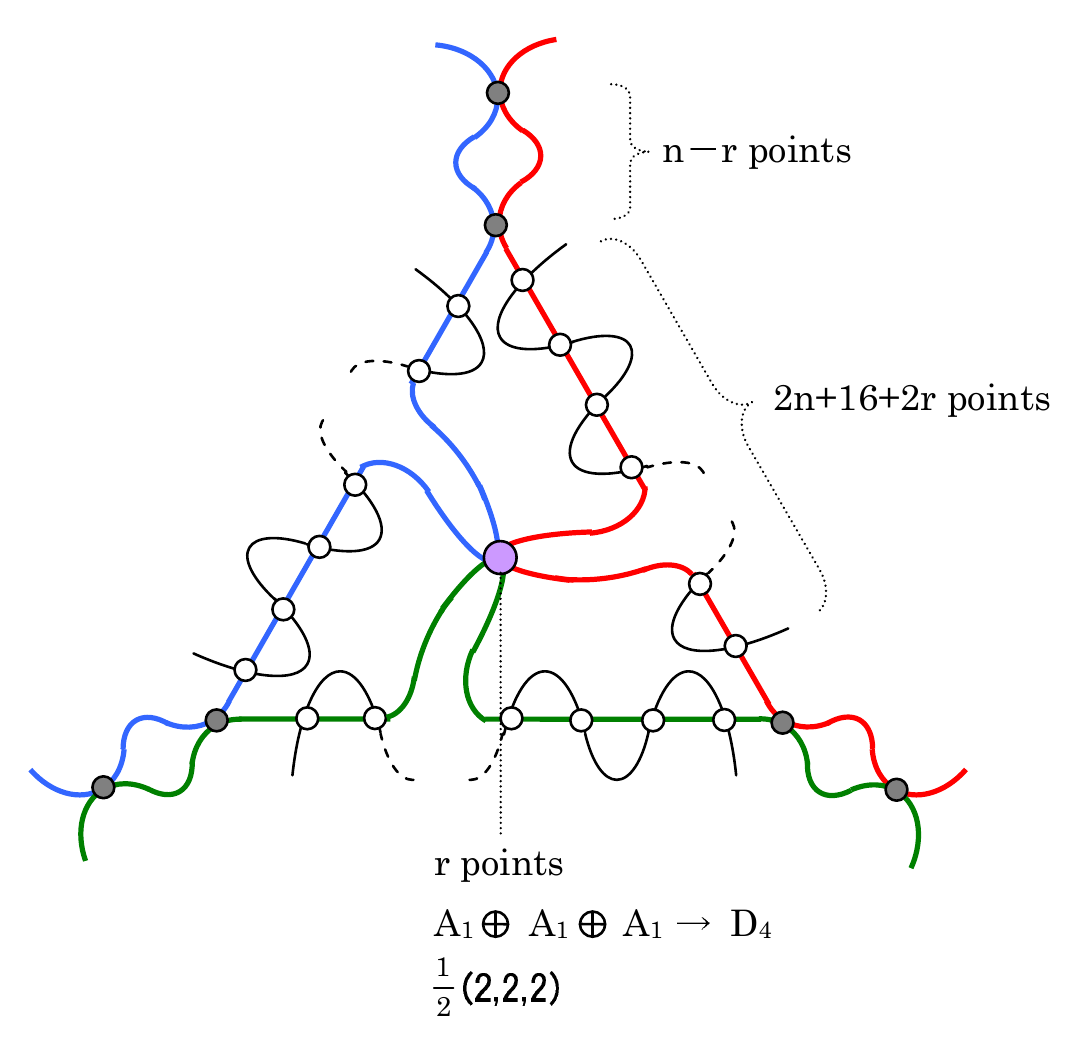}
        \end{center}
      \end{minipage}
    \end{tabular}
    \label{fig:A1A1A1}
  \end{center}
\end{figure}

A brief sketch for the structure of the symmetry enhancement is depicted in Figure \ref{fig:A1A1A1}.
The curves express the discriminant locus $\Dres=0$ (the shapes are not accurate). 
Each matter is localized at each intersection point.

\subsection{Looijenga's theorem}

Our geometry not only reproduces the heterotic spectrum but also 
encodes the structure of the moduli space $\mathcal{M}(H)$ of the 
gauge bundle $H$. 
On the heterotic side, Looijenga's theorem claims that the moduli space $\mathcal{M}(H)$ is parameterized by
the sections (we consider the six-dimensional case)
\beq
  \omega_k \in \Gamma(\mathcal{L}^{-d_k}\times \mathcal{N}^{s_k}) \quad \quad (k=0,\ldots , \mbox{rank}\,G).
\eeq
Here $d_k$ is the degree of the independent Casimir of $H$ ($d_0 \equiv 0$ for $k=0$), and
$s_k$ is the coefficient of the $k$-th coroot when the lowest root $-\theta$ is expanded ($s_0 \equiv 1$ for $k=0$).
$\mathcal{L}$ is the anti-canonical line bundle of the base $\PP^1$ of the heterotic K3 and
$\mathcal{N}$ is the ``twisting" line bundle over $\PP^1$.
Explicitly,
\bes
  \mathcal{L} &= \mathcal{O}(2), \\
  \mathcal{N} &= \mathcal{O}(p),
\ees
where $p$ corresponds to the instanton number. 
Therefore, degrees of $\omega_k$ with respect to the coordinate $z'$ of $\PP^1$ is given by
\beq
 \mbox{deg}(\omega_k) = p\, s_k -2 \, d_k.
\eeq
$z'$ is identified with the coordinate of the base $\PP^1$ of $F_n$, {\it i.e.}, the variable of polynomials $a_{ij}$.

In the present case, $H=SO(8)\times SU(2)$ with instanton numbers $(8+n-r,4+r)$.
Then we find that the sections $\omega_k$ in Looijenga's theorem exactly match the 7 independent polynomials 
\eqref{eq:su2su2su2instpoly} describing the geometry on the F-theory side (see Table \ref{tab:Looijenga}). 
\begin{table}[htb]
\caption{Looijenga's theorem}
\begin{center}
\begin{tabular}{|c|c|@{\hspace{0.4cm}}lll|l|}
\hline 
$H$      &   $p$     & $s_k$  & $d_k$  & deg$(\omega_k)$ & Polynomial           \\ \hline \hline
$SO(8)$  &  $8+n-r$  &   $1$  &  $0$   &  $8+n-r$   &  $j_{8+n-r}$         \\
         &           &   $1$  &  $2$   &  $4+n-r$   &  $k_{4+n-r}$         \\
         &           &   $1$  &  $4$   &  $n-r$     &  $h_{n-r}$               \\
         &           &   $1$  &  $4$   &  $n-r$     &  $h'_{n-r}$              \\
         &           &   $2$  &  $6$   &  $4+2n-2r$ &  $s_{4+2n-2r}$       \\   \hline
$SU(2)$  &  $4+r$    &   $1$  &  $0$   &  $4+r$     &  $q_{4+r}$           \\
         &           &   $1$  &  $2$   &  $r$       &  $t_r$               \\   \hline
\end{tabular}   
\label{tab:Looijenga}
\end{center}
\end{table}

This correspondence is also valid for $r=0$.
When $r=0$, the order of the polynomial $t_r$ becomes 0
and the independent polynomials are reduced to those of the $r=0$ case \eqref{eq:su2su2su2poly}.
In other words, the geometry for the $r=0$ case 
already captures the structure of the moduli space of the bundle $H$.
In the following discussion, we concentrate on the $r=0$ case.

\section{Geometry for non-Cartan $SU(7)$ series}
\label{sec:SU7series}

We have constructed the CY${}_3$ for $G=SU(2)\times SU(2)\times SU(2)$ in the previous section.
This is No.7 and is not contained in Table \ref{tab:MWLsu7su8},
but by unHiggsing the first $SU(2)$ successively, we obtain the CY${}_3$ for No.12 in the $SU(7)$ series and the CY${}_3$ for No.22 in the $SU(8)$ series as
\beq
  SU(2) \times SU(2) \times SU(2)_{(7)} \rightarrow  SU(3) \times SU(2) \times SU(2)_{(12)} \rightarrow SU(4) \times SU(2) \times SU(2)_{(22)}.
\eeq
CY${}_3$'s for other cases can be obtained by further unHiggsing them.
Once we obtain a CY${}_3$ for each case, we then map it to a RES-fibered geometry,
extract the matter spectrum and compare it with the heterotic result given in Section \ref{sec:heteroindex}, Appendix A and Appendix B.

\subsection{$G=SU(3)\times SU(2)\times SU(2)$ (No.12)}
\label{sec:No12CY}

Let us enhance the first $SU(2)$ of $G=SU(2)\times SU(2)\times SU(2)$ to $SU(3)$.
The discriminant for $G = SU(2)\times SU(2)\times SU(2)$ with $r=0$ is given by $\Delta = z^2(z+h_n)^2(z+h'_n)^2\, D_{6n+24}$ 
(see \eqref{eq:su2su2su2Delta}).
To enhance the first factor, we impose that $D_{6n+24}$ is factorized as
\beq
D_{6n+24}=z\cdot E_{5n+24},
\label{eq:su2su3enhance}
\eeq
where $E_{5n+24}$ should be an irreducible polynomial. 
For this, we require that the constant term of $D_{6n+24}$ vanishes. From \eqref{eq:su2su2su2Dfactor}, the condition is
\beq
  A_{20}^2\,(4A_{20}A_{22}^3-27A_{43}^2+36A_{20}A_{22}\tilde{A}_{44}+108A_{20}\tilde{A}_{66}) = 0.
\eeq
A solution is $A_{20}=0$, but it is not suitable. 
As seen from the explicit form of $f$ and $g$ \eqref{eq:su2su2su2fg}, $A_{20}=0$ enhances the orders not only of $\Delta$ but also of 
$f$ and $g$ near $z=0$ to be $(\mbox{ord}(f),\mbox{ord}(g),\mbox{ord}(\Delta))=(1,2,3)$. The fiber degeneracy enhances as $I_2 \rightarrow III$,
but the singularity is kept fixed as $A_1 \rightarrow A_1$, not enhance to $A_2$.
Thus we have to require the other factor vanishes. Namely,
\beq
 -27A_{43}^2+4 A_{20}(A_{22}^3+9A_{22}\tilde{A}_{44}+27\tilde{A}_{66}) = 0.
\eeq
Since it contains the squire $A_{43}^2$, the remaining part
$
  4 A_{20}(A_{22}^3+9A_{22}\tilde{A}_{44}+27\tilde{A}_{66})
$
should be a perfect square.
There are two solutions. 
First one is
$
 \tilde{A}_{66}=-\frac{1}{27}(A_{22}^3+9A_{22}\tilde{A}_{44})+A_{20}\, p_{4-n}^2,
$
and then $A_{43}$ is solved as
$
 A_{43}= 2A_{20}\,p_{4-n}.
$
However, one can check that the split condition is not satisfied by this solution.
The split condition for $SU(3)$ says that the coefficient of $z^3$ in $\Delta$ should be 
proportional to
$
  a_{10}^3
$
(see Table \ref{tab:Tate}), but in this solution it is proportional to 
$
 A_{20}^3,
$
as one can see by calculating the next to leading order term of $D_{6n+24}$ in \eqref{eq:su2su2su2Dfactor}.
The other solution is 
\bes
 A_{20} & = p_{2+n}^2, \\
 \tilde{A}_{66} & = -\frac{1}{27}(A_{22}^3+9A_{22}\tilde{A}_{44})+q_6^2,
\label{eq:A20A66}
\ees
and then
\beq
 A_{43} = 2\,p_{2+n}\,q_6.
\label{eq:A43}
\eeq
In this case, the coefficient of $z^3$ in the expansion of $\Delta$ is proportional to $p_{2+n}^3$, 
and hence the split condition is fulfilled.
Substituting \eqref{eq:A20A66} and \eqref{eq:A43} into \eqref{eq:su2su2su2Dfactor}, we obtain the explicit form of $E_{5n+24}$ \eqref{eq:su2su3enhance} as
\bes
 E_{5n+24} &= \frac{4}{3} p_{2 + n}^3 (- A_{22}^2 p_{2 + n}^2 q_6 +  3 A_{21} p_{2 + n} q_6^2
             - 3 \tilde{A}_{44} p_{2 + n}^2 q_6 +A_{22} A_{45} p_{2 + n}^3 + 3 \tilde{A}_{67} p_{2 + n}^3  - 3  q_6^3 \sigma_2)  \\
           &+o(z).
\label{eq:E}
\ees
One can check that $E_{5n+24}$ is irreducible and the gauge symmetry is in fact $SU(3)\times SU(2)\times SU(2)$.

Mapping $f$ and $g$ to $f'$ and $g'$, and  
setting the polynomials \eqref{eq:truncatedsu2su2su2} to zero in $f'$ and $g'$, 
one obtains the RES-fibered geometry with $G=SU(3)\times SU(2)\times SU(2)$.
The explicit forms of $\fres$, $\gres$ and $\Dres$ are given in Appendix~C.2.  
In this way, we obtain the geometry for the non-Cartan Mordell-Weil lattice $E(K) = \Lambda_{12}$ in Table \ref{tab:MWLsu7su8}.

Let us extract the matter spectrum from the resulting geometry.
We first notice that $\fres$ and $\gres$ contain the following six independent polynomials:
\beq
\begin{array}{|c|c|}
\hline 
\mbox{Polynomial}  & \mbox{Degree}  \\ \hline \hline
h_n            &   n      \\
h'_n           &   n      \\
p_{2+n}        &   2+n    \\
A_{21}         &   4+n    \\
A_{22}         &   4      \\
q_6            &   6      \\ \hline                                                    
\end{array}   
\label{eq:su3su2su2poly}
\eeq
For counting the number of singlets, it should be noted that the middle polynomial $\tilde{A}_{66}$ is 
written in terms of these polynomials as in \eqref{eq:A20A66}.
In other words, these six polynomials  include 
the degrees of freedom of $\tilde{A}_{66}$, which are the
geometric moduli of K3 on the heterotic side.
In order to focus on the gauge bundle moduli only, we subtract $13$ degrees of freedom corresponding to $\tilde{A}_{66}$ 
and return it to the geometric moduli of K3. 
As a result, we obtain 
\beq
 n(H_0) = n+n+(2+n)+(4+n)+(4)+(6)+6-1-13 = 4n+8.
\label{eq:su3su2su2singlet}
\eeq

The derivation of the charged matter spectrum is given in Appendix~C.2 and the result is summarized in \eqref{eq:su3su2su2charge}.
Together with the singlets \eqref{eq:su3su2su2singlet}, the full spectrum is given by
\bes
 4n+8~  &:~ (\bf{1},\bf{1},\bf{1})  \\
 2n+18~ &:~ (\bf{3},\bf{1},\bf{1})  \\
  n+16~ &:~ (\bf{1},\bf{2},\bf{1})  \\
  n+16~ &:~ (\bf{1},\bf{1},\bf{2})  \\
  n   ~ &:~ (\bf{3},\bf{2},\bf{1})  \\
  n   ~ &:~ (\bf{3},\bf{1},\bf{2})  \\
  n   ~ &:~ (\bf{1},\bf{2},\bf{2}) 
\label{eq:No12Fspec}
\ees
This F-theoretic spectrum is equivalent to the heterotic spectrum (see Appendix~A.2) given in Table~\ref{spectrumNo.12} with $r'_{U(1)}=2$,
or the one given in Table~\ref{spectrumNo.12(2)} with $(r,\tilde{r},r_{(17)})$ for $k=2$ in \eqref{eq:rrtr17}.

\subsection{$G=SU(5)\times SU(2)$ (No.17)}
\label{sec:su5su2}

The discriminant of CY${}_3$ for $G=SU(3)\times SU(2)\times SU(2)$ is 
$\Delta = z^3\,(z+h_n)^2\,(z+h'_n)^2\, E_{5n+24}$ (see \eqref{eq:su2su3enhance}).
One expect that tuning the parameters $h_n,\,h'_n$ causes the unHigssing process as follows:
\bes
   &  h'_n = 0      \hspace{1.1cm} \rightarrow ~ G=SU(5)\times SU(2), \\
   &  h'_n = h_n   \hspace{0.9cm} \rightarrow ~ G=SU(3)\times SU(4), \\
   &  h'_n = h_n = 0   ~  \rightarrow ~  G=SU(7).
\ees
In the first case, the tuning gives the expected enhancement.
However, in the second and third cases, na\"{i}ve tunings give a non-split fiber and 
unresolvable singularities, respectively, and hence we need some modifications of the geometry.

In this subsection, we consider the $G=SU(5)\times SU(2)$ case. The CY${}_3$ is obtained by merely setting
\beq
h'_n=0
\eeq
in the formulae obtained in the previous section.
%

We then map $f$ and $g$ to $f'$ and $g'$. The resulting $f'$ and $g'$ contain polynomials 
\beq
 A_{45}, A_{46}, A_{47}, \tilde{A}_{67}, \tilde{A}_{68},
\label{eq:su5su2hot}
\eeq
which constitute the coefficients of the terms higher than $o(z^4)$ in $f$ and $o(z^6)$ in $g$.
After setting them to zero, one obtains the RES-fibered geometry with $G=SU(5)\times SU(2)$.
The explicit forms of $\fres$, $\gres$ and $\Dres$ are given in Appendix~C.3.  
In this way, we obtain the geometry for the non-Cartan Mordell-Weil lattice $E(K) = \Lambda_{17}$ in Table \ref{tab:MWLsu7su8}.

The number of singlets is reduced by $n+1$ (of $h'_n$) from \eqref{eq:su3su2su2singlet}:
\beq
 n(H_0) = 4n+8-(n+1)= 3n+7.
\label{eq:su5su2singlet}
\eeq
The charged matter spectrum is derived in Appendix~C.3 and given in \eqref{eq:su5su2charge}.
Together with the singlets \eqref{eq:su5su2singlet}, the full spectrum is given by
\bes
 3n+7~  &:~ (\bf{1},\bf{1})  \\
 n+16~  &:~ (\bf{5},\bf{1})  \\
  n+16~ &:~ (\bf{1},\bf{2})  \\
  n+2~  &:~ (\bf{10},\bf{1})  \\
  n   ~ &:~ (\bf{5},\bf{2})
\ees
This F-theoretic spectrum is equivalent to the heterotic spectrum (see Section \ref{sec:No17hetero}) given in Table~\ref{spectrumNo.17} 
with $(r,\tilde{r})$ for $k=2$ in \eqref{eq:rrtNo17}.\footnote{This F-theoretic spectrum does not coincide with 
the heterotic spectrum given 
in Table~\ref{spectrumNo.17(2)}, which is coarser than the one in Table~\ref{spectrumNo.17}.
In order to obtain the corresponding geometry, 
we have to deform the one we found here in response 
to the change of the distribution of instantons, 
as we did for $G=SU(2)\times SU(2)\times SU(2)$, 
but so far, such a geometry has not been found.}

\subsection{$G = SU(3)\times SU(4)$ (No.19)}
\label{sec:su3su4}

Let us impose on the CY${}_3$ with $G=SU(3)\times SU(2)\times SU(2)$ the condition
\beq
h'_n=h_n
\label{eq:su3su4cond1}
\eeq
to obtain a CY${}_3$ with $G = SU(3)\times SU(4)$.
This tuning, however, breaks the split condition.
By the replacement $h'_n = h_n$, one expects the $SU(4)$ singularity appears at $\tilde{z}=z+h_n=0$.
One can show that the leading expansion of $\Delta$ near $\tilde{z}=0$ takes the form 
\beq
  \Delta =(p_{2+n}^2-A_{21}h_n+A_{22}h_n^2)^2 \, h_n^3 \, \tilde{T}_{16+n}\, \tilde{z}^4 + \cdots,
\label{eq:su3su4split}
\eeq
which is different from the split condition for $SU(4)$ (see Table \ref{tab:Tate})
\beq
 \Delta = -\frac{1}{16}\, a_{10}^4\, b_{84} \, \tilde{z}^4+\cdots.
\eeq
To fulfill this condition, we have to require the perfect square form for the factor in \eqref{eq:su3su4split} as
\beq
p_{2+n}^2-A_{21}h_n+A_{22}h_n^2 = \tilde{p}_{2+n}^2
\label{eq:su4split}
\eeq
with some order $2+n$ polynomial $\tilde{p}_{2+n}$.
For this, we need a term linear of $p_{2+n}$ in the l.h.s. Let us write
\beq
 A_{21} = p_{n+2}\,s_2+h_n r_4.
\label{eq:su3su4cond2}
\eeq
Then the l.h.s of \eqref{eq:su4split} is
perfect squire if and only if
\beq
 r_4 = A_{22}- \frac{1}{4} s_2^2,
\label{eq:su3su4cond3}
\eeq
and $\tilde{p}_{2+n}$ is solved as 
\beq
    \tilde{p}_{2+n} = p_{2+n}-\frac{1}{2}s_2 h_n.
\eeq

Therefore, unHiggsing to $SU(3)\times SU(4)$ is given by imposing on the CY${}_3$ of $SU(3)\times SU(2)\times SU(2)$ 
not only \eqref{eq:su3su4cond1} but also \eqref{eq:su3su4cond2} with \eqref{eq:su3su4cond3}.

Mapping $f$ and $g$ to $f'$ and $g'$,
and setting the polynomials (5.19) to zero in $f'$ and $g'$, one obtains the RES-fibered geometry with $G=SU(3)\times SU(4)$.
The explicit forms of $\fres$, $\gres$ and $\Dres$ are given in Appendix~C.4.  
In this way, we obtain the geometry for the non-Cartan Mordell-Weil lattice $E(K) = \Lambda_{19}$ in Table \ref{tab:MWLsu7su8}.

In the table \eqref{eq:su3su2su2poly} of the independent polynomials, $h'_n$ is eliminated and $A_{21}$ is replaced by $s_2$,
which reduce the degrees of freedom by $(n+1)+(5+n-3) = 2n+3$ from \eqref{eq:su3su2su2singlet}.
The number of singlets is hence given by
\beq
  n(H_0)=4n+8-(2n+3) = 2n+5.
\label{eq:su3su4singlet}
\eeq

The charged matter spectrum is derived in Appendix~C.4 and given in \eqref{eq:su3su4charge}.
Together with the singlets \eqref{eq:su3su4singlet}, the full spectrum is given by
\bes
 2n+5~  &:~ (\bf{1},\bf{1})  \\
 2n+18~ &:~ (\bf{3},\bf{1})  \\
  n+16~ &:~ (\bf{1},\bf{4})  \\
  n+2~  &:~ (\bf{1},\bf{6})  \\
  n   ~ &:~ (\bf{3},\bf{4})
\ees
This F-theoretic spectrum is equivalent to the heterotic spectrum (see Appendix~A.1)
given in Table~\ref{spectrumNo.19} with $(r,\tilde{\tilde{r}})$ for $k=2$ in \eqref{eq:rrtNo19},
or the one given in Table~\ref{spectrumNo.19(2)} with $r'_{U(1)}=1$.

\subsection{$G = SU(7)$ (No.25)}
\label{sec:su7}

To obtain a CY${}_3$ for $G=SU(7)$, we impose on the CY${}_3$ for $G=SU(3)\times SU(4)$ derived above a further condition 
\beq
  h_n=0.
\label{eq:su7cond1}
\eeq
However, an explicit calculation tells us that it contains unresolvable singularities.
Namely, $(\mbox{ord}(f),\mbox{ord}(g),\mbox{ord}(\Delta))=(4,6,12)$ at $p_{2+n}=0$.

To make the singularity milder, we have to do some redefinition of polynomials and ``absorb" the factors of $p_{2+n}$.
For this, look at the explicit form of $f$ :
\bes
  f &= -\frac{1}{3}p_{2+n}^4 -\frac{2}{3}p_{2+n}^3s_2  z-\frac{1}{3}(2A_{22}p_{2+n}^2+p_{2+n}^2s_2^2)z^2
     +\frac{2}{3}(3p_{2+n}q_6-A_{22}p_{2+n}s_2)z^3  \\
    &+\tilde{A}_{44}z^4+A_{45}z^5+A_{46}z^6+A_{47}z^7+A_{48}z^8,
\ees
where the order is enhanced to $\mbox{ord}(f)=4$ at $p_{2+n}=0$. 
Here, one can notice that $s_2$ always appears as a product form $p_{2+n}s_2$.
We can hence introduce a new polynomial $r_{4+n}$ and replace $s_2$ as
\beq
 s_2 = \frac{r_{4+n}}{p_{2+n}}.
\label{eq:su7cond2}
\eeq
The expansion of $f$ is rewritten as
\beq
 f  = -\frac{1}{3}p_{2+n}^4 -\frac{2}{3}p_{2+n}^2 r_{4+n} z-\frac{1}{3}(2A_{22}p_{2+n}^2+r_{4+n}^2)z^2+\cdots,
\label{eq:su7f}
\eeq
whose order at $p_{2+n}=0$ is suppressed to $\mbox{ord}(f)=2$.
Similar calculation for $g$ and $\Delta$ shows that the singularity at $p_{2+n}=0$ gets milder as 
$(\mbox{ord}(f),\mbox{ord}(g),\mbox{ord}(\Delta))= (2,3,9)$, 
which is $D_5$ singularity and can be resolved without any problem.

Therefore, unHiggsing to $G=SU(7)$ is obtained by imposing \eqref{eq:su7cond1} and \eqref{eq:su7cond2} 
on the CY${}_3$ for $G=SU(3)\times SU(4)$.\footnote{This CY${}_3$ is equivalent to the one constructed in \cite{AGRT} 
with $r=0$ ($\delta=1$ and $\zeta_1=0$).
The dictionary is as follows: 
$p_{2 + n} =\frac{1}{2} \xi$,
$r_{4+n} = \zeta_2$,
$A_{22} = \omega $,
$q_6 = -3 \lambda_1$,
$\tilde{A}_{44} = \psi_4 - \frac{1}{12} \omega^2$,
$A_{4j}=f_j$ ($5\leq j \leq 8$),
$\tilde{A}_{67}=g_7+\frac{1}{3}\zeta_2 f_6+ \frac{1}{12}\xi^2 f_7 $,
$\tilde{A}_{68}=g_8+\frac{1}{3}\zeta_2 f_7+ \frac{1}{12}\xi^2 f_8 $,
$\tilde{A}_{69}=g_9+\frac{1}{3}\zeta_2 f_8 $,
$\tilde{A}_{6,10}=g_{10}$ and
$A_{6j}=g_{j}$ ($j=10,11$).}

We then map $f$ and $g$ to $f'$ and $g'$. 
The resulting $f'$ and $g'$ contain polynomials $A_{45}$ and $A_{46}$, 
which constitute the coefficients of the terms higher than $o(z^4)$ in $f$ and $o(z^6)$ in $g$.
After setting them to zero, one obtains a RES-fibered geometry with $G=SU(7)$.
The explicit forms of $\fres$, $\gres$ and $\Dres$ are given in Appendix~C.5.  
In this way, we obtain the geometry for the non-Cartan Mordell-Weil lattice $E(K) = \Lambda_{25}$ in Table \ref{tab:MWLsu7su8}.

Among the independent polynomials for $G=SU(3)\times SU(4)$, $h_n$ is eliminated and 
$s_2$ is replaced by $r_{4+n}$. The degrees of freedom \eqref{eq:su3su4singlet} is reduced by $(n+1)+\{3-(5+n)\}=-1$, that is, increases by 1:
\beq
 n(H_0) = 2n+5+1 = 2n+6.
\label{eq:su7singlet}
\eeq
The charged matter spectrum is derived in Appendix~C.5 and given in \eqref{eq:su7charge}.
%
%
Together with the singlets \eqref{eq:su7singlet}, the full spectrum is given by
\bes
 2n+6~  &:~ \bf{1}  \\
  n+16~ &:~ \bf{7}  \\
  n+2~ &:~  \bf{21}
\ees
This F-theoretic spectrum is equivalent to the heterotic spectrum (see section~\ref{sec:No25hetero})
given in Table~\ref{spectrumNo.25} with $r'=1$.

\section{Geometry for non-Cartan $SU(8)$ series}

\subsection{$G=SU(4)\times SU(2)\times SU(2)$ (No.22)}
\label{sec:su4su2su2}

In section \ref{sec:No12CY}, we constructed a CY${}_3$ for $G=SU(3)\times SU(2)\times SU(2)$.
The discriminant has the form
$
 \Delta = z^3(z+h_n)^2(z+h'_n)^2 E_{5n+24}.
$
The enhancement $SU(3)\rightarrow SU(4)$ occurs if it factorizes as
\beq
 E_{5n+24}=z\cdot F_{4n+24},
\label{eq:su3tosu4}
\eeq
where $F_{4n+24}$ should be an irreducible polynomial.
The explicit form of $E_{5n+24}$ is given in \eqref{eq:E}.
One way to achieve (\ref{eq:su3tosu4}) is to set 
$p_{2+n}=0$ so that the $o(1)$ term of \eqref{eq:E} vanishes. 
This does not, however, lead to the enhancement to $SU(4)$ since 
in this case we have $(\mbox{ord}(f),\mbox{ord}(g),\mbox{ord}(\Delta))=(2,2,4)$,
which means that the fiber type is $IV$, and hence 
the singularity 
does not change $A_2\rightarrow A_2$. 
Thus we must require the other factor of the $o(1)$ term to vanish:
\beq
 - A_{22}^2 p_{2 + n}^2 q_6 +  3 A_{21} p_{2 + n} q_6^2
             - 3 \tilde{A}_{44} p_{2 + n}^2 q_6 +A_{22} A_{45} p_{2 + n}^3 + 3 \tilde{A}_{67} p_{2 + n}^3  - 3  q_6^3 h_n h'_n=0.
\label{eq:su4su2su2cond}
\eeq
Here we used $\sigma_2=h_nh'_n$. 

There are two solutions to \eqref{eq:su4su2su2cond}.
One solution is given by
\bes
 q_6 & = t_{4-n}p_{2+n}, \\
 \tilde{A}_{67}& = -\frac{1}{3}(A_{22} A_{45} - A_{22}^2 t_{4-n} - 3 \tilde{A}_{44} t_{4-n} 
                   + 3 A_{21} t_{4-n}^2 - 3 h_n h'_n t_{4-n}^3 ).
\label{eq:su4su2su2another}
\ees
Let us call this CY${}_3$, ``$\mbox{CY}_3^{(1)}"$.
In this solution, however, $\tilde A_{67}$, which is a coefficient of a term higher than $o(z^6)$ in $g$, contains terms 
written by only the polynomials which are needed to 
express $\fres$ and $\gres$.
This means that, to achieve the singularities, it is not sufficient 
to constrain the terms of $o(z^4)$ or lower in $f$ and 
of $o(z^6)$ or lower in $g$, but  we also need to impose 
conditions on the higher order terms, which are supposed 
to be independently tuned to describe the instanton 
bundle of the ``other" $E_8$ gauge group of the dual heterotic theory.
Thus, in the same way as in the $G=SU(8)$ case \eqref{eq:su8fg0} discussed in section 4.2, the map from $\mbox{CY}_3^{(1)}$
to a RES-fibered geometry changes the singularity.

The other solution can be found as follows.
From the form of \eqref{eq:su4su2su2cond}, we can see that it is
solved for $A_{21}$ by requiring that all the terms have 
a common factor of $p_{2+n}q_6^2$.
This is possible if\,\footnote{We can use $h'_n =r_{n-4}\, s'_4$ instead of the second equation. The resulting spectrum is equivalent.} 
\bes
  p_{2+n} & = r_{n-4}\, q_6,  \\
  h_n     & = r_{n-4}\, s_4.
\label{eq:su4su2su2sol1}
\ees
The solution is given by
\beq
 A_{21} =  \frac{1}{3} \Big(A_{22}^2 + 3 \tilde{A}_{44} -A_{22} A_{45} r_{n-4} - 3 \tilde{A}_{67} r_{n-4} \Big) r_{n-4} +  h'_n s_4.
\label{eq:su4su2su2sol2}
\eeq

This CY${}_3$, which we call $\mbox{CY}_3^{(2)}$, can be mapped to a RES-fibered geometry, where the singularity is unchanged. 
Mapping $f$ and $g$ to $f'$ and $g'$ and setting the polynomials \eqref{eq:truncatedsu2su2su2} contained in $f'$ and $g'$,
one obtains the RES-fibered geometry with $G=SU(4)\times SU(2)\times SU(2)$.
The explicit form of $\fres$, $\gres$ and $\Dres$ are given in Appendix~C.6.  
In this way, we obtain the geometry for the non-Cartan Mordell-Weil lattice $E(K) = A^*_1\oplus A^*_1\oplus \left< 1/4 \right>$ 
in Table \ref{tab:MWLsu7su8}.

The independent polynomials for $G=SU(3)\times SU(2)\times SU(2)$ \eqref{eq:su3su2su2poly} are reduced 
by \eqref{eq:su4su2su2sol1} and \eqref{eq:su4su2su2sol2}
to the following 5 polynomials:
\beq
\begin{array}{|c|c|}
\hline 
\mbox{Polynomial}  & \mbox{Degree}  \\ \hline \hline
r_{n-4}        &   n-4    \\
h'_n           &   n      \\
s_4            &   4      \\
A_{22}         &   4      \\
q_6            &   6      \\ \hline                                                    
\end{array}   
\label{eq:su4su2su2poly}
\eeq
The number of singlets is given by
\beq
 n(H_0) = (n-4)+n+(4)+(4)+(6)+5-1-13 = 2n+1.
\label{eq:su4su2su2singlet}
\eeq
Here we subtracted the degree 13 of $\tilde{A}_{66}$ for the same reason as in \eqref{eq:su3su2su2singlet}.

The charged matter spectrum is derived in Appendix~C.6 and given in \eqref{eq:su4su2su2charge}.
Together with the singlets \eqref{eq:su4su2su2singlet}, the full spectrum is given by
\bes
 2n+1~  &:~ (\bf{1},\bf{1},\bf{1})  \\
 2n+8~  &:~ (\bf{4},\bf{1},\bf{1})  \\
  6~    &:~ (\bf{6},\bf{1},\bf{1})  \\
 n+12~  &:~ (\bf{1},\bf{2},\bf{1})  \\
 16   ~ &:~ (\bf{1},\bf{1},\bf{2})  \\
  4   ~ &:~ (\bf{4},\bf{2},\bf{1})  \\
  n   ~ &:~ (\bf{4},\bf{1},\bf{2})  \\
  n   ~ &:~ (\bf{1},\bf{2},\bf{2})  \\
 n-4  ~ &:~ \frac{1}{2}(\bf{6},\bf{2},\bf{1})  
\label{eq:su4su2su2CYspec}
\ees
This F-theoretic spectrum is equivalent to the heterotic spectrum (see Appendix~B.2)
given in Table~\ref{spectrumNo.22} with $(r,\tilde{r})=(4,n)$.

Two comments are in order.
First, it is difficult to construct this geometry in Tate's form.
Recall the construction for $G=SU(2)\times SU(2)\times SU(2)$ given in section \ref{sec:su2su2su2}.
As seen in \eqref{eq:A1A1A1curveTate}, $a_i(z)$ contains $a_{ij}$
with orders greater than or equal to $\mbox{ord}(a_i) = (0,0,3,3,6)$. 
For example, $a_4(z)$ contains $a_{43},a_{44},\ldots,a_{48}$. 
These orders are determined by the sum of the orders of each $SU(2)$ factor.
Since an $SU(2)$ singularity has $\mbox{ord}(a_i) = (0,0,1,1,2)$, $SU(2)\times SU(2) \times SU(2)$ gives 
$\mbox{ord}(a_i) = (0,0,3,3,6)$.
Similar counting for $SU(4)\times SU(2) \times SU(2)$ gives $\mbox{ord}(a_i) = (0,1,4,4,8)$.
These orders are equal to those for $SU(8)$. In particular, $\mbox{ord}(a_3)$ and $\mbox{ord}(a_6)$ exceed 3 and 6.
As we have explained in section \ref{sec:conditions}, it leads to a factorized form of the discriminant $\Dres = a_4^2(4a_4-b_2^2)$,
yielding unwanted additional singularities.\footnote{
As for the $SU(7)$ series, they can be constructed by Tate's form.
For example, the counting of the orders for $SU(3)\times SU(2)\times SU(2)$ gives $\mbox{ord}(a_i)=(0,1,3,4,7)$,
and hence $a_{33}$ remains non-zero in $\fres$ and $\gres$.
Additional singularities do not arise in such cases.
Therefore, the advantage of the Weierstrass model starts from $SU(8)$ series. }

Second comment is : $\mbox{CY}_3^{(1)}$ of the first solution \eqref{eq:su4su2su2another}, which cannot be 
mapped to a RES-fibered geometry, is related to the known CY${}_3$ \eqref{eq:su8fg0} for $G=SU(8)$ \cite{AGRT}.
In fact, setting $h_n=h'_n=0$ in $\mbox{CY}_3^{(1)}$ causes the unHiggsing process $SU(4)\times SU(2)\times SU(2)\rightarrow SU(8)$,
and the resulting CY${}_3$ precisely coincides with \eqref{eq:su8fg0}.\footnote{The dictionary is as follows: 
$p_{2+n}=\frac{1}{2}\tau$, $A_{21}=\zeta_3$, $A_{22}=\omega_1$,
$t_{4-n}= -6\lambda_2$, $\tilde{A}_{44}=\phi_4 - \frac{1}{3} (18 \zeta_3 \lambda_2  + \omega_1^2)$, $A_{4j}=f_j$ ($5\leq j\leq 8$),
$\tilde{A}_{68}= g_8 + \frac{1}{3} \zeta_3 f_7 +\frac{1}{12}\tau^2 f_8$, $\tilde{A}_{69}=g_9 +\frac{1}{3}\zeta_3 f_8$,
$\tilde{A}_{6,10}=g_{10}$ and $A_{6j}=g_j$ ($j=11,12$). }
In other words, the $\mbox{CY}_3^{(2)}$ we have constructed in this paper, without which one can never make a connection with the theory 
of the Mordell-Weil lattice, belongs to a new branch different from the one connected to the known CY${}_3$
with $G=SU(8)$ of \cite{AGRT}.\footnote{Unlike $\mbox{CY}_3^{(1)}$, it seems that $\mbox{CY}_3^{(2)}$ cannot be
unHiggsed to any CY${}_3$ with $G=SU(8)$ (see the next subsection).}

\subsection{$G=SU(6)\times SU(2)$ (No.29)}
\label{sec:su6su2}

As seen from \eqref{eq:su3tosu4} and \eqref{eq:su4su2su2sol1}, the discriminant of CY${}_3$ for $G=SU(4)\times SU(2) \times SU(2)$ is 
$\Delta = z^4\, (z+r_{n-4}s_4)^2\,(z+h'_n)^2\,F_{2n+24} $.
By tuning the polynomials $r_{n-4}$, $s_4$ and $h'_n$ suitably, one may expect to continue similar unHiggsing processes.

Among them, we find that setting 
\beq
s_4 = 0
\eeq
yields a CY${}_3$ with $G=SU(6)\times SU(2)$.
We then map $f$ and $g$ to $f'$ and $g'$,
and set the polynomials \eqref{eq:su5su2hot} contained in $f'$ and $g'$ to zero. It gives a RES-fibered geometry with $G=SU(6)\times SU(2)$.
The explicit forms of $\fres$, $\gres$ and $\Dres$ are given in Appendix~C.7.  
In this way, we obtain the geometry for the non-Cartan Mordell-Weil lattice $E(K) = A_1^*\oplus \left< 1/6 \right>$ in Table \ref{tab:MWLsu7su8}.

Among the independent polynomials \eqref{eq:su4su2su2poly}, we have required $s_4=0$,
and hence the number of singlets is reduced by $4+1=5$ from \eqref{eq:su4su2su2singlet}, 
\beq
 n(H_0) = 2n+1-5= 2n-4.
\label{eq:su6su2singlet}
\eeq

The charged matter spectrum is derived in Appendix~C.7 and given in \eqref{eq:su6su2charge}.
Together with the singlets \eqref{eq:su6su2singlet}, the full spectrum is given by
\bes
 2n-4~  &:~ (\bf{1},\bf{1})  \\
 n+12~  &:~ (\bf{6},\bf{1})  \\
    6~  &:~ (\bf{15},\bf{1})  \\
 n-4~   &:~ \frac{1}{2}(\bf{20},\bf{1})  \\
  16~   &:~ (\bf{1},\bf{2})  \\
  n   ~ &:~ (\bf{6},\bf{2})
\label{eq:su6su2CYspec}
\ees
This F-theoretic spectrum is equivalent to the heterotic spectrum (see Appendix~B.1)
given in Table~\ref{spectrumNo.29} with $r=12$.

\subsection{Other cases}
\label{sec:othercases}

We expect that the other tunings give the other models in the $SU(8)$ series, but we have not obtained corresponding RES-fibered geometries yet.
For example, setting 
\beq
 h'_n = r_{n-4}s_4
\eeq
is expected to give a CY${}_3$ with $G=SU(4)\times SU(4)$, but it leads to unresolvable singularities.
For the enhancement $SU(3)\times SU(2)\times SU(2)\rightarrow SU(3)\times SU(4)$ discussed in section 
\ref{sec:su3su4}, we were able to deform such singularities to resolvable ones.
In the present case, however, we could  not 
find how to deform them suitably. Since we have not obtained a CY${}_3$ yet, we do not have a RES-fibered geometry, either.
There are two possibilities. First, there {\em are} such geometries 
but we just haven't found them yet. 
Second, one can prove that there are no such geometries in these cases.\footnote{The argument given in \cite{JohnsonTaylor} may be valid
to analyze these cases.}
It would be interesting to find which is correct, 
but we leave it for future work.
Also, we could not find the way of realizing the enhancement $SU(3)\times SU(2)\times SU(2)\rightarrow SU(3)\times SU(3)\times SU(2)$.
We also leave this issue for future investigations.

\section{Conclusions}

In this paper, we studied how the non-Cartan Mordell-Weil lattices are realized in 
the six-dimensional heterotic/F-theory duality.
We focused on the $SU(7)$ series and $SU(8)$ series given in Table \ref{tab:MWLsu7su8}.

On the heterotic string side, we give a procedure to derive the massless matter spectrum by using index theorem.
Except No.45, the gauge bundle $H$ is a product of a semi-simple group and one or more $U(1)$ group(s).
The total $12+n$ instantons are distributed into each group factor.
We found that the $U(1)$ instanton numbers are a multiple of some particular non-one integer $r$
and the resulting number of each matter is integral.
We showed that this integer $r$ is determined by the lattice spacing of the $U(1)$ direction in the $E_8$ lattice,
which is orthogonal to the lattice of the semi-simple part of $H$ and the singularity lattice $T$. 
For $\mbox{rank}(E(K))>2$ cases, more than one choices of the number of $U(1)$ direction(s) in $H$ are possible for the given MW lattice.
In these cases, more than one matter spectra are obtained for a gauge symmetry $G$.
We found that if the number of the $U(1)$ factor(s) is larger, the spectrum is finer.

On the dual F-theory side, we examined how to construct a geometry for a given non-Cartan MW lattice $E(K)$.
We first construct CY${}_3$ with singularity $G$ of the lattice $T$ for the given $E(K)$.
It starts from $G=SU(2)\times SU(2)\times SU(2)$ case, and successive unHiggsing processes 
are carefully performed. 
Then we construct the map from these CY${}_3$ to RES-fibered threefolds.
In order that the singularity $G$ is unchanged by the map, we generically have to do slight modifications on the geometry.
As a result, we obtained the explicit forms of RES-fibered geometries for all the $SU(7)$ series and for some cases in 
the $SU(8)$ series ($G=SU(4)\times SU(2)\times SU(2)$ (No.22) and $G=SU(6)\times SU(2)$ (No.29)).
Note that the geometries for the $SU(8)$ series we have constructed cannot be obtained from the (Higgsing of the) known CY${}_3$ with $G=SU(8)$ of \cite{AGRT},
because mapping that CY${}_3$ to a RES-fibered geometry inevitably changes the singularity.
We found that the F-theoretic spectrum derived from each of these geometries is completely identical to the 
heterotic spectrum with a specific distribution of instantons.

We have not succeeded to construct CY${}_3$'s (and hence RES-fibered threefolds) for $SU(8)$ series other than 
the above two cases yet.
It would be an interesting problem 
to understand whether there are no such geometries
in principle, or we merely have not found them yet for some technical reason;
we leave it for future investigations.
Also, each RES-fibered geometry for $SU(7)$ and $SU(8)$ series we have constructed in this paper 
corresponds only to a particular distribution of instantons.
It is expected that the deformation to the generic distributions can be done similarly 
to the case of $G=SU(2)\times SU(2)\times SU(2)$, though it is technically much more elaborated.
We also leave this issue for future work.

\section*{\small{Acknowledgements}}
We thank K.~Hori, Y.~Kimura, Y.~Yamada and T.~Watari for discussions. 
T.~T. would also like to thank Kavli Institute for the Physics and Mathematics of the Universe (Kavli IPMU) 
for hospitality, where most of his work was done. 
The work of S.~M. is supported by Grant-in-Aid for Scientific Research  
(C) \#16K05337 
from The Ministry of Education, Culture, Sports, Science
and Technology of Japan.


\newpage

\section*{A ~~Heterotic index computations for the $SU(7)$ series}
\setcounter{subsection}{0}
\renewcommand\thesection{A}

\subsection{$G=SU(4)\times SU(3)$ $(T=A_3\oplus A_2)$ (No.19)}

\subsubsection{The case when $H=SU(2)\times U(1)\times U(1)$}
The junction lattice is
\beqa
\Lambda_{(19)}^{-1}&=&
\left(
\begin{array}{crr}
  2& 0  &-1    \\
  0& 2 &-1 \\
  -1 & -1 & 4
\end{array}
\right).
\eeqa

As we did in the case No.17 in Section~\ref{sec:No17hetero}, we can break $SU(7)$ to $SU(4)\times SU(3)$ by further introducing 
instantons in another $U(1)(=\widetilde{\widetilde{U}}(1))$. 
We assume $12+n-r-\tilde{\tilde r}$, $r$ and $\tilde{\tilde r}$ instantons for each 
factor of $H=SU(2)\times U(1)\times {\widetilde{\widetilde{U}}(1)}$.

The decomposition of {\bf 248} is as shown in Table \ref{No19decomposition}. 
\begin{table}[htp]
\caption{Decomposition of ${\bf 248}$.}
\begin{center}
\begin{tabular}{|c|ccc|}
\hline
Rep. of $SU(4)\times SU(3)$ &Rep. of $SU(2)$&$\frac23 h$~charge&$\tilde{\tilde h}$~charge\\
\hline
\hline
$({\bf 15},{\bf 1})$&{\bf 1}&0&0\\
$({\bf 1},{\bf 8})$&{\bf 1}&0&0\\
$({\bf 4},\overline{\bf 3})$&{\bf 1}&0&7\\
$({\overline{\bf 4}},{\bf 3})$&{\bf 1}&0&$-7$\\
$({\bf 1},{\bf 1})$ ($\tilde{\tilde h}$)&{\bf 1}&0&0\\
&&&\\
$({\bf 4},{\bf 1})$&{\bf 2}&$3$&$3$\\
$({\bf 4},{\bf 1})$&{\bf 1}&$-4$&$3$\\
$({\bf 1},{\bf 3})$&{\bf 2}&$3$&$-4$\\
$({\bf 1},{\bf 3})$&{\bf 1}&$-4$&$-4$\\
$(\overline{\bf 4},{\bf 1})$&{\bf 2}&$-3$&$-3$\\
$(\overline{\bf 4},{\bf 1})$&{\bf 1}&$4$&$-3$\\
$({\bf 1},\overline{\bf 3})$&{\bf 2}&$-3$&$4$\\
$({\bf 1},\overline{\bf 3})$&{\bf 1}&$4$&$4$\\
&&&\\
$(\overline{\bf 4},{\bf 1})$&{\bf 1}&$2$&$9$\\
$({\bf 6},{\bf 3})$&{\bf 1}&$2$&$2$\\
$({\bf 4},\overline{\bf 3})$&{\bf 1}&$2$&$-5$\\
$({\bf 1},{\bf 1})$&{\bf 1}&$2$&$-12$\\
$({\bf 4},{\bf 1})$&{\bf 1}&$-2$&$-9$\\
$({\bf 6},\overline{\bf 3})$&{\bf 1}&$-2$&$-2$\\
$(\overline{\bf 4},{\bf 3})$&{\bf 1}&$-2$&$5$\\
$({\bf 1},{\bf 1})$&{\bf 1}&$-2$&$12$\\
&&&\\
$({\bf 6},{\bf 1})$&{\bf 2}&$-1$&$6$\\
$({\bf 4},{\bf 3})$&{\bf 2}&$-1$&$-1$\\
$({\bf 1},\overline{\bf 3})$&{\bf 2}&$-1$&$-8$\\
$({\bf 6},{\bf 1})$&{\bf 2}&$1$&$-6$\\
$(\overline{\bf 4},\overline{\bf 3})$&{\bf 2}&$1$&$1$\\
$({\bf 1},{\bf 3})$&{\bf 2}&$1$&$8$\\
$({\bf 1},{\bf 1})$ ($\frac23 h$)&{\bf 1}&0&0\\
$({\bf 1},{\bf 1})$&{\bf 3}&0&0\\
\hline
\end{tabular}
\end{center}
\label{No19decomposition}
\end{table}%
In this case
\beqa
\sum_{E_8}(U(1)\mbox{~charge})^2&=&14\cdot 60,\n
\sum_{E_8}( {\tilde{\tilde U}}(1)\mbox{~charge})^2&=&84\cdot 60.
\eeqa
Therefore
\beqa
\frac1{8\pi^2}\int F^a\wedge F^b~\mbox{Tr}\tau^a \tau^b&=&12+n-r-\tilde{\tilde r},
\n
\frac1{8\pi^2}\int F^{U(1)}\wedge F^{U(1)}&=&\frac r{14},
\n
\frac1{8\pi^2}\int F^{\widetilde{\widetilde{U}}(1)}\wedge F^{\widetilde{\widetilde{U}}(1)}&=&\frac{\tilde{\tilde r}}{84}.
\eeqa 

The computed spectrum is shown in Table \ref{spectrumNo.19}.
\begin{table}[htp]
\caption{The spectrum for the configuration No.19.}
\begin{center}
\begin{tabular}{|c|c|}
\hline
Representation&Multiplicity\\
\hline
\hline
$({\bf 15},{\bf 1})$&$-1$($SU(4)$ vector)\\
$({\bf 1},{\bf 8})$&$-1$($SU(3)$ vector)\\
$({\bf 1},{\bf 1})$&$2n+17-6\left(\frac27 r + \frac1{21}{\tilde {\tilde r}}\right)$\\
$({\bf 4},{\bf 3})$&$n+4-2\left(\frac27 r + \frac1{21}{\tilde {\tilde r}}\right)$\\
$({\bf 4},{\bf 1})$&$n+4+6\left(\frac27 r + \frac1{21}{\tilde {\tilde r}}\right)$\\
$({\bf 1},{\bf 3})$&$2n+14+2\left(\frac27 r + \frac1{21}{\tilde {\tilde r}}\right)$\\
$({\bf 6},{\bf 3})$&$-2+\left(\frac27 r + \frac1{21}{\tilde {\tilde r}}\right)$\\
$({\bf 6},{\bf 1})$&$n+8-3\left(\frac27 r + \frac1{21}{\tilde {\tilde r}}\right)$\\
\hline
\end{tabular}
\end{center}
\label{spectrumNo.19}
\end{table}%
%
%
The multiplicity of 
$({\bf 4},{\bf 3})$ is a sum of
\beqa
n_{(\bf 4,\bf 3)}+n_{(\overline{\bf 4},\overline{\bf 3})}
&=&-4+\frac27 r+\frac{37}{42}\tilde{\tilde r}
\eeqa
and
\beqa
n_{(\overline{\bf 4},\bf 3)}+n_{(\bf 4,\overline{\bf 3})}
&=&n+8-\frac67 r-\frac{41}{42}\tilde{\tilde r}.
\eeqa
The general condition for 
$\frac27 r + \frac1{21}{\tilde {\tilde r}}$ to be an integer is
\beqa
(r,\tilde {\tilde r})&=&(3,3)k + (-1,6)l~~~(k,l\in{\ZZ}),
\label{eq:rrtNo19}
\eeqa
and then
\beqa
\frac27 r +\frac1{21}{\tilde {\tilde r}}&=&k~\in{\ZZ}.
\eeqa

The orthogonal decomposition of the junction lattice (inverse the MW lattice) is 
\beqa
\left(
\begin{array}{ccc}
  7& ~~1& ~~2      \\
  0& ~~1& ~~0     \\
0& ~~0& ~~1  
\end{array}
\right)
\left(
\begin{array}{crr}
  2& 0  &-1    \\
  0& 2 &-1 \\
  -1 & -1 & 4
\end{array}
\right)
\left(
\begin{array}{ccc}
  7& ~~0& ~~0      \\
  1& ~~1& ~~0     \\
2& ~~0& ~~1 
\end{array}
\right)
&=&
\left(
\begin{array}{ccc}
  84& ~0  & ~0   \\
  0& ~2 & -1\\
0& -1 & ~4
\end{array}
\right).\n
\eeqa
The right-bottom block is $\Lambda_{(25)}^{-1}$.

\subsubsection{The case when $H=SU(2)\times SU(2) \times U(1)$}

$\Lambda_{(19)}^{-1}$
can also be regarded as an extension of 
the semi-simple $SU(2)\times SU(2)$ Cartan matrix.
$E_8$ is decomposed into representations of 
$SU(6)\times SU(3) \times SU(2)$ as
\beqa
{\bf 248}&=&
~~~({\bf 35},{\bf 1},{\bf 1})~\oplus~
~({\bf 1},{\bf 8},{\bf 1})~~\oplus~
~({\bf 1},{\bf 1},{\bf 3})~\n
&&\oplus~({\bf 15},{\bf 3},{\bf 1})
~\oplus~(\overline{\bf 15},\overline{\bf 3},{\bf 1})~\oplus~
({\bf 20},{\bf 1},{\bf 2})~\oplus~
~({\bf 6},\overline{\bf 3},{\bf 2})~\oplus~
~({\bf 6},{\bf 3},{\bf 2}).
\eeqa
By further breaking this 
$SU(6)$ into $SU(4)\times SU(2)(\equiv SU(2)')$ by an $U(1)(\equiv U(1)_{(19)})$
one obtains the decomposition in representations of 
$SU(4)\times SU(3)\times SU(2)\times SU(2)' \times U(1)_{(19)}$ 
 (Table \ref{No19decomposition2}).
 The instanton numbers are assumed to be 
 $12+n-r-r_{U(1)}$, $r$ and $r_{U(1)}$ for $SU(2)$,  $SU(2)'$ and  $U(1)_{(19)}$.
The sum of ($U(1)_{(19)}$charge$){}^2$ is
\beqa
\sum_{E_8}(U(1)_{(19)}\mbox{charge})^2&=&12\cdot 60,
\eeqa
so that
\beqa
\frac1{8\pi^2}\int F^a\wedge F^b~\mbox{Tr}\tau^a\tau^b&=&12+n-r-r_{U(1)},
\n
\frac1{8\pi^2}\int F'^a\wedge F'^b~\mbox{Tr}\tau'^a\tau'^b&=&r,
\n
\frac1{8\pi^2}\int F^{U(1)}\wedge F^{U(1)}&=&\frac {r_{U(1)}}{12}.
\eeqa 
%
\begin{table}[htp]
\caption{Decomposition of ${\bf 248}$ in reps. of $SU
(4)\times SU(3)\times SU(2)\times SU(2)' \times U(1)$.}
\begin{center}
\begin{tabular}{|ccccc|}
\hline
Rep. of $SU(4)$ 
&Rep. of $SU(3)$ 
&Rep. of $SU(2)$ 
&Rep. of $SU(2)'$ 
&$U(1)_{(19)}$~charge\\
\hline
\hline
${\bf 15}$&${\bf 1}$&${\bf 1}$&${\bf 1}$&$0$\\
${\bf 1}$&${\bf 1}$&${\bf 1}$&${\bf 1}$& $0$\\
${\bf 1}$&${\bf 1}$&${\bf 1}$&${\bf 3}$& $0$\\
${\bf 4}$&${\bf 1}$&${\bf 1}$&${\bf 2}$& $3$\\
$\overline{\bf 4}$&${\bf 1}$&${\bf 1}$&${\bf 2}$& $-3$\\

$\overline{\bf 4}$&${\bf 1}$&${\bf 2}$&${\bf 1}$& $3$\\
${\bf 6}$&${\bf 1}$&${\bf 2}$&${\bf 2}$& $0$\\
${\bf 4}$&${\bf 1}$&${\bf 2}$&${\bf 1}$& $-3$\\

${\bf 6}$&${\bf 3}$&${\bf 1}$&${\bf 1}$& $2$\\
${\bf 4}$&${\bf 3}$&${\bf 1}$&${\bf 2}$& $-1$\\
${\bf 1}$&${\bf 3}$&${\bf 1}$&${\bf 1}$& $-4$\\

${\bf 6}$&$\overline{\bf 3}$&${\bf 1}$&${\bf 1}$& $-2$\\
$\overline{\bf 4}$&$\overline{\bf 3}$&${\bf 1}$&${\bf 2}$& $1$\\
${\bf 1}$&$\overline{\bf 3}$&${\bf 1}$&${\bf 1}$& $4$\\

${\bf 4}$&$\overline{\bf 3}$&${\bf 2}$&${\bf 1}$& $1$\\
${\bf 1}$&$\overline{\bf 3}$&${\bf 2}$&${\bf 2}$& $-2$\\
$\overline{\bf 4}$&${\bf 3}$&${\bf 2}$&${\bf 1}$& $-1$\\
${\bf 1}$&${\bf 3}$&${\bf 2}$&${\bf 2}$& $2$\\

${\bf 1}$&${\bf 8}$&${\bf 1}$&${\bf 1}$& $0$\\
${\bf 1}$&${\bf 1}$&${\bf 3}$&${\bf 1}$& $0$\\

\hline
\end{tabular}
\end{center}
\label{No19decomposition2}
\end{table}%
%


\begin{table}[htp]
\caption{The spectrum for the configuration No.19. $H=SU(2)\times SU(2)\times U(1)$.}
\begin{center}
\begin{tabular}{|c|c|}
\hline
Representation&Multiplicity\\
\hline
\hline
$({\bf 15},{\bf 1})$&$-1$($SU(4)$ vector)\\
$({\bf 1},{\bf 8})$&$-1$($SU(3)$ vector)\\
$({\bf 1},{\bf 1})$&$2n+17-12r'_{U(1)}$\\

$({\bf 4},{\bf 3})$&$-4+r+r'_{U(1)}$\\
$({\bf 4},\overline{\bf 3})$&$n+8-r-5r'_{U(1)}$\\

$({\bf 4},{\bf 1})$&$n+4+12r'_{U(1)}$\\
$({\bf 1},{\bf 3})$&$2n+14+4r'_{U(1)}$\\
$({\bf 6},{\bf 3})$&$-2+2r'_{U(1)}$\\
$({\bf 6},{\bf 1})$&$n+8-6r'_{U(1)}$\\
\hline
\end{tabular}
\end{center}
\label{spectrumNo.19(2)}
\end{table}%
The spectrum in this case is summarized in Table \ref{spectrumNo.19(2)},
where
\beqa
r_{U(1)}=6r'_{U(1)}.
\eeqa
This coincides with the result shown in Table.\ref{spectrumNo.19} if 
$({\bf 4},{\bf 3})$ and 
$({\bf 4},\overline{\bf 3})$ are identified and the replacement 
\beqa
\frac27 r +\frac1{21}{\tilde {\tilde r}}&=&k=~2r'_{U(1)}
\eeqa
is made. Again, the spectrum for $SU(2)\times U(1)\times U(1)$ is 
more general than that for $SU(2)\times SU(2)\times U(1)$ as $k$ is 
restricted to even in the latter.
The orthogonal decomposition of the junction lattice in this case is 
\beqa
\left(
\begin{array}{crr}
  2& ~0  &~0    \\
  0& ~2 &~0 \\
  1 & ~1 & ~2
\end{array}
\right) 
\left(
\begin{array}{crr}
  2& 0  &-1    \\
  0& 2 &-1 \\
  -1 & -1 & 4
\end{array}
\right) 
\left(
\begin{array}{crr}
  2& ~0  &~1    \\
  0& ~2 &~1 \\
  0 & ~0 & ~2
\end{array}
\right) &=&
\left(
\begin{array}{crr}
  2& ~0  &~0   \\
  0& ~2 &~0\\
  0 & ~0 & ~12
\end{array}
\right) ,\n
\eeqa
this agrees with the fact that the $U(1)$ instanton number 
is a multiple of 6.

\subsection{$G=SU(3)\times SU(2)\times SU(2)$ $(T=A_2\oplus A_1\oplus A_1)$ (No.12)}

The inverse of the MW lattice is
\beqa
\Lambda_{(12)}^{-1}&=&
\left(
\begin{array}{cccc}
  4& -1  &0 &1    \\
  -1& 2 &-1 &0\\
  0 & -1 & 2 &-1\\
  1&0&-1&2
\end{array}
\right).
\eeqa
It has 
the $SU(4)$ Cartan matrix in the right-bottom block, so we first take 
$SU(4)\times U(1)$ as $H$ and compute the spectrum. 
Since $G=SU(3)\times SU(2)\times SU(2)$, we can use the 
decomposition Table \ref{No19decomposition2} for No.19.
Assuming the instanton distribution $12+n-r_{U(1)}$, $r_{U(1)}$ 
for $SU(4)$, $U(1)$, respectively, we find the spectrum as shown 
in Table \ref{spectrumNo.12}
($r_{U(1)}=3r'_{U(1)}$). 

\begin{table}[htp]
\caption{The spectrum for the configuration No.12. $H=SU(4)\times U(1)$.}
\begin{center}
\begin{tabular}{|c|c|}
\hline
Representation&Multiplicity\\
\hline
\hline
$({\bf 8},{\bf 1},{\bf 1})$&$-1$($SU(3)$ vector)\\
$({\bf 1},{\bf 3},{\bf 1})$&$-1$($SU(2)$ vector)\\
$({\bf 1},{\bf 1},{\bf 3})$&$-1$($SU(2)'$ vector)\\
$({\bf 1},{\bf 1},{\bf 1})$&$4n+32-12 r'_{U(1)}$\\
$({\bf 1},{\bf 1},{\bf 2})$&$n+4 +6 r'_{U(1)}$\\
$({\bf 1},{\bf 2},{\bf 1})$&$n+4 +6 r'_{U(1)}$\\
$({\bf 1},{\bf 2},{\bf 2})$&$n+6 -3 r'_{U(1)}$\\
$({\bf 3},{\bf 1},{\bf 1})$&$2n+10 +4 r'_{U(1)}$\\
$({\bf 3},{\bf 1},{\bf 2})$&$n+4 -2 r'_{U(1)}$\\
$({\bf 3},{\bf 2},{\bf 1})$&$n+4 -2 r'_{U(1)}$\\
$({\bf 3},{\bf 2},{\bf 2})$&$-2+ r'_{U(1)}$\\
\hline
\end{tabular}
\end{center}
\label{spectrumNo.12}
\end{table}%

Orthogonal decomposition of the inverse of the MW lattice:
\beqa
\left(
\begin{array}{cccc}
  2& ~~1& ~~0& -1      \\
  0& ~~1& ~~0 & ~~0    \\
0& ~~0& ~~1 & ~~0 \\
0& ~~0& ~~0 & ~~1
\end{array}
\right)
\left(
\begin{array}{cccc}
  4& -1  &0 &1    \\
  -1& 2 &-1 &0\\
  0 & -1 & 2 &-1\\
  1&0&-1&2
\end{array}
\right)
 \left(
\begin{array}{cccc}
  2& ~~0& ~~0& 0      \\
  1& ~~1& ~~0 & ~~0    \\
0& ~~0& ~~1 & ~~0 \\
-1& ~~0& ~~0 & ~~1
\end{array}
\right)
 &=&
 \left(
\begin{array}{cccc}
  12& 0  &0 &0   \\
  0& 2 &-1 &0\\
  0 & -1 & 2 &-1\\
  0&0&-1&2
\end{array}
\right).\n
\eeqa

This spectrum can also be 
computed by breaking $SU(5)\times SU(2)$ for No.17 
to $SU(3)\times SU(2)\times SU(2)$ by giving   
instantons to $U(1)_{(17)}$.
In this case the multiplicities can be simply obtained by adding 
\beqa
\frac{r_{(17)}}{30}\cdot (\mbox{$U(1)_{(17)}$ charge})^2\cdot(\mbox{the dimension of the representation of $H$ ($2$ or $1$)})
\eeqa
for each $SU(3)\times SU(2)$ representation in the $SU(5)$ decomposition in No.17.
Assuming the instanton numbers to be 
$12+n-r-{\tilde r}-r_{(17)}$, $r$, ${\tilde r}$ and $r_{(17)}$ for 
$SU(2)$, $U(1)$  $\widetilde{U(1)}$ and
$U(1)_{(17)}$, respectively, we obtain the result as shown in Table  \ref{spectrumNo.12(2)},
which coincides with Table \ref{spectrumNo.12} provided that
\beqa
r'_{U(1)}&=&\frac27 r + \frac1{70}{\tilde r}+\frac1{30}r_{(17)}.
\eeqa

 \begin{table}[htp]
\caption{The spectrum for the configuration No.12. $H=SU(2)\times U(1)^3$.}
\begin{center}
\begin{tabular}{|c|c|}
\hline
Representation&Multiplicity\\
\hline
\hline
$({\bf 8},{\bf 1},{\bf 1})$&$-1$($SU(3)$ vector)\\
$({\bf 1},{\bf 3},{\bf 1})$&$-1$($SU(2)$ vector)\\
$({\bf 1},{\bf 1},{\bf 3})$&$-1$($SU(2)'$ vector)\\
$({\bf 1},{\bf 1},{\bf 1})$&$4n+32-12 \left(\frac27 r + \frac1{70}{\tilde r}+\frac1{30}r_{(17)}\right)$\\
$({\bf 1},{\bf 1},{\bf 2})$&$n+4 +6 \left(\frac27 r + \frac1{70}{\tilde r}+\frac1{30}r_{(17)}\right)$\\
$({\bf 1},{\bf 2},{\bf 1})$&$n+4 +6 \left(\frac27 r + \frac1{70}{\tilde r}+\frac1{30}r_{(17)}\right)$\\
$({\bf 1},{\bf 2},{\bf 2})$&$n+6 -3 \left(\frac27 r + \frac1{70}{\tilde r}+\frac1{30}r_{(17)}\right)$\\
$({\bf 3},{\bf 1},{\bf 1})$&$2n+10 +4 \left(\frac27 r + \frac1{70}{\tilde r}+\frac1{30}r_{(17)}\right)$\\
$({\bf 3},{\bf 1},{\bf 2})$&$n+4 -2 \left(\frac27 r + \frac1{70}{\tilde r}+\frac1{30}r_{(17)}\right)$\\
$({\bf 3},{\bf 2},{\bf 1})$&$n+4 -2 \left(\frac27 r + \frac1{70}{\tilde r}+\frac1{30}r_{(17)}\right)$\\
$({\bf 3},{\bf 2},{\bf 2})$&$-2+ \left(\frac27 r + \frac1{70}{\tilde r}+\frac1{30}r_{(17)}\right)$\\
\hline
\end{tabular}
\end{center}
\label{spectrumNo.12(2)}
\end{table}%

For 
$\frac27 r + \frac1{70}{\tilde r}+\frac1{30}r_{(17)}$ 
to be integer, the general solution is
\beqa
(r,{\tilde r},r_{(17)})&=&(3,3,3)k + (-1,20,0)l+(0,-7,3)m~~~(k,l,m\in{\ZZ}),
\label{eq:rrtr17}
\eeqa
in which
\beqa
\frac27 r + \frac1{70}{\tilde r}+\frac1{30}r_{(17)}&=&k~\in{\ZZ}.
\eeqa

\section*{B ~~Heterotic index computations for the $SU(8)$ series}
\setcounter{subsection}{0}
\renewcommand\thesection{B}

\subsection{$G=SU(6)\times SU(2)$ $(T=A_5\oplus A_1)$ (No.29)}

%
There are two $A_7=SU(8)$ embeddings in the $E_8$ root lattice, No.45 and No.44.
The former $SU(8)$ is the one in the $SU(9)$ sublattice of $E_8$, while the 
latter is the one in the $E_7$ sublattice.
Correspondingly there are two $G=A_5\oplus A_1 (SU(6)\times SU(2))$ cases,
No.29 and No.28. The present case No.29 is the one descended from No.45.

In this case $E(K)=A_1^* \oplus \langle 1/6\rangle$, and thus 
$H=SU(2)\times U(1)$. 
Taking 
\beqa
{\bf e}_1-{\bf e}_2,~ {\bf e}_2-{\bf e}_3,~ {\bf e}_3-{\bf e}_4, ~{\bf e}_4-{\bf e}_5, ~{\bf e}_5-{\bf e}_6 
\eeqa
and 
\beqa
{\bf e}_7-{\bf e}_8 
\eeqa
as the simple roots 
of  $G=SU(6)\times SU(2)$ in the notation (\ref{SU(9)roots})(\ref{3formroots}),
the simple root of the $SU(2)$ in $H$ is 
\beqa
{\bf e}_7+{\bf e}_8+{\bf e}_9-\frac13\sum_{l=1}^9 {\bf e}_l&\equiv&{\bf e}_{789},
\eeqa
whereas the $E_8$ root corresponding to the $U(1)$ in $H$ 
$(\equiv U(1)_{\langle 1/6\rangle})$ is
\beqa
{\bf e}_7+{\bf e}_8-2{\bf e}_9&\equiv&\mbox{\boldmath $\omega$}_{\langle 1/6\rangle}.
\eeqa
Indeed, we have ${\bf e}_{789}^2=2$, $\mbox{\boldmath $\omega$}_{\langle 1/6\rangle}^2=6$ 
and ${\bf e}_{789}\cdot\mbox{\boldmath $\omega$}_{\langle 1/6\rangle}=0$.
The decomposition of $E_8$ in this $G\times H$ is as shown 
in Table \ref{No29decomposition}.

\begin{table}[htp]
\caption{Decomposition of ${\bf 248}$.}
\begin{center}
\begin{tabular}{|c|cc|}
\hline
Rep. of $SU(6)\times SU(2)$ &Rep. of $SU(2)\subset H$ &$U(1)_{\langle 1/6\rangle}$charge\\
\hline
\hline
$({\bf 35},{\bf 1})$ $(SU(6))$&{\bf 1}&$0$\\
$({\bf 6},{\bf 2})$&{\bf 2}&$-1$\\
$(\overline{\bf 6},{\bf 2})$&{\bf 2}&$1$\\
$({\bf 1},{\bf 3})$ $(SU(2))$&{\bf 1}&$0$\\

$({\bf 6},{\bf 1})$&{\bf 2}&$2$\\
$(\overline{\bf 6},{\bf 1})$&{\bf 2}&$-2$\\

$({\bf 1},{\bf 2})$&{\bf 1}&$3$\\
$({\bf 1},{\bf 2})$&${\bf 1}$&$-3$\\

&&\\
$({\bf 20},{\bf 1})$&${\bf 2}$&$0$\\
$({\bf 15},{\bf 2})$&{\bf 1}&$1$\\
$(\overline{\bf 15},{\bf 2})$&{\bf 1}&$-1$\\

$({\bf 15},{\bf 1})$&{\bf 1}&$-2$\\
$(\overline{\bf 15},{\bf 1})$&{\bf 1}&$2$\\
&&\\
$({\bf 1},{\bf 1})$ $(SU(2)\subset H)$&{\bf 3}&$0$\\
$({\bf 1},{\bf 1})$ $(U(1)_{\langle 1/6\rangle})$&${\bf 1}$&$0$\\

\hline
\end{tabular}
\end{center}
\label{No29decomposition}
\end{table}%

Assuming the instantons in $12+n-r$ and $r$ 
in $SU(2)$ and $U(1)_{\langle 1/6\rangle}$, 
we obtain the spectrum of No.29 as shown in Table \ref{spectrumNo.29}.
\begin{table}[htp]
\caption{The spectrum for the configuration No.29.}
\begin{center}
\begin{tabular}{|c|c|}
\hline
Representation&Multiplicity\\
\hline
\hline
$({\bf 35},{\bf 1})$&$-1$($SU(6)$ vector)\\
$({\bf 1},{\bf 3})$&$-1$($SU(2)$ vector)\\
$({\bf 6},{\bf 2})$&$n-\frac23 r +8$\\
$({\bf 6},{\bf 1})$&$n+\frac13 r +8$\\
$({\bf 1},{\bf 2})$&$-2+\frac32 r$\\
$({\bf 20},{\bf 1})$&$\frac12(n-r+8)$\\

$({\bf 15},{\bf 2})$&$-2+\frac16 r$\\

$({\bf 15},{\bf 1})$&$-2+\frac23 r$\\

$({\bf 1},{\bf 1})$&$2n-2r+20$\\

\hline
\end{tabular}
\end{center}
\label{spectrumNo.29}
\end{table}%

\newpage

\subsection{$G=SU(4)\times SU(2)\times SU(2)$ $(T=A_3\oplus A_1\oplus A_1)$ (No.22)}

%
This is one of the two $G=SU(4)\times SU(2)\times SU(2)$ cases, 
the one descended from No.29.
$E(K)=A_1^* \oplus A_1^* \oplus\langle 1/4\rangle$, and thus 
$H=SU(2)\times SU(2)\times U(1)$. 
Taking 
\beqa
{\bf e}_1-{\bf e}_2,~ {\bf e}_2-{\bf e}_3,~ {\bf e}_3-{\bf e}_4, ~{\bf e}_4-{\bf e}_5
\eeqa
as the simple roots 
of  $SU(4)\subset G$ and 
\beqa
{\bf e}_5-{\bf e}_6,~{\bf e}_7-{\bf e}_8 
\eeqa
as those of $SU(2)\times SU(2) (\equiv SU(2)_{56}\times SU(2)_{78})\subset G$,
the simple roots of 
$SU(2)\times SU(2) \subset H$ $(\equiv SU(2)_{569}\times SU(2)_{789})$ are 
\beqa
&&{\bf e}_5+{\bf e}_6+{\bf e}_9-\frac13\sum_{l=1}^9 {\bf e}_l,\\
&&{\bf e}_7+{\bf e}_8+{\bf e}_9-\frac13\sum_{l=1}^9 {\bf e}_l,
\eeqa
whereas 
the $U(1)$ in $H$ $(\equiv U(1)_{\langle 1/4\rangle})$ is associated with the $E_8$ root
\beqa
-\frac13\left(
{\bf e}_1+{\bf e}_2+{\bf e}_3+{\bf e}_4
\right)
+\frac23\left(
{\bf e}_5+{\bf e}_6+{\bf e}_7+{\bf e}_8
\right)
-\frac43{\bf e}_9.
\eeqa

The decomposition of $E_8$  is 
in Table \ref{No22decomposition}.
\begin{table}[htp]
\caption{Decomposition of ${\bf 248}$.}
\begin{center}
\begin{tabular}{|c|ccc|}
\hline
Rep. of $SU(5)\times SU(2)_{56}\times SU(2)_{78}$ &Rep. of $SU(2)_{569}$&Rep. of $SU(2)_{789}$&$U(1)_{\langle 1/4\rangle}$charge\\
\hline
\hline
$({\bf 15},{\bf 1},{\bf 1})$ $(SU(4))$&{\bf 1}&{\bf 1}&$0$\\

$({\bf 4},{\bf 2},{\bf 1})$ &{\bf 2}&{\bf 1}&$-1$\\
$(\overline{\bf 4},{\bf 2},{\bf 1})$ &{\bf 2}&{\bf 1}&$1$\\

$({\bf 1},{\bf 3},{\bf 1})$ $(SU(2)_{56})$&{\bf 1}&{\bf 1}&$0$\\
$({\bf 1},{\bf 1},{\bf 1})$ $(U(1)_{\langle 1/4\rangle})$&{\bf 1}&{\bf 1}&$0$\\

$({\bf 4},{\bf 1},{\bf 2})$ &{\bf 1}&{\bf 2}&$-1$\\
$(\overline{\bf 4},{\bf 1},{\bf 2})$ &{\bf 1}&{\bf 2}&$1$\\

$({\bf 1},{\bf 2},{\bf 2})$ &{\bf 2}&{\bf 2}&$0$\\

$({\bf 1},{\bf 1},{\bf 3})$ $(SU(2)_{78})$&{\bf 1}&{\bf 1}&0\\

$({\bf 4},{\bf 1},{\bf 1})$ &{\bf 2}&{\bf 2}&$1$\\
$({\bf 1},{\bf 2},{\bf 1})$ &{\bf 1}&{\bf 2}&$2$\\
$({\bf 1},{\bf 1},{\bf 2})$ &{\bf 2}&{\bf 1}&$2$\\

$(\overline{\bf 4},{\bf 1},{\bf 1})$ &{\bf 2}&{\bf 2}&$-1$\\
$({\bf 1},{\bf 2},{\bf 1})$ &{\bf 1}&{\bf 2}&$-2$\\
$({\bf 1},{\bf 1},{\bf 2})$ &{\bf 2}&{\bf 1}&$-2$\\

$({\bf 6},{\bf 2},{\bf 1})$ &{\bf 1}&{\bf 2}&$0$\\
$({\bf 6},{\bf 1},{\bf 2})$ &{\bf 2}&{\bf 1}&$0$\\

$({\bf 4},{\bf 2},{\bf 2})$ &{\bf 1}&{\bf 1}&$1$\\
$(\overline{\bf 4},{\bf 2},{\bf 2})$ &{\bf 1}&{\bf 1}&$-1$\\

$({\bf 6},{\bf 1},{\bf 1})$ &{\bf 1}&{\bf 1}&$-2$\\
$({\bf 6},{\bf 1},{\bf 1})$ &{\bf 1}&{\bf 1}&$2$\\

$({\bf 1},{\bf 1},{\bf 1})$ $(SU(2)_{569})$&{\bf 3}&{\bf 1}&0\\
$({\bf 1},{\bf 1},{\bf 1})$ $(SU(2)_{789})$&{\bf 1}&{\bf 3}&0\\
\hline
\end{tabular}
\end{center}
\label{No22decomposition}
\end{table}%
In this case we assume instantons $r$, $\tilde r$ 
and $12+n-r-\tilde r$  
in $SU(2)_{56}$, $SU(2)_{78}$ and $U(1)_{\langle 1/4\rangle}$ 
so that the $SU(2)$'s in $H$ are treated in a symmetric way. 
The spectrum of No.29 is as shown in Table \ref{spectrumNo.22}.
(Again, representations with $\overline{\phantom{A}}$ and without  $\overline{\phantom{A}}$ 
are identified.)
\begin{table}[htp]
\caption{The spectrum for the configuration No.22.}
\begin{center}
\begin{tabular}{|c|c|}
\hline
Representation&Multiplicity\\
\hline
\hline
$({\bf 15},{\bf 1},{\bf 1})$&$-1$($SU(4)$ vector)\\
$({\bf 1},{\bf 3},{\bf 1})$&$-1$($SU(2)_{56}$ vector)\\
$({\bf 1},{\bf 3},{\bf 1})$&$-1$($SU(2)_{78}$ vector)\\

$({\bf 4},{\bf 2},{\bf 1})$&$\frac12(n+r-\tilde r)+2$\\
$({\bf 4},{\bf 1},{\bf 2})$&$\frac12(n-r+\tilde r)+2$\\

$({\bf 1},{\bf 2},{\bf 2})$&$r+\tilde r-4$\\

$({\bf 4},{\bf 1},{\bf 1})$&$n+r+\tilde r+4$\\

$({\bf 1},{\bf 2},{\bf 1})$&$2n-2r-\tilde r+20$\\
$({\bf 1},{\bf 1},{\bf 2})$&$2n-r-2\tilde r+20$\\

$({\bf 6},{\bf 2},{\bf 1})$&$\frac{\tilde r}2-2$\\
$({\bf 6},{\bf 1},{\bf 2})$&$\frac{r}2-2$\\

$({\bf 4},{\bf 2},{\bf 2})$&$\frac14(n-r-\tilde r)+1$\\

$({\bf 6},{\bf 1},{\bf 1})$&$n-r-\tilde r+10$\\

$({\bf 1},{\bf 1},{\bf 1})$&$2r+2\tilde r-7$\\

\hline
\end{tabular}
\end{center}
\label{spectrumNo.22}
\end{table}%

\newpage

\section*{C ~~Charged matter spectrum from F-theory geometry}
\setcounter{subsection}{0}
\renewcommand\thesection{C}

In this appendix, we present the explicit form of the series expansions of $\fres,\gres$ and $\Dres$ 
and the resulting charged matter spectrum.

\subsection{$G=SU(2)\times SU(2)\times SU(2)$ ($r\neq 0$)}
%
The construction of the geometry is given in section \ref{sec:No7instCY}. The series expansions of $\fres,\gres$ and $\Dres$ 
near each of the three lines $z=0,\tilde{z}=0$ and $\hat{z}=0$ are obtained as follows~:
\bes
 \fres &= -\frac{1}{3}\, t_r^2\, K_{4+2n-r}^2+(\cdots)\,t_r\, z+o(z^2) \\
   &= -\frac{1}{3}\, t_r^2\, \tilde{K}_{4+2n-r}^2+(\cdots)\,t_r\, \tilde{z}+o(\tilde{z}^2) \\
   &= -\frac{1}{3}\, t_r^2\, \hat{K}_{4+2n-r}^2+(\cdots)\,t_r\, \hat{z}+o(\hat{z}^2),  
\ees
\bes
 \gres &= \frac{2}{27}\, t_r^3\, K_{4+2n-r}^3+(\cdots)\,t_r^2\, K_{4+2n-r}\,z+(\cdots)\,t_r\,z^2+o(z^3) \\
   &= \frac{2}{27}\, t_r^3\, \tilde{K}_{4+2n-r}^3+(\cdots)\,t_r^2\, \tilde{K}_{4+2n-r}\,\tilde{z}+(\cdots)\,t_r\,\tilde{z}^2+o(\tilde{z}^3) \\
   &= \frac{2}{27}\, t_r^3\, \hat{K}_{4+2n-r}^3+(\cdots)\,t_r^2\, \hat{K}_{4+2n-r}\,\hat{z}+(\cdots)\,t_r\,\hat{z}^2+o(\hat{z}^3),
\ees
\bes
\mbox{$\Dres$} 
& \mbox{ $=t_r^4\, h_{n-r}^2 {h'}_{n-r}^2\,K_{4+2n-r}^2\, Q_{16+2n+2r}\,z^2+(\cdots)\,t_r^3\,h_{n-r}h'_{n-r}\,z^3+(\cdots)\,t_r^2\,z^4
        +(\cdots)\,t_r\,z^5+o(z^6)$}\\
& \mbox{ $=t_r^4\, h_{n-r}^2 (h_{n-r}-{h'}_{n-r})^2\, \tilde{K}_{4+2n-r}^2\, \tilde{Q}_{16+2n+2r}\,\tilde{z}^2
        +(\cdots)\,t_r^3\,h_{n-r}(h_{n-r}-h'_{n-r})\,\tilde{z}^3
        +(\cdots)\,t_r^2\, \tilde{z}^4 $}\\
& \quad +(\cdots)\,t_r \,\tilde{z}^5+o(\tilde{z}^6)  \\
& \mbox{ $=t_r^4\, {h'}_{n-r}^2 (h_{n-r}-{h'}_{n-r})^2\, \hat{K}_{4+2n-r}^2\, \hat{Q}_{16+2n+2r}\,\hat{z}^2
        +(\cdots)\,t_r^3\,h'_{n-r}(h_{n-r}-h'_{n-r})\, \hat{z}^3
        +(\cdots)\,t_r^2\, \hat{z}^4 $}\\
& \quad +(\cdots)\,t_r \, \hat{z}^5+o(\hat{z}^6),
\label{eq:su2su2su2instDexp}
\ees
where 
\bes
 K_{4+2n-r}&=s_{4+2n-2r}\,t_r+\frac{1}{3}q_{4+r}\, \hat{\sigma}_2, \\
 \tilde{K}_{4+2n-r}&=s_{4+2n-2r}\,t_r-k_{4+n-r}\, t_r\, h_{n-r}+\frac{1}{3}q_{4+r}\, h_{n-r}\, (h_{n-r}-h'_{n-r}), \\
 \hat{K}_{4+2n-r}& =s_{4+2n-2r}\,t_r-k_{4+n-r}\, t_r\, h'_{n-r}+\frac{1}{3}q_{4+r}\, h'_{n-r}\, (h'_{n-r}-h_{n-r}).
\ees
$Q_{16+2n+2r},\tilde{Q}_{16+2n+2r}$ and $\hat{Q}_{16+2n+2r}$ are degree $16+2n+2r$ irreducible polynomials.
The resulting charged matter spectrum is given as follows:
\beq
\begin{array}{l|c|c@{\hspace{0.2cm}}c@{\hspace{0.2cm}}c|c||c} \hline 
\mbox{\quad Zero}  & \mbox{Degree} & {\small \mbox{ord$(\fres)$}}
                                             & {\small \mbox{ord$(\gres)$}}
                                                   & {\small \mbox{ord$(\Dres)$}} & \mbox{Enhancement}  & \mbox{Matter} \\  \hline
K,\tilde{K},\hat{K} & 4+2n-r      & 1   &  2   &  3    & \hspace{0.0cm} A_1 \rightarrow A_1 (I_2\rightarrow III)    &  \mbox{none} \\ \hline
Q_{16+2n+2r}         & 16+2n+2r   & 0   &  0   &  3   & \hspace{0.0cm} A_1 \rightarrow A_2 &  (\bf{2},\bf{1},\bf{1}) \\
\tilde{Q}_{16+2n+2r} & 16+2n+2r   & 0   &  0   &  3   & \hspace{0.0cm} A_1 \rightarrow A_2 &  (\bf{1},\bf{2},\bf{1}) \\
\hat{Q}_{16+2n+2r}   & 16+2n+2r   & 0   &  0   &  3   & \hspace{0.0cm} A_1 \rightarrow A_2 &  (\bf{1},\bf{1},\bf{2}) \\ \hline
h_{n-r}              &   n-r      & 0   &  0   &  4   & \hspace{0.0cm} A_1\oplus A_1 \rightarrow A_3 &  (\bf{2},\bf{2},\bf{1}) \\  
h'_{n-r}             &   n-r      & 0   &  0   &  4   & \hspace{0.0cm} A_1\oplus A_1 \rightarrow A_3 &  (\bf{2},\bf{1},\bf{2}) \\   
h_{n-r}-h'_{n-r}     &   n-r      & 0   &  0   &  4   & \hspace{0.0cm} A_1\oplus A_1 \rightarrow A_3 &  (\bf{1},\bf{2},\bf{2}) \\ \hline
t_r                  &    r       & 2   &  3   &  6   & \hspace{0.0cm} A_1\oplus A_1\oplus A_1\rightarrow D_4 &  \frac{1}{2}(\bf{2},\bf{2},\bf{2}) \\ \hline
\end{array}   
\label{eq:su2su2su2instcharge}
\eeq

\subsection{$G=SU(3)\times SU(2)\times SU(2)$ (No.12)}
%
The construction of the geometry is given in section \ref{sec:No12CY}. The series expansions of $\fres,\gres$ and $\Dres$ 
near each singularity are obtained as follows~:
\bes
\fres&=-\frac{1}{3}\, p_{2+n}^4+(\cdots)\,p_{2+n}\,z + o(z^2)  \\
 &= -\frac{1}{3}{\tilde{K}_{4+2n}}^2+o(\tilde{z}) \\
 &= -\frac{1}{3}{\hat{K}_{4+2n}}^2+o(\hat{z}), 
\label{eq:su3su2su2f}
\ees
\bes
\gres&=\frac{2}{27}\, p_{2+n}^6+ (\cdots)\, p_{2+n}^3\,z +o(z^2) \\
 &=\frac{2}{27}{\tilde{K}_{4+2n}}^3+(\cdots)\tilde{K}_{4+2n}\,\tilde{z}+o(\tilde{z}^2) \\
 &=\frac{2}{27}{\hat{K}_{4+2n}}^3+(\cdots)\hat{K}_{4+2n}\,\hat{z}+o(\hat{z}^2), 
\label{eq:su3su2su2g}
\ees
\bes
\Dres & = h_n^2 \, {h'_n}^2 \, p_{2 + n}^3 \, q_6 \, R_{12+2n}\, z^3+(\cdots)\, h_n\, {h'_n}\,z^4+o(z^5) \\
       & = {\tilde{K}_{4+2n}}^2 \, h_n^3 \, (h_n-h'_n)^2 \tilde{S}_{16+n}\, \tilde{z}^2
          +(\cdots)\, h_n^2\, (h_n-h'_n)\, \tilde{z}^3+(\cdots)\, h_n \, \tilde{z}^4+o(\tilde{z}^5) \\
       & = {\hat{K}_{4+2n}}^2 \,{h'_n}^3\, (h_n-h'_n)^2 \hat{S}_{16+n}\,  \hat{z}^2
          +(\cdots){h'_n}^2 \, (h_n-h'_n)\, \hat{z}^3+(\cdots)\, h'_n \, \hat{z}^4+  o(\hat{z}^5).
\label{eq:su3su2su2D}
\ees
Here
\bes
 \tilde{K}_{4+2n} &= p_{2+n}^2-A_{21}h_n +A_{22}h_n^2,  \\
 \hat{K}_{4+2n}   &= p_{2+n}^2-A_{21}h'_n +A_{22}{h'_n}^2. 
\label{eq:su3su2su2K}
\ees
$R_{12+2n},\tilde{S}_{16+n}$ and $\hat{S}_{16+n}$ are irreducible polynomials of orders $12+2n$, $16+n$ and $16+n$, respectively.
The charged matter spectrum is then read
\beq
\begin{array}{l|c|c@{\hspace{0.2cm}}c@{\hspace{0.2cm}}c|c||c} \hline 
\mbox{\, Zero}  & \mbox{Degree} & {\small \mbox{ord$(\fres)$}}
                                             & {\small \mbox{ord$(\gres)$}}
                                                   & {\small \mbox{ord$(\Dres)$}} & \mbox{Enhancement}  & \mbox{Matter} \\  \hline
p_{2+n}           & 2+n     & 2   &  2   &  4    & \hspace{0.0cm} A_2 \rightarrow A_2 (I_3\rightarrow IV)    &  \mbox{none} \\ 
q_{6}             & 6       & 0   &  0   &  4    & \hspace{0.0cm} A_2 \rightarrow A_3                        &   (\bf{3},\bf{1},\bf{1}) \\ 
R_{12+2n}         & 12+2n   & 0   &  0   &  4    & \hspace{0.0cm} A_2 \rightarrow A_3                        &   (\bf{3},\bf{1},\bf{1}) \\   \hline
\tilde{K}_{4+2n}  & 4+2n    & 1   &  2   &  3    & \hspace{0.0cm} A_1 \rightarrow A_1 (I_2\rightarrow III)   &  \mbox{none} \\ 
\tilde{S}_{16+n}  & 16+n    & 0   &  0   &  3    & \hspace{0.0cm} A_1 \rightarrow A_2                        &  (\bf{1},\bf{2},\bf{1}) \\     \hline
\hat{K}_{4+2n}    & 4+2n    & 1   &  2   &  3    & \hspace{0.0cm} A_1 \rightarrow A_1 (I_2\rightarrow III)   &  \mbox{none} \\ 
\hat{S}_{16+n}    & 16+n    & 0   &  0   &  3    & \hspace{0.0cm} A_1 \rightarrow A_2                        &  (\bf{1},\bf{1},\bf{2}) \\     \hline
h_n               &   n     & 0   &  0   &  5    & \hspace{0.0cm} A_2\oplus A_1 \rightarrow A_4 &  (\bf{3},\bf{2},\bf{1}) \\  
h'_n              &   n     & 0   &  0   &  5    & \hspace{0.0cm} A_2\oplus A_1 \rightarrow A_4 &  (\bf{3},\bf{1},\bf{2}) \\   
h_n-h'_n          &   n     & 0   &  0   &  4    & \hspace{0.0cm} A_1\oplus A_1 \rightarrow A_3 &  (\bf{1},\bf{2},\bf{2}) \\ \hline
\end{array}   
\label{eq:su3su2su2charge}
\eeq

\subsection{$G=SU(5)\times SU(2)$ (No.17)}

The construction of the geometry is given in section \ref{sec:su5su2}. 
The series expansions near the two lines $z=0$ and $\tilde{z}=z+h_n=0$ are given by~
\bes
\fres&=-\frac{1}{3}\, p_{2+n}^4+(\cdots)\,p_{2+n}^2\,z + o(z^2)  \\
 &= -\frac{1}{3}{\tilde{K}_{4+2n}}^2+o(\tilde{z}), 
\label{eq:su5su2f}
\ees
\bes
\gres&=\frac{2}{27}\, p_{2+n}^6  + (\cdots)\, p_{2+n}^4\,z +(\cdots)\, p_{2+n}^2\,z^2 + o(z^3) \\
 &=\frac{2}{27}{\tilde{K}_{4+2n}}^3+(\cdots)\tilde{K}_{4+2n}\,\tilde{z}+o(\tilde{z}^2), 
\label{eq:su5su2g}
\ees
\bes
\Dres & = h_n^2 \, p_{2 + n}^4 \, q_6 \, R_{10+n}\, z^5+(\cdots)\, h_n\, p_{n+2}^2 \,z^6+o(z^7) \\
       & ={\tilde{K}_{4+2n}}^2 \, h_n^5 \, \tilde{S}_{16+n}\, \tilde{z}^2+(\cdots)\, h_n^4\, \tilde{z}^3+\cdots+ (\cdots)\, h_n \, \tilde{z}^6+o(\tilde{z}^7).
\label{eq:su5su2D}
\ees
Here $R_{10+n}$ is a degree $10+n$ irreducible polynomial. $\tilde{S}_{16+n}$ is the same one contained in \eqref{eq:su3su2su2D} and 
$\tilde{K}_{4+2n}$ is given in \eqref{eq:su3su2su2K}.
Charged matter spectrum is obtained as follows:
\beq
\begin{array}{l|c|c@{\hspace{0.2cm}}c@{\hspace{0.2cm}}c|c||c} \hline 
\mbox{Zero}  & \mbox{Degree} & {\small \mbox{ord$(\fres)$}}
                                             & {\small \mbox{ord$(\gres)$}}
                                                   & {\small \mbox{ord$(\Dres)$}} & \mbox{Enhancement}  & \mbox{Matter} \\  \hline
p_{2+n}           & 2+n     & 2   &  3   &  7    & \hspace{0.0cm} A_4 \rightarrow D_5                        &  (\bf{10},\bf{1})  \\ 
q_{6}             & 6       & 0   &  0   &  6    & \hspace{0.0cm} A_4 \rightarrow A_5                        &   (\bf{5},\bf{1}) \\ 
R_{10+n}          & 10+n    & 0   &  0   &  6    & \hspace{0.0cm} A_4 \rightarrow A_5                        &   (\bf{5},\bf{1}) \\   \hline
\tilde{K}_{4+2n}  & 4+2n    & 1   &  2   &  3    & \hspace{0.0cm} A_1 \rightarrow A_1 (I_2\rightarrow III)   &  \mbox{none} \\ 
\tilde{S}_{16+n}  & 16+n    & 0   &  0   &  3    & \hspace{0.0cm} A_1 \rightarrow A_2                        &  (\bf{1},\bf{2}) \\     \hline
h_n               &   n     & 0   &  0   &  7    & \hspace{0.0cm} A_4\oplus A_1 \rightarrow A_6 &  (\bf{5},\bf{2})  \\ \hline
\end{array}   
\label{eq:su5su2charge}
\eeq

\subsection{$G=SU(3)\times SU(4)$ (No.19)}

The construction of the geometry is given in section \ref{sec:su3su4}. 
The expansions near the two lines $z=0$ and $\tilde{z}=z+h_n=0$ are given by~
\bes
\fres&=-\frac{1}{3}\, p_{2+n}^4+(\cdots)\,p_{2+n}\,z  + o(z^2)  \\
 &= -\frac{1}{3}\, \tilde{p}_{2+n}^4 + (\cdots)\,\tilde{p}_{2+n}^2 \,\tilde{z} + o(\tilde{z}^2), 
\label{eq:su3su4f}
\ees
\bes
\gres&=\frac{2}{27}\, p_{2+n}^6  + (\cdots)\, p_{2+n}^3\,z  + o(z^2) \\
 &=\frac{2}{27}\, \tilde{p}_{2+n}^6 +(\cdots)\, \tilde{p}_{2+n}^4\,\tilde{z}+ (\cdots)\, \tilde{p}_{2+n}^2\, \tilde{z}^2+ o(\tilde{z}^3), 
\label{eq:su3su4g}
\ees
\bes
\Dres & = h_n^4 \, p_{2 + n}^3 \, q_6 \, R_{12+2n}\, z^3+(\cdots)\, h_n^3 \,z^4 +(\cdots)\, h_n^2 \,z^5+(\cdots)\, h_n \,z^6+o(z^7) \\
       & = h_n^3 \, \tilde{p}_{2+n}^4 \, \tilde{S}_{8+n}\,\tilde{S}'_{8} \,\tilde{z}^4+(\cdots)\, h_n^2\,\tilde{p}_{2+n}^2 \, \tilde{z}^5
          +(\cdots)\, h_n \, \tilde{z}^6+o(\tilde{z}^7). 
\label{eq:su3su4D}
\ees
Charged matter spectrum is obtained as follows:
\beq
\begin{array}{l|c|c@{\hspace{0.2cm}}c@{\hspace{0.2cm}}c|c||c} \hline 
\mbox{Zero}  & \mbox{Degree} & {\small \mbox{ord$(\fres)$}}
                                             & {\small \mbox{ord$(\gres)$}}
                                                   & {\small \mbox{ord$(\Dres)$}} & \mbox{Enhancement}  & \mbox{Matter} \\  \hline
p_{2+n}           & 2+n     & 2   &  2   &  4    & \hspace{0.0cm} A_2 \rightarrow A_2 (I_3\rightarrow IV)    &  \mbox{none}   \\ 
q_{6}             & 6       & 0   &  0   &  4    & \hspace{0.0cm} A_2 \rightarrow A_3                        &   (\bf{3},\bf{1}) \\ 
R_{12+2n}         & 12+2n   & 0   &  0   &  4    & \hspace{0.0cm} A_2 \rightarrow A_3                        &   (\bf{3},\bf{1}) \\   \hline
\tilde{p}_{2+n}   & 2+n     & 2   &  3   &  6    & \hspace{0.0cm} A_3 \rightarrow D_4                        &   (\bf{1},\bf{6}) \\ 
\tilde{S}_{8+n}   & 8+n     & 0   &  0   &  5    & \hspace{0.0cm} A_3 \rightarrow A_4                        &   (\bf{1},\bf{4}) \\ 
\tilde{S}'_{8}     & 8       & 0   &  0   &  5    & \hspace{0.0cm} A_3 \rightarrow A_4                        &   (\bf{1},\bf{4}) \\    \hline
h_n               &   n     & 0   &  0   &  7    & \hspace{0.0cm} A_2\oplus A_3 \rightarrow A_6              &   (\bf{3},\bf{4}) \\  \hline
\end{array}   
\label{eq:su3su4charge}
\eeq

\subsection{$G=SU(7)$ (No.25)}

The construction of the geometry is given in section \ref{sec:su7}. 
The series expansions are given by~
\bes
 \fres & = -\frac{1}{3}p_{2+n}^4 -\frac{2}{3}p_{2+n}^2 r_{4+n} z +o(z^2)   \\
 \gres & = \frac{2}{27} p_{2 + n}^6 + \frac{2}{9} p_{2 + n}^4 r_{4 + n} z
      +(\cdots)\,p_{2+n}^2  z^2 +o(z^3), \\
 \Dres & = -\frac{4}{3} p_{2 + n}^4 q_6 R_{10+n} z^7
           +(\cdots)\,p_{2+n}^2 \, z^8 + o(z^9), 
\label{eq:su7gD}  
\ees
with
\beq
 R_{10+n}\equiv A_{22}^2 p_{2 + n} + 3 \tilde{A}_{44} p_{2 + n} - 3 q_6 r_{4 + n}.
\eeq

Charged matter spectrum is obtained as follows:
\beq
\begin{array}{l|c|c@{\hspace{0.2cm}}c@{\hspace{0.2cm}}c|c||c} \hline 
\mbox{Zero}  & \mbox{Degree} & {\small \mbox{ord$(\fres)$}}
                                             & {\small \mbox{ord$(\gres)$}}
                                                   & {\small \mbox{ord$(\Dres)$}} & \mbox{Enhancement}  & \mbox{Matter} \\  \hline
p_{2+n}           & 2+n     & 2   &  3   &  9    & \hspace{0.0cm} A_6 \rightarrow D_7                        &  \bf{21}   \\ 
q_{6}             & 6       & 0   &  0   &  8    & \hspace{0.0cm} A_6 \rightarrow A_7                        &  \bf{7}  \\ 
R_{10+n}          & 10+n    & 0   &  0   &  8    & \hspace{0.0cm} A_6 \rightarrow A_7                        &  \bf{7}  \\  \hline
\end{array}   
\label{eq:su7charge}
\eeq

\subsection{$G=SU(4)\times SU(2) \times SU(2)$ (No.22)}

The construction of the geometry is given in section \ref{sec:su4su2su2}. 
The expansions near each line of the singularity are given by
\bes
\fres&=-\frac{1}{3}\, r_{n-4}^4\, q_6^4+(\cdots)\,r_{n-4}^2 \,q_6^2 \, z + o(z^2)  \\
 &= -\frac{1}{3}\, r_{n-4}^2 \, {\tilde{L}_{8+n}}^2+ (\cdots)\, r_{n-4}\, \tilde{z}+o(\tilde{z}^2) \\
 &= -\frac{1}{3}{\hat{K}_{4+2n}}^2+o(\hat{z}), 
\label{eq:su4su2su2f}
\ees
\bes
\gres&=\frac{2}{27}\, r_{n-4}^6 \, q_6^6+ (\cdots)\, r_{n-4}^4\, q_6^4 \, z + (\cdots)\, r_{n-4}^2\, q_6^2 \, z^2 + o(z^3) \\
 &=\frac{2}{27}\, r_{n-4}^3 \, {\tilde{L}_{8+n}}^3+(\cdots)\, r_{n-4}^2\tilde{L}_{8+n}\, \tilde{z}+ (\cdots)\, r_{n-4}\, \tilde{z}^2 + o(\tilde{z}^3) \\
 &=\frac{2}{27}{\hat{K}_{4+2n}}^3+(\cdots)\hat{K}_{4+2n}\, \hat{z}+o(\hat{z}^2), 
\label{eq:su4su2su2g}
\ees
\bes
\Dres & =  s_4^2 \, {h'_n}^2 \, r_{n-4}^4 \, q_6^4 \, T_{8+2n}\, z^4+(\cdots)\, s_4 \, {h'_n}\, r_{n-4}^2 \, q_6^2\, z^5+(\cdots)\, r_{n-4}\, z^6 + o(z^7) \\
       & = {\tilde{L}_{8+n}}^2 \, s_4^4 \, r_{n-4}^5 \, (r_{n-4}s_4-h'_n)^2 \, \tilde{U}_{12+n}\, \tilde{z}^2
          +(\cdots)\,s_4^3 \, r_{n-4}^4 \, (r_{n-4}s_4-h'_n) \, \tilde{z}^3   \\
       & \quad  +(\cdots)\,s_4^2 \, r_{n-4}^3 \, \tilde{z}^4
          +(\cdots)\,s_4 \, r_{n-4}^2 \, \tilde{z}^5 + (\cdots)\, r_{n-4} \, \tilde{z}^6 + o(\tilde{z}^7) \\
       & = {\hat{K}_{4+2n}}^2 \,{h'_n}^4\, (r_{n-4}s_4 - h'_n)^2 \, \hat{V}_{16}\,  \hat{z}^2+(\cdots){h'_n}^3 \, (r_{n-4}s_4 -  h'_n)\, \hat{z}^3
          +(\cdots)\, {h'_n}^2 \, \hat{z}^4   \\
       & \quad  +(\cdots)\, {h'_n} \, \hat{z}^5+  o(\hat{z}^6),
\label{eq:su4su2su2D}
\ees
where 
\bes
 \tilde{L}_{8+n} &= r_{n-4}q_6^2 
 -\Big( \frac{1}{3}A_{22}^2 +  \tilde{A}_{44}\Big ) r_{n-4} s_4  - h'_n s_4^2 +A_{22}r_{n-4}s_4^2,  \\
 \hat{K}_{4+2n}   &=  r_{n-4}^2 q_6^2   - \Big( \frac{1}{3}A_{22}^2 + \tilde{A}_{44}\Big)r_{n-4} h'_n -s_4 {h'_n}^2  + A_{22} {h'_n}^2.  
\label{eq:su4su2su2LK}
\ees
$T_{8+n},\tilde{U}_{12+n}$ and $\hat{V}_{16}$ are irreducible polynomials with degrees $8+n$, $12+n$ and $16$.

Charged matter spectrum can be read from these expansions. The result is
\beq
\begin{array}{l|c|c@{\hspace{0.2cm}}c@{\hspace{0.2cm}}c|c||c} \hline 
\mbox{\quad \, Zero}  & \mbox{Degree} & {\small \mbox{ord$(\fres)$}}
                                             & {\small \mbox{ord$(\gres)$}}
                                                   & {\small \mbox{ord$(\Dres)$}} & \mbox{Enhancement}  & \mbox{Matter} \\  \hline
q_{6}             & 6       & 2   &  3   &  6    & \hspace{0.0cm} A_3 \rightarrow D_4                        &   (\bf{6},\bf{1},\bf{1}) \\ 
T_{8+2n}          & 8+2n    & 0   &  0   &  5    & \hspace{0.0cm} A_3 \rightarrow A_4                        &   (\bf{4},\bf{1},\bf{1}) \\   \hline
\tilde{L}_{8+n}   & 8+n     & 1   &  2   &  3    & \hspace{0.0cm} A_1 \rightarrow A_1 (I_2\rightarrow III)   &    \mbox{none} \\ 
\tilde{U}_{12+n}  & 12+n    & 0   &  0   &  3    & \hspace{0.0cm} A_1 \rightarrow A_2                        &   (\bf{1},\bf{2},\bf{1}) \\     \hline
\hat{K}_{4+2n}    & 4+2n    & 1   &  2   &  3    & \hspace{0.0cm} A_1 \rightarrow A_1 (I_2\rightarrow III)   &   \mbox{none} \\ 
\hat{V}_{16}      & 16      & 0   &  0   &  3    & \hspace{0.0cm} A_1 \rightarrow A_2                        &   (\bf{1},\bf{1},\bf{2}) \\     \hline
s_4               & 4       & 0   &  0   &  6    & \hspace{0.0cm} A_3\oplus A_1 \rightarrow A_5              &   (\bf{4},\bf{2},\bf{1}) \\   
r_{n-4}           &   n-4   & 2   &  3   &  7    & \hspace{0.0cm} A_3\oplus A_1 \rightarrow D_5              & \frac{1}{2}(\bf{6},\bf{2},\bf{1}) \\  \hline 
h'_n              &   n     & 0   &  0   &  6    & \hspace{0.0cm} A_3\oplus A_1 \rightarrow A_5              &   (\bf{4},\bf{1},\bf{2}) \\     \hline
r_{n-4}s_4-h'_n   &   n     & 0   &  0   &  4    & \hspace{0.0cm} A_1\oplus A_1 \rightarrow A_3              &   (\bf{1},\bf{2},\bf{2}) \\     \hline 
\end{array}   
\label{eq:su4su2su2charge}
\eeq
In the third line from the bottom, enhancement $SU(4)\times SU(2)\rightarrow SO(10)$ occurs
and a half-hypermultiplet $\frac{1}{2}(\bf{6},\bf{2},\bf{1})$ appears. (The third $SU(2)$ is not concerned with this enhancement.)
The reason is as follows. The maximal embedding corresponding to this enhancement is $SO(10)\supset SU(4)\times SU(2) \times SU(2)$, whose 
branching is given by
\beq
  \bf{45}= (\bf{15},\bf{1},\bf{1})\oplus (\bf{1},\bf{3},\bf{1})\oplus (\bf{1},\bf{1},\bf{3})\oplus(\bf{6},\bf{2},\bf{2}).
\eeq
As explained below Eq.\,\eqref{eq:so8branching}, the representation coupled to $\bf{2}$ is pseudo-real,
yielding a half-hypermultiplet. In this case, it is $\frac{1}{2}(\bf{6},\bf{2})$.

\subsection{$G=SU(6)\times SU(2)$ (No.29)}

The construction of the geometry is given in section \ref{sec:su6su2}. 
The expansions are given by
\bes
\fres&=-\frac{1}{3}\, r_{n-4}^4\, q_6^4 +(\cdots)\,r_{n-4}^3\, q_6^2\, z +(\cdots)\,r_{n-4}\, z^2 + o(z^3)  \\
 &= -\frac{1}{3}{\hat{K}_{4+2n}}^2+o(\hat{z}), 
\label{eq:su5su2f}
\ees
\bes
\gres&=\frac{2}{27}\, r_{n-4}^6\, q_6^6  + (\cdots)\, r_{n-4}^5\, q_6^4 \, z + (\cdots)\, r_{n-4}^3\, q_6^2 \, z^2  
  + (\cdots)\, r_{n-4}^2 \, z^3  + o(z^4) \\
 &=\frac{2}{27}{\hat{K}_{4+2n}}^3+(\cdots)\tilde{K}_{4+2n}\, \hat{z}+o(\hat{z}^2),
\label{eq:su5su2g}
\ees
\bes
\Dres & = r_{n-4}^3 \, q_6^4  \, {h'_n}^2 \, R_{12+n}\, z^6 + (\cdots)r_{n-4}^2 \, q_6^2  \, {h'_n} \,z^7 + o(z^8) \\
       & = {\hat{K}_{4+2n}}^2 \, {h'_n}^6 \, \hat{S}_{16}\, \hat{z}^2+(\cdots)\, {h'_n}^5\, \hat{z}^3+\cdots+ (\cdots)\, h'_n \, \hat{z}^7+o(\hat{z}^8), 
\label{eq:su6su2D}
\ees
where $R_{12+n}$ and $\hat{S}_{16}$ are degree $12+n$ and $16$ irreducible polynomials. 
$\hat{K}_{4+2n}$ is the one given in \eqref{eq:su4su2su2LK} with $s_4$ being set to zero.

Charged matter spectrum is given as follows:
\beq
\begin{array}{l|c|c@{\hspace{0.2cm}}c@{\hspace{0.2cm}}c|c||c} \hline 
\mbox{Zero}  & \mbox{Degree} & {\small \mbox{ord$(\fres)$}}
                                             & {\small \mbox{ord$(\gres)$}}
                                                   & {\small \mbox{ord$(\Dres)$}} & \mbox{Enhancement}  & \mbox{Matter} \\  \hline
r_{n-4}           & n-4     & 3   &  4   &  8    & \hspace{0.0cm} A_5 \rightarrow E_6                        &  \frac{1}{2}(\bf{20},\bf{1})  \\ 
q_{6}             & 6       & 2   &  3   &  8    & \hspace{0.0cm} A_5 \rightarrow D_6                        &   (\bf{15},\bf{1}) \\ 
R_{12+n}          & 12+n    & 0   &  0   &  7    & \hspace{0.0cm} A_5 \rightarrow A_6                        &   (\bf{6},\bf{1}) \\   \hline
\hat{K}_{4+2n}    & 4+2n    & 1   &  2   &  3    & \hspace{0.0cm} A_1 \rightarrow A_1 (I_2\rightarrow III)   &  \mbox{none} \\ 
\hat{S}_{16}      & 16      & 0   &  0   &  3    & \hspace{0.0cm} A_1 \rightarrow A_2                        &  (\bf{1},\bf{2}) \\     \hline
h'_n              &   n     & 0   &  0   &  8    & \hspace{0.0cm} A_5\oplus A_1 \rightarrow A_7              &  (\bf{6},\bf{2}) \\ \hline
\end{array}   
\label{eq:su6su2charge}
\eeq
In the first line, at $r_{n-4}=0$, enhancement $SU(6)\rightarrow E_6$ occurs.
The corresponding maximal embedding is $E_6 \supset SU(6)\times SU(2)$ and the branching is given by
\beq
 \bf{78}=(\bf{35},\bf{1})\oplus(\bf{3},\bf{1})\oplus(\bf{20},\bf{2}).
\eeq
Therefore a half-hypermultiplet $\frac{1}{2}\bf{20}$ appears at this point.

\clearpage

\end{document}